# ALUMINUM-, CALCIUM- AND TITANIUM-RICH OXIDE STARDUST IN ORDINARY CHONDRITE METEORITES

Larry R. Nittler[1], Conel M. O'D. Alexander[1], Roberto Gallino[2,3], Peter Hoppe[4], Ann N. Nguyen[1], Frank J. Stadermann[5], Ernst K. Zinner[5]

[1]Department of Terrestrial Magnetism, Carnegie Institution of Washington, 5241 Broad Branch Road, NW Washington DC, 20015, USA

[2]Dipartimento di Fisica Generale, Università di Torino, Via P. Giuria 1, I-10125 Torino, Italy

[3]Centre for Stellar and Planetary Astrophysics, Monash University, PO Box 28M, Clayton, VIC 3800 Australia

[4]Max-Planck-Institut für Chemie, Abteilung Partikelchemie, P.O. Box 3060, D-55020 Mainz, Germany

[5]Laboratory for Space Sciences and the Physics Department, Washington University, One Brookings Drive, St. Louis, MO 63130, USA

We report O-, Al-Mg, K, Ca, and Ti isotopic data for a total of 96 presolar oxide grains found in residues of several unequilibrated ordinary chondrite meteorites. Identified grain types include $Al_2O_3$, $MgAl_2O_4$, hibonite ($CaAl_{12}O_{19}$) and Ti oxide. This work greatly increases the presolar hibonite database, and is the first report of presolar Ti oxide. O-isotopic compositions of the grains span previously observed ranges and indicate an origin in red giant and asymptotic giant branch (AGB) stars of low mass (<2.5 $M_{\odot}$) for most grains. Cool bottom processing in the parent AGB stars is required to explain isotopic compositions of many grains. Potassium-41 enrichments in hibonite grains are attributable to *in situ* decay of now-extinct $^{41}$Ca. Inferred initial $^{41}$Ca/$^{40}$Ca ratios are in good agreement with model predictions for low-mass AGB star envelopes, provided that ionization suppresses $^{41}$Ca decay. Stable Mg and Ca isotopic ratios of most of the hibonite grains reflect primarily the initial compositions of the parent stars and are generally consistent with expectations for Galactic chemical evolution, but require some local interstellar chemical inhomogeneity. Very high $^{17}$O/$^{16}$O or $^{25}$Mg/$^{24}$Mg ratios suggest an origin for some grains in binary star systems where mass transfer from an evolved companion has altered the parent star compositions. A supernova origin for the hitherto enigmatic $^{18}$O-rich Group 4 grains is strongly supported by multi-element isotopic data for two grains. The Group 4 data are consistent with an origin in a single supernova in which variable amounts of material from the deep $^{16}$O-rich interior mixed with a unique end-member mixture of the outer layers. The Ti oxide grains primarily formed in low-mass AGB stars. They are smaller and rarer than presolar $Al_2O_3$, reflecting the lower abundance of Ti than Al in AGB envelopes.

## 1. Introduction

Presolar grains of stardust are identified in meteorites, interplanetary dust particles and cometary dust on the basis of highly unusual isotopic compositions (Clayton & Nittler 2004; McKeegan et al. 2006; Messenger et al. 2003; Zinner 2003, 2007). These compositions point to nuclear processes in stars and indicate that the grains condensed in the outflows and ejecta of evolved stars and supernovae. The isotopic and elemental compositions of the grains as well as their microstructures provide a great deal of information on astrophysical processes, including Galactic chemical evolution, stellar evolution and nucleosynthesis, circumstellar dust formation processes, interstellar dust processing and solar system processes. Presolar stardust studies can be consideed a new sort of observational astronomy, carried out in microanalytical laboratories.

Since the discoveries of presolar diamond, SiC and graphite in the late 1980s (Amari et al. 1990; Bernatowicz et al. 1987; Lewis et al. 1987), much of the focus of presolar grain research has been on these carbonaceous phases, largely because techniques to isolate them are well developed and there is not a large background in primitive meteorites of these minerals that formed in the Solar System. O-rich stardust, in contrast, has been more difficult to isolate and identify, mainly due to the fact that most of the mass of primitive meteorites is in the form of silicate and oxide minerals that formed in the early Solar System. However, O-rich presolar grains are of keen scientific interest since most present-day interstellar dust is believed to be in the form of silicates and both oxides and silicates are believed to form in O-rich evolved stars and supernovae. Following the first discovery of a presolar $Al_2O_3$ grain in the Orgueil meteorite (Hutcheon et al. 1994), automated techniques played a key role in the identification of significant numbers of presolar oxides (Choi, Huss, & Wasserburg 1998; Choi, Wasserburg, & Huss 1999; Nittler et al. 1994; Nittler et al. 1997). The vast majority of grains found in these studies were alumina ($Al_2O_3$), with only a very small number of spinel ($MgAl_2O_4$) and hibonite ($CaAl_{12}O_{19}$) grains. Note that presolar alumina grains often have been referred to as "corundum" as this is the stable mineral form of $Al_2O_3$ known on Earth. However, microstructural studies indicate that presolar $Al_2O_3$ exists in a variety of forms (Stroud et al. 2004; Stroud et al. 2007) not just as corundum. More recently, the development of the high-sensitivity Cameca NanoSIMS and ims-1270 ion microprobes has led to the discovery of many more presolar spinel grains (Zinner et al. 2003) and, more significantly, the discovery and characterization of presolar silicate minerals (Floss et al. 2006; Messenger et al. 2003; Mostefaoui & Hoppe 2004; Nagashima, Krot, & Yurimoto 2004; Nguyen et al. 2007; Nguyen & Zinner 2004; Vollmer et al. 2007). The latter discoveries are significant both because of the dominance of silicates in dusty astrophysical environments, and because they have shown that presolar grains can be identified *in situ* in extraterrestrial materials without the harsh chemical and physical techniques used to concentrate phases like SiC, graphite and $Al_2O_3$.

Comparison of the isotopic data for these grains with observations and models of dust-producing stars has led to the conclusion that most grains formed in low-mass red giant stars and asymptotic giant branch stars, with a small fraction having formed in supernova ejecta (Choi et al. 1998; Nittler et al. 1997; Nittler et al. 1998). Their compositions have lent strong support to the idea that an extra mixing process, not predicted by standard stellar evolution models, occurs in some low-mass AGB stars (Wasserburg, Boothroyd, & Sackmann 1995a), their microstructures have provided information on dust formation in AGB stars (Stroud et al. 2004) and their compositions have also been used to constrain the age of the Galaxy in a new way (Nittler & Cowsik 1997). Despite the useful astrophysical information presolar oxides have provided, there remain important unsolved problems (e.g., the origin of grains with $^{18}O$ enrichments) and the number of grains for which multiple elements have been analyzed for their isotopic compositions is still limited.

We report here isotopic data for presolar oxide stardust grains from acid-resistant residues of several unequilibrated ordinary chondrites (UOCs). We focused on UOCs for this study because previous work has indicated relatively lower abundances of solar-system-derived refractory oxides in these meteorites relative to primitive carbonaceous chondrites. The grains were identified by an automated particle isotopic measurement system previously reported by Nittler & Alexander (2003). This work greatly expands the previously limited isotopic database of presolar hibonite grains, and reports the first discovery of presolar titanium oxide. Preliminary reports of these data have been given at a number of conferences (Nittler & Alexander 1999; Nittler et al. 2005; Nittler, Alexander, & Tera 2001). Moreover, some isotopic and structural data for a subset of $Al_2O_3$ and $MgAl_2O_4$ grains reported here have been published previously (Stroud et al. 2004; Zinner et al. 2005). Following descriptions of experimental techniques and results, we discuss in detail many astrophysical implications of the data.

## 2. Experimental

We report presolar oxide grains identified in four separate unequilibrated ordinary chondrite acid residues prepared at different times. The first residue ("OC") was prepared by combining small HF/HCl residues of Semarkona (LL3.0), Tieschitz (H/L3.6) Bishunpur (LL3.1) and Krymka (LL3.1) prepared in an earlier study (Alexander 1993) and treating the combined residue in hot perchloric acid to destroy chromite (Nittler & Alexander 1999). For the second residue (Nittler et al. 2001), the aqueous fluoride salt technique of Cody, Alexander & Tera (2002) was used to dissolve a 1 gram sample of Tieschitz. The sample was first treated with $NH_4F$, followed by perchloric acid. The aqueous fluoride salt (in this case, CsF) technique was also used to prepare the third residue, of an 8 gram sample of Krymka (Nittler et al. 2005) and the fourth ("UOC"), a mixed residue of the primitive Antarctic chondrites QUE 97008, WSG 95300 and MET 00452. For all samples, centrifugation was used to produce size separates of nominal size 1-5 μm which were deposited in a suspension of isopropanol/water onto sputter-cleaned Au foils. Scanning electron microscopy (SEM) analysis indicated that all mounts consisted primarily of $Al_2O_3$, $MgAl_2O_4$, and SiC, with smaller amounts of other oxides, especially hibonite and $TiO_2$.

A fully automated isotopic analysis ("mapping") system was used to analyze grains on the mounts for their O-isotopic composition. The system has been described in detail elsewhere (Nittler & Alexander 2003), so we will only provide a brief summary here. The system, implemented on a Cameca ims-6f ion microprobe, uses a scanning ion image of an area of a mount to locate individual grains. The primary ion beam is then deflected to each grain in turn and an isotopic measurement is performed. To minimize sample consumption, each grain measurement is monitored and stopped if (i) a preset statistical precision is reached, (ii) the grain is isotopically anomalous to a preset level of uncertainty, or (iii) the secondary ion signal decays too rapidly indicating the grain is being sputtered away. After all grains in an area have been analyzed, the sample stage is moved and the procedure is repeated on a new area. Because the searches described here took place over several years and many improvements were made to the system in that time, analytical conditions for the different samples were not strictly identical. However, for all of the samples a 30-50 pA $Cs^+$ beam was used, and oxide grains were identified in $^{16}O^-$ secondary ion images and measured for $^{16}O^-$, $^{17}O^-$, $^{18}O^-$ and $^{27}Al^{16}O^-$ ions. Mass-resolving powers of ~4,500 to 6,000 were used to eliminate isobaric interferences, especially $^{16}OH^-$ from $^{17}O^-$. The typical image raster size was 100×100 μm$^2$, though some measurements were done with 75×75 μm$^2$ and 120×120 μm$^2$ fields.

As shown by Nittler & Alexander (2003, see their Figure 5), the estimated intrinsic reproducibility of $^{17}O/^{16}O$ and $^{18}O/^{16}O$ ratios measured with the mapping system is 1-1.5%, sufficient to identify isotopically anomalous presolar grains. Of course, the uncertainty of many measurements, especially of the smallest grains, is dominated by counting statistics and the precision can be worse than the value quoted above. To take into account the dependence of precision on count rate (and thus grain size) in order to accurately identify grains with statistically significant isotopic anomalies, we calculated the uncertainty as a function of count rate for each set of data acquired in a single session of isotopic mapping. The procedure is illustrated in Figure 1. For a given data set, results were binned according to total ion counts of the rare isotopes and the standard deviation of isotopic ratios ($^{17}O/^{16}O$ or $^{18}O/^{16}O$) was determined for each bin. Obvious outliers were excluded from this calculation. A simple functional fit of the resulting standard deviations of the different bins then provided a measure of uncertainty as a function of the number of counted ions. In general, the total uncertainty described by this function is close to the error expected from counting statistics combined with an additional uncertainty of 1-1.5% due to non-statistical effects like sample topography, matrix effects, etc. Any grain that lies more than 1$\sigma$ (according to the grain's internal error) outside the calculated $\pm 3\sigma$ limits (solid curves on Fig. 1) is then identified as a presolar grain (open symbols on Fig. 1). Note that although we have plotted the $^{16}O$ count rate on Figure 1 in order to have a common scale for all plots, the calculations themselves were done in terms of the total counts of $^{17}O$ or $^{18}O$, as appropriate for calculating counting-statistical errors on $^{17}O/^{16}O$ or $^{18}O/^{16}O$, respectively.

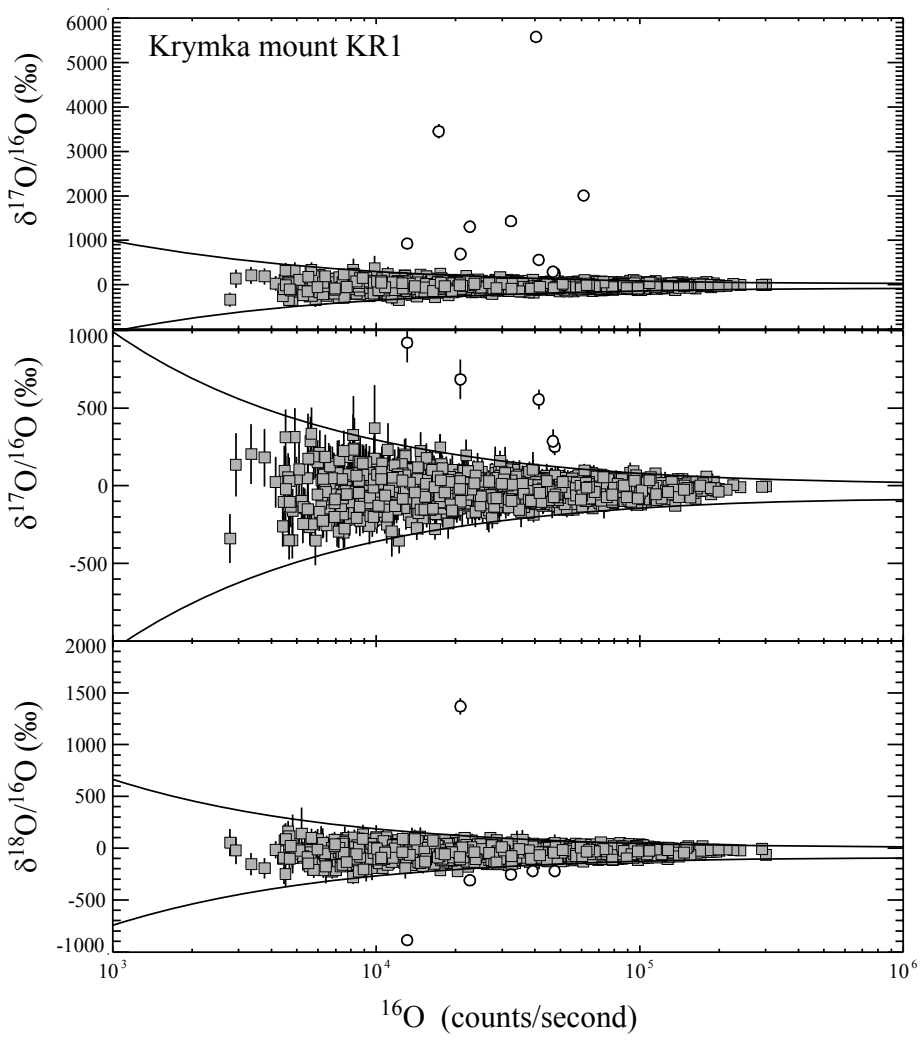

**Figure 1:** Oxygen isotopic ratios plotted against $^{16}O$ count rate for 1717 μm-sized oxide grains from the Krymka meteorite (sample mount KR1). Isotopic ratios are expressed as δ-values, the ‰ deviations from a standard value: $\delta R=(R_m/R_o-1) \times 10^3$ where $R_m$ is a measured isotopic ratio and $R_o$ is a standard isotopic ratio. Standard ratios for O isotopes are given in Table 2. Middle panel is expanded view of top panel with reduced vertical scale. Solid curves are ±3-σ range of data as a function of count rate, derived from the data. Measurements that lie outside these ranges are inferred to be presolar grains (open symbols). Error bars in this and subsequent figures are 1σ.

Once identified as presolar, most grains were re-located and examined by scanning electron microscopy to determine their mineralogy and size. Over the course of the work, three SEMs were used: A JEOL 840A at the Smithsonian Institution, a LEO 1540 at the Naval Research Laboratory and a JEOL 6500F at the Carnegie Institution, all equipped with energy dispersive x-ray spectrometers for chemical analysis.

SEM analysis indicated that some of the `grains' with significant O-isotopic anomalies consisted in fact of two or more individual grains sitting adjacent to each other which were not resolved at the spatial resolution of the ims-6f. In these cases, the presolar grain was so anomalous that even dilution from the neighboring grain(s) did not erase the isotopic effects. Some of these samples were re-measured by isotopic imaging in Cameca NanoSIMS ion microprobes at the Max Planck Institute for Chemistry in Mainz, Washington University and the Carnegie Institution. For these measurements, a ~100 nm $Cs^+$ beam was rastered over the grains with simultaneous detection of $^{16}O^-$, $^{17}O^-$, $^{18}O^-$, and $^{27}Al^{16}O^-$ secondary ions. For some grains, $^{24}Mg^{16}O^-$ was also measured. Spinel grains on the sample mounts were used as isotopic standards and assumed to have the typical chondritic spinel composition of $\delta^{17}O/^{16}O = \delta^{18}O/^{16}O = -40$‰ (Clayton 1993); see Figure 1 for the definition of δ-values. Isotopic ratios for individual grains were determined from the ion images by use of custom software; in each case it was obvious which of the analyzed grains was presolar.

Following O-isotopic analysis, many of the identified presolar oxide grains were subsequently analyzed for their isotopic compositions of other elements. Several of the OC and Tieschitz grains were analyzed with the Carnegie ims-6f for Mg-Al isotopic compositions. For these measurements, a 0.1 – 0.2 nA $O^-$ beam was used to produce secondary ions of $^{24}Mg^+$, $^{25}Mg^+$, $^{26}Mg^+$, and $^{27}Al^+$ and the mass resolution was sufficient to resolve important isobaric interferences. A Burma spinel standard was used to correct for instrumental mass fractionation and to determine the relative sensitivity factor, Γ, reflecting the different secondary ion yields of Mg and Al. Large excesses in $^{26}Mg$ observed in many grains are attributed to *in situ* decay of extinct $^{26}Al$ ($T_{1/2}$=735,000 years). The initial $^{26}Al/^{27}Al$ ratios were determined from the measured count rates by use of the equation: $^{26}Al/^{27}Al = (^{26}Mg_{meas} - ^{24}Mg_{meas} \times std)/(^{27}Al_{meas} \times \Gamma)$, where "std" is the terrestrial $^{26}Mg/^{24}Mg$ ratio (0.13932, Catanzaro et al. 1966). The value of Γ was variable from measurement session to session, but was generally between 1 and 2.

Additional isotopic measurements were made with NanoSIMS ion probes, with a ~200-400 nm $O^-$ primary ion beam. The isotopic compositions of K and Ca were determined for Krymka presolar hibonite grains with the Washington University NanoSIMS; the combined peak-jumping/multicollection mode of data acquisition was used (Stadermann et al. 2005). The magnetic field was repeatedly cycled over two settings. In the first step, secondary ions of $^{27}Al^+$, $^{39}K^+$, $^{41}K^+$, $^{43}Ca^+$ and mass 48 (including both $^{48}Ca^+$ and $^{48}Ti^+$) were measured on the five NanoSIMS 50 electron multipliers. We did not attempt to measure $^{46}Ca$ due to its low abundance (terrestrial $^{46}Ca/^{40}Ca=3.2 \times 10^{-5}$) and potential interference from $^{46}Ti$. In the second step, $^{40}Ca^+$, $^{42}Ca^+$ and $^{44}Ca^+$ were analyzed on the middle three detectors. Terrestrial amazonite ($KAlSi_3O_8$) and perovskite ($CaTiO_3$) were used as standards for K and Ca isotopes, respectively. Large excesses in $^{41}K$ observed in several grains are attributed to *in situ* decay of extinct $^{41}Ca$ ($T_{1/2}$=105,000 years). An equation analogous to that given above for $^{26}Al$ was used to determine the initial $^{41}Ca/^{40}Ca$ ratios. We took the terrestrial $^{41}K/^{39}K$ ratio to be 0.07217 (Garner et al. 1975) and the relative sensitivity factor for Ca and K to be 3.0 (Hinton 1990).

Mg isotopes for presolar spinel grains from the "OC" sample, measured with the Mainz NanoSIMS, have been reported previously (Zinner et al. 2005). We report additional NanoSIMS Mg and Al data for several presolar $Al_2O_3$ and hibonite grains from Krymka and spinel and hibonite grains from the mixed "UOC" residue. For the Krymka grains, measured in St. Louis, secondary ions of $^{40}Ca^{++}$, $^{24}Mg^+$, $^{25}Mg^+$, $^{26}Mg^+$, and $^{27}Al^+$ were detected in multicollection mode. For the UOC grains, the Carnegie NanoSIMS was used in imaging mode and only the Mg and Al isotopes were measured. For these measurement series, grains of Burma spinel as well as Mg-rich non-presolar grains from the sample mounts were used as isotopic standards and to determine the Mg/Al relative sensitivity factor. Initial $^{26}Al/^{27}Al$ ratios were determined for grains with large $^{26}Mg$ excesses as described above.

We report Ti isotopic data for three presolar Ti oxide grains, obtained with three different instruments. Grain OC13 was measured with the Carnegie ims-6f ion probe. A ~2 μm $O^-$ beam was focused on the grain and positive secondary ions of $^{27}Al$, $^{46}Ti$, $^{47}Ti$, $^{48}Ti$, $^{49}Ti$ , $^{50}Ti$ and $^{52}Cr$ were counted. A mass-resolving power of ~5,000 was used to resolve $^{48}TiH$ from $^{49}Ti$. Micron-sized grains of synthetic TiC were used to correct for instrumental mass fractionation. Grain KT-2 was measured with the Washington University NanoSIMS 50 in the combined peak-jumping/multicollection mode. Synthetic $TiO_2$ was used as a standard. Three magnetic field steps were used. In the first step, $^{46}Ti^+$, $^{48}Ti^+$ and $^{50}Ti^+$ were measured on the three highest-mass detectors, in the second $^{47}Ti^+$, $^{49}Ti^+$ and $^{51}V^+$ were measured on the same detectors, and in the final step we measured $^{27}Al^+$, $^{40}Ca^+$, $^{48}Ti^+$, $^{50}Ti^+$ and $^{52}Cr^+$ on the five detectors. Titanium measurements were not possible in hibonite grains because of the unresolved interference of $^{48}Ca$ with $^{48}Ti$. Grain UOC-T1 was measured with the Carnegie NanoSIMS 50L. This instrument has a larger magnet than the NanoSIMS 50 and two additional detectors. These features allow the simultaneous collection of all Ti isotopes and two additional species without having to switch between different magnetic field settings. Here, positive secondary ions of $^{27}Al$, $^{46,47,48,49,50}Ti$ and $^{52}Cr$ were simultaneously acquired. Grains of synthetic TiC were used as an isotopic standard and gave reproducibility of about 5 ‰ for the Ti isotopic ratios. For all three grains, possible $^{50}Cr$ interference on $^{50}Ti$ was corrected for by assuming a terrestrial $^{50}Cr/^{52}Cr$ ratio, but this correction was very small (<0.5%). Vanadium and calcium interferences were completely negligible for KT-2 and this was likely the case for the other two grains as well.

## 3. Results

### *3.1. Identified Presolar Grains*

The numbers of grains analyzed along with the numbers and types of presolar grains identified in each sample are summarized in Table 1. Note that clumping of grains often occurs on mounts and in many cases the automated system actually analyzed multiple grains instead of single grains. In such cases, an anomalous isotopic signature from a presolar grain can be diluted by the signal from adjacent grains, such that the grain is not identified as presolar. Thus, the fraction of grains identified as presolar on each mount by automated mapping is a lower limit. Because all of the identified grains were significantly sputtered by the ims-6f $Cs^+$ beam during the isotopic mapping, it is difficult to assess the original grain sizes. However, all grains probably had original sizes in the range of 0.5 – 5 μm, with most being < 1.5 μm in diameter. The distribution of presolar grain types is not the same for the different samples. For example, hibonite makes up about one third of the identified presolar grains in the Krymka sample and was identified in the UOC residue, but no presolar hibonite grains were found in Tieschitz. Spinel constitutes only a small fraction of the identified Krymka and Tieschitz presolar oxides, but comprises about half of the OC and UOC grains. A few grains were completely sputtered away during the O-isotopic measurements and their mineralogy thus could not be determined; these are listed as "Unknown" in Tables 1 and 2.

## 3.2. Isotopic Data

Isotopic data for the presolar oxide grains are given in Tables 2-4. Oxygen isotopic ratios are based on ims-6f measurements, except in a few cases (noted) where NanoSIMS measurements were used to resolve neighboring grains. The O and Mg isotopic data for the "OC" spinel grains can be found in Zinner et al. (2005). We also report O and Mg data for a spinel grain, M16, from the Murray meteorite originally reported by Zinner et al. (2005). Re-examination of this grain revealed that the original measurement included material from closely adjacent grains on the sample mount, diluting to some extent the anomalous isotope signatures. Re-measurement of this grain with the Washington University NanoSIMS, following sputtering of the interfering grains, yielded the more anomalous values given in Table 2.

### *Oxygen*

The O-isotopic compositions of the new presolar oxides grains are compared to previous data for presolar oxide and silicate grains in Figure 2. Nittler et al. (1994; 1997) previously divided the presolar oxide O data into four groups to aid in discussions. These groups are indicated on Figure 2, but it should be noted that they are somewhat arbitrarily defined and the groups certainly merge into one another. It is thus difficult in many cases to unambiguously associate a given grain with a specific group, but as discussed in later sections, the groupings do reflect differences in the properties of the parent stars. As seen in previous work, the presolar oxide distribution is dominated by Group 1 grains, with $^{17}O$ enrichments, relative to solar, and solar to sub-solar $^{18}O/^{16}O$ ratios. The smaller populations of Group 2 and Group 3 grains are characterized by strong $^{18}O$ depletions and $^{16}O$ enrichments, respectively, while the Group 4 grains are enriched in both $^{17}O$ (typically) and $^{18}O$.

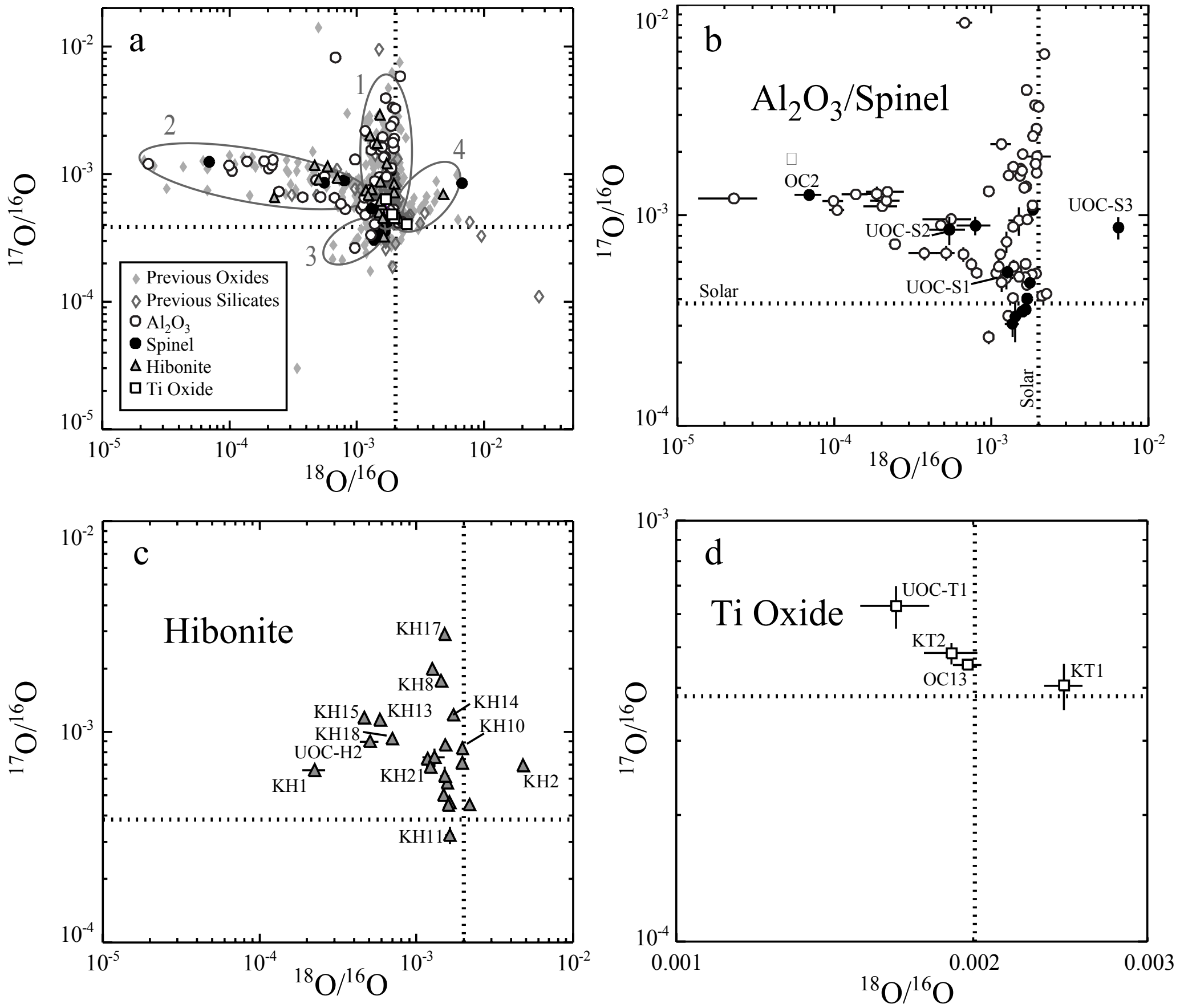

**Figure 2:** O-isotopic ratios of presolar oxides found in this study and of previous presolar oxide and silicate grains. Dotted lines indicate the terrestrial (assumed to be solar) O- isotopic ratios. Indicated on panel (a) are the grain Groups defined by Nittler et al. (1994, 1997). Previous data are from (Choi et al. 1998; Choi et al. 1999; Floss et al. 2006; Messenger et al. 2005; Messenger et al. 2003; Mostefaoui & Hoppe 2004; Nguyen et al. 2007; Nguyen & Zinner 2004; Nittler et al. 1994; Nittler et al. 1997; Nittler et al. 1998; Zinner et al. 2003).

***Magnesium-Aluminum***

Magnesium isotopic data for the "OC" spinel grains have previously been reported by Zinner et al. (2005) and discussed by Lugaro et al. (2007). Here, we report additional Mg and Al data for 28 $Al_2O_3$, 18 hibonite, and 3 spinel grains (Table 2). The Mg isotopic ratios are plotted versus the $^{27}Al/^{24}Mg$ ratios in Figure 3. A wide range of $^{25}Mg/^{24}Mg$ ratios was observed (Fig. 3a). The hibonite grains exhibit $\delta^{25}Mg/^{24}Mg$ values ranging from –320 ‰ to +318 ‰ and a relatively narrow range of $^{27}Al/^{24}Mg$ ratios from 37 to 284, indicating that they contain significantly more Mg than do the $Al_2O_3$ grains ($^{27}Al/^{24}Mg$ from ~100 to ~20,000, with a median value of 300. Note that an Al/Mg ratio of 300 would be considered low for terrestrial or meteoritic $Al_2O_3$, but is consistent with previous observations of presolar grains, perhaps reflecting reaction of $Al_2O_3$ with gaseous Mg in a stellar outflow, e.g., Choi et al. 1998 ). The $Al_2O_3$ grains span a range of $\delta^{25}Mg/^{24}Mg$ values similar to that of the hibonite grains, but errors are larger due to lower Mg contents. The three UOC spinel grains also show a wide range of $\delta^{25}Mg/^{24}Mg$ values from –235 ‰ to +256 ‰. Large excesses in $^{26}Mg$ were observed in a majority of the measured hibonite and $Al_2O_3$ grains (Fig. 3b). These excesses can be attributed to *in situ* decay of $^{26}Al$, with inferred initial $^{26}Al/^{27}Al$ ratios ranging from $1.4 \times 10^{-6}$ to 0.08 (see solid curves on Fig. 3b). However, two of the measured hibonites and nine of the $Al_2O_3$ grains showed no excess $^{26}Mg$, indicating that they condensed without containing appreciable amounts of $^{26}Al$. This result is comparable to previous studies that found that ~30% of presolar $Al_2O_3$ grains do not have measurable $^{26}Al$ (Choi et al. 1998; Nittler et al. 1994; Nittler et al. 1997). The three UOC spinel grains also show $^{26}Mg$ excesses, but because two of these grains also show $^{25}Mg$ anomalies of comparable magnitude (Table 2) one cannot uniquely ascribe the $^{26}Mg$ excesses to $^{26}Al$ decay. For these grains, $^{26}Al/^{27}Al$ ratios were inferred by projecting the Mg isotopic composition onto a theoretical relationship between $^{25}Mg/^{24}Mg$ and $^{26}Mg/^{24}Mg$ for AGB nucleosynthesis and ascribing any remaining $^{26}Mg$ excess to $^{26}Al$. This procedure is discussed in detail by Zinner et al. (2005).

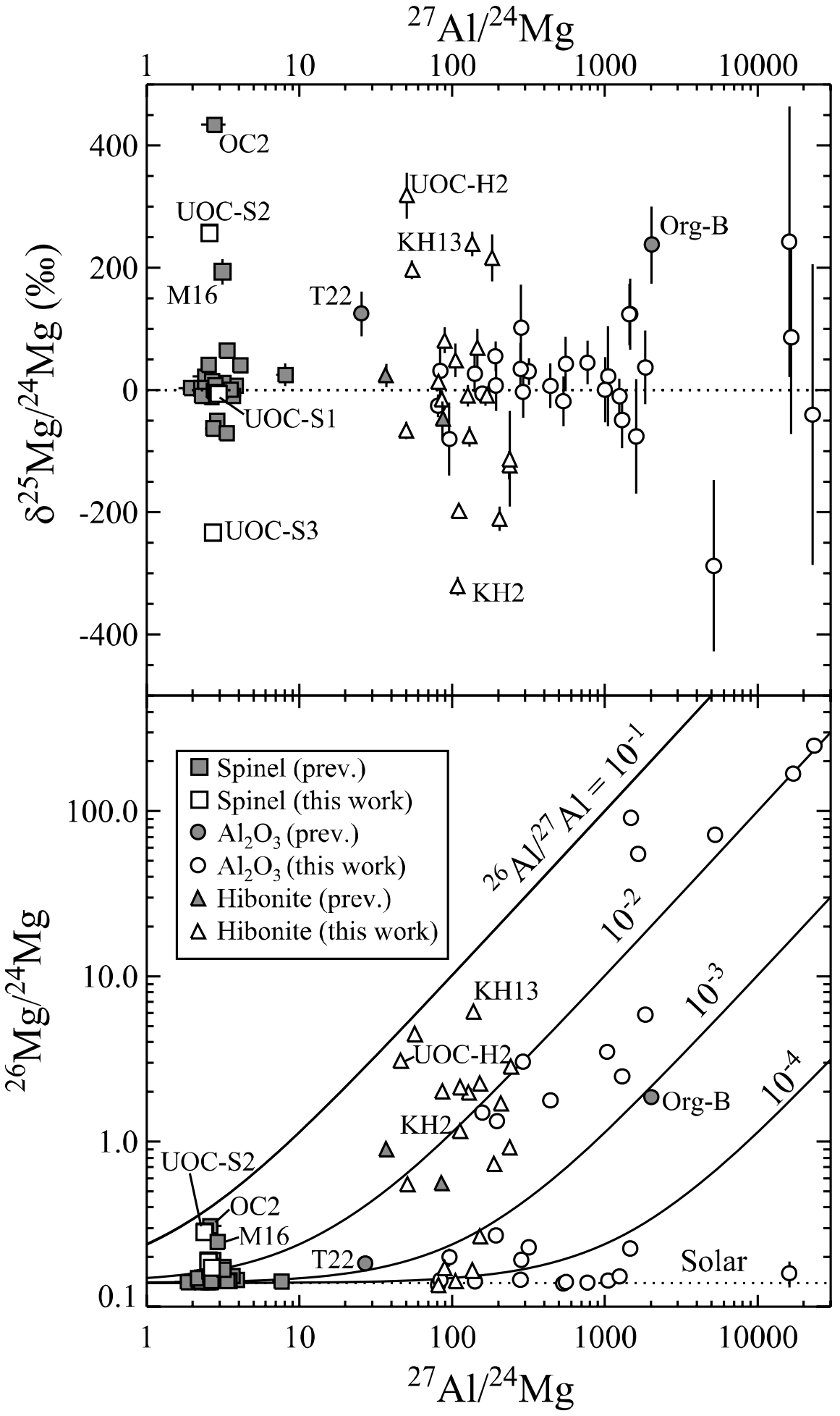

**Figure 3:** Mg isotopic ratios, expressed as δ-values (see caption to Figure 1), plotted against $^{27}Al/^{24}Mg$ ratios for presolar oxide grains. Normalizing ratios are 0.12663 and 0.13932 for $^{25}Mg/^{24}Mg$ and $^{26}Mg/^{24}Mg$, respectively (Catanzaro et al. 1966). Solid lines on lower panel indicate constant inferred initial $^{26}Al/^{27}Al$ ratios. The presolar oxide grains show a wide range of Mg isotopic compositions and inferred $^{26}Al/^{27}Al$ ratios. Previous data are from Huss et al. (1995) Nittler et al. (1997), Choi et al. (1999) and Zinner et al. (2005). The plotted values for Murray spinel grain M16 are based on a new analysis (Table 2) and are more anomalous than those reported by Zinner et al. (2005).

### *Potassium-Calcium*

Potassium and Ca isotopes were measured in sixteen presolar hibonite grains from Krymka (Table 3). Ten of the measured grains showed large excesses of $^{41}K$ (Fig. 4) attributable to *in situ* decay of $^{41}Ca$. The inferred initial $^{41}Ca/^{40}Ca$ ratios range from $1.7 \times 10^{-5}$ to $4.3 \times 10^{-4}$; this range contains the value of $1.5 \times 10^{-4}$ reported for one presolar hibonite grain by Choi et al. (1999). Ca/K ratios were higher in the Krymka hibonites than in the grains reported by Choi et al. (1999), perhaps reflecting lower K contamination levels from sample substrates when using the much smaller ion beam of the NanoSIMS compared to the previous ims-3f measurements. Figure 4 suggests a detection limit for $^{41}Ca/^{40}Ca$ of about $10^{-6}$ for µm-sized presolar hibonite grains.

Eight of the measured hibonite grains have significant (>2σ) isotopic anomalies in one or more ratios of stable Ca isotopes. Determination of $^{48}Ca/^{40}Ca$ ratios was not possible due to the presence of Ti in the grains. These grains show a wide variety of Ca-isotopic patterns, as shown in Figure 5. Grains KH11, KH12 and KH14 all show enrichments in $^{42}Ca$ and $^{43}Ca$, with smaller excesses or deficits of $^{44}Ca$. This pattern is qualitatively similar to predictions for the envelope of AGB stars, where neutron capture enhances the abundances of these three Ca isotopes (see, e.g., the last panel of Fig. 5, discussed below). Three grains, KH2, KH8 and KH13 show $^{44}Ca$ excesses that are larger than effects in the other stable Ca isotopes. Grain KH10 has larger excesses in $^{42}Ca$ and $^{43}Ca$ than observed in the other grains, but a solar $^{44}Ca/^{40}Ca$ ratio. KH17 shows a $^{42}Ca$ excess similar to that previously seen in a presolar hibonite grain by Choi et al. (1999).

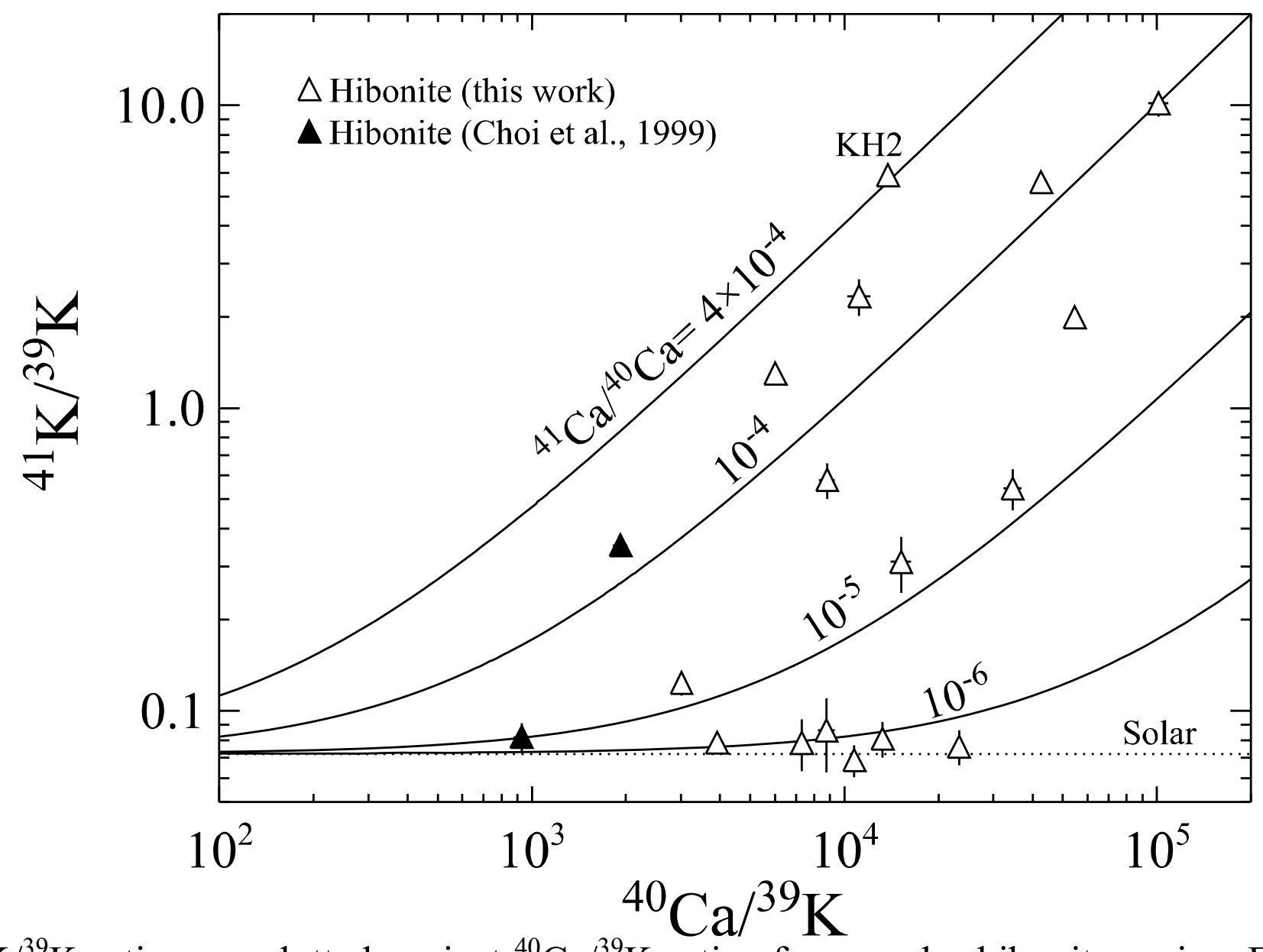

**Figure 4:** The $^{41}K/^{39}K$ ratios are plotted against $^{40}Ca/^{39}K$ ratios for presolar hibonite grains. Excesses in $^{41}K$ are attributed to in situ decay of $^{41}Ca$. Solid lines indicate constant inferred initial $^{41}Ca/^{40}Ca$ ratios. The solar $^{41}K/^{39}K$ ratio of 0.07217 (Garner et al. 1975) is indicated.

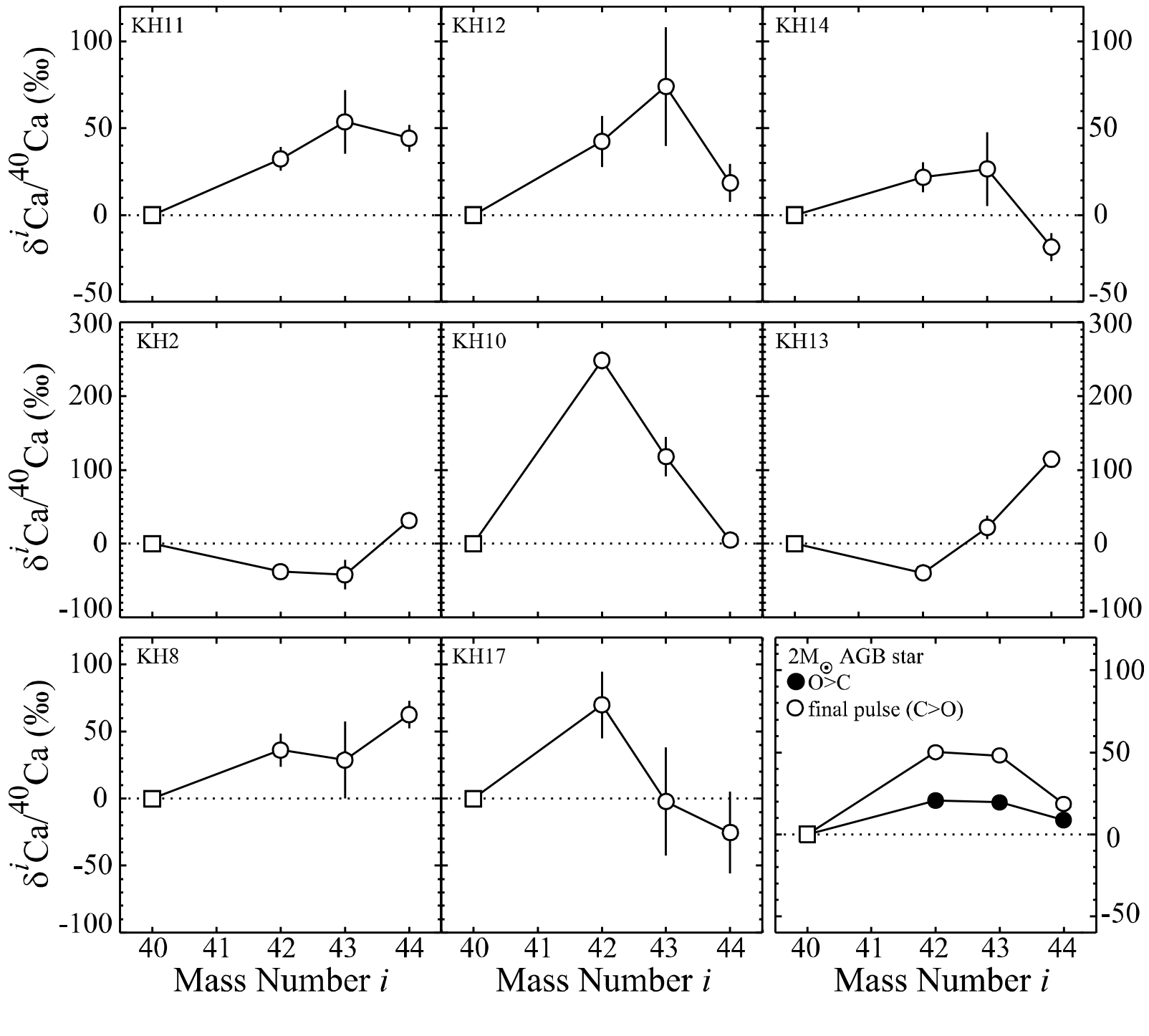

**Figure 5:** Patterns of stable Ca-isotopic ratios (expressed as δ-values, see caption for Figure 1) for 8 presolar hibonite grains with at least one isotopic ratio being more than 2σ away from solar. The final panel shows the isotopic pattern expected for the envelope of a $2M_\odot$ AGB star of solar metallicity after third dredge-up; solid symbols indicate the composition at the last thermal pulse while the stellar envelope is still O-rich and the open symbols indicate the final composition of the envelope.

### *Titanium*

Figure 6 shows Ti isotopic patterns for three of the four identified presolar titanium oxide grains; the fourth (KT-1) was completely sputtered away before it could be measured. All three grains show Ti isotopic ratios close to the solar values, with a few marginal (~2σ) anomalies.

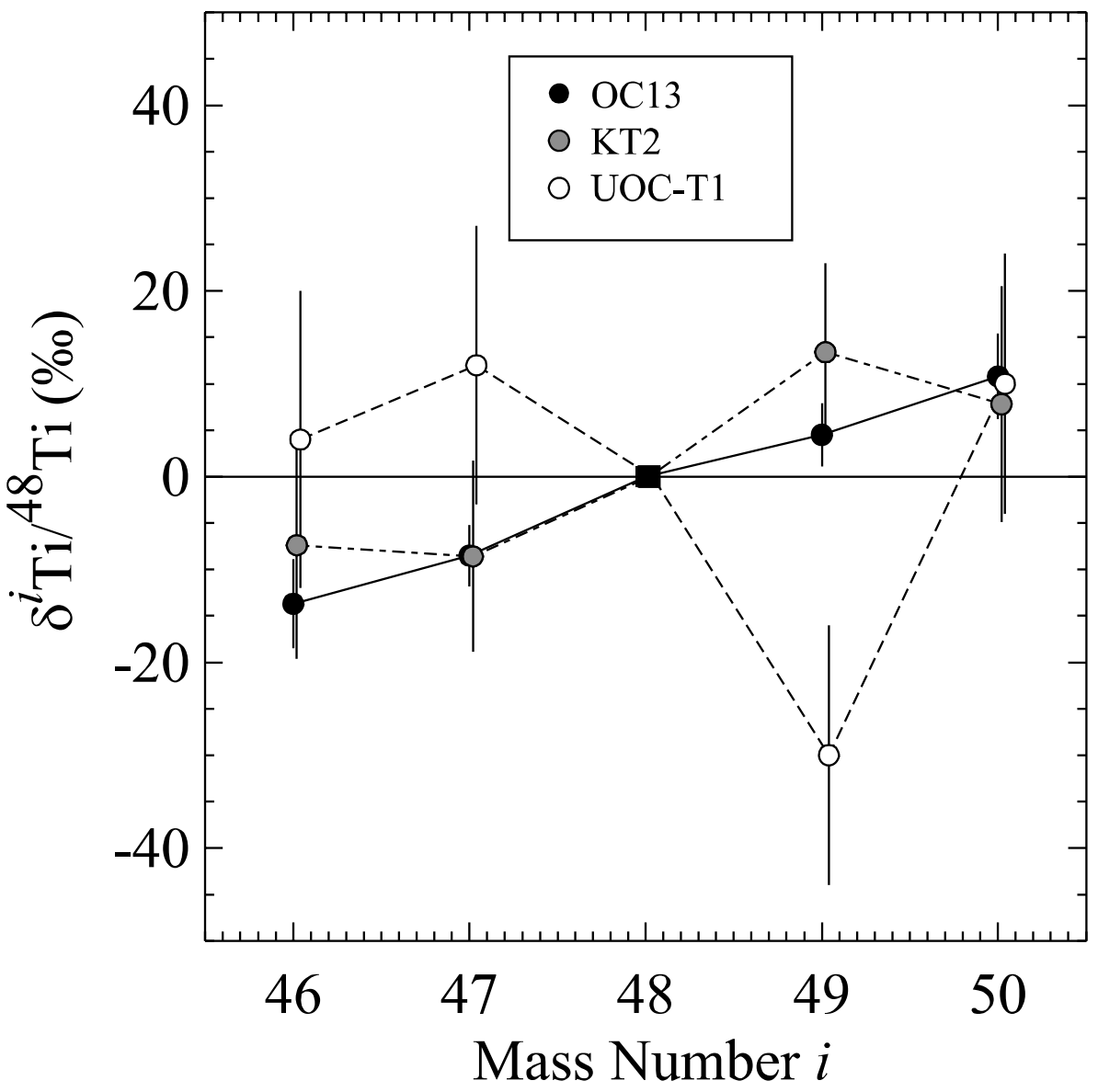

**Figure 6:** Ti isotopic compositions (expressed as δ-values, see caption to Figure 1) of three presolar titanium oxide grains. Normalizing ratios are given in Table 3. All measured isotopic ratios are close to solar, though OC13 has small depletions in $^{46}Ti$ and $^{47}Ti$, and UOC-T1 has a $^{49}Ti$ depletion.

## 4. Grains from Red Giant Branch and Asymptotic Giant Branch Stars

Most presolar oxide grains (and silicates), those belonging to Groups 1-3 (Fig. 2), have been argued to originate in low-mass red giants and asymptotic giant branch (AGB) stars (Huss et al. 1994; Nittler et al. 1994; Nittler et al. 1997; Wasserburg et al. 1995a). The surface compositions of these stars depend both on their initial compositions and the nucleosynthetic and mixing processes that occur as they evolve. We thus review the processes that affect these compositions as a prelude to discussing the grain data in detail.

### *4.1. Evolution of Low- and Intermediate-mass Stars*

The initial isotopic and elemental composition of every star is determined by the chemical evolution of the Galaxy (GCE). GCE describes the process by which the elemental and isotopic composition of a galaxy varies in time and place according to the nuclear history of the material present (Nittler & Dauphas 2006; Pagel 1997). As succeeding generations of stars are born and die, they return freshly synthesized nuclei to the interstellar medium. In this way, the abundances of the heavy elements have increased throughout the history of the Galaxy, and stars formed at different times (and places) have, on average, different chemical and isotopic compositions. For example, the fraction of elements heavier than He (metallicity) has increased on average over time. In terms of isotopes, one can make a rough distinction between primary ones – those whose nucleosynthesis is independent of metallicity – and secondary ones – those that require pre-existing heavy elements in the star for their production. The ratio of a secondary isotope to a primary isotope is expected to increase roughly linearly with metallicity (Clayton 1988; Timmes & Clayton 1996). In general, for the elements discussed here, the major isotopes (e.g., $^{16}O$, $^{24}Mg$, $^{40}Ca$) are considered to be primary and the minor isotopes (e.g., $^{17}O$, $^{25}Mg$, etc.) secondary, such that stars of lower metallicity are expected to have lower initial $^{17}O/^{16}O$, $^{25}Mg/^{24}Mg$, etc., ratios than stars of higher metallicity. Moreover, as discussed further below, "α" isotopes (those whose nuclei consist of integer multiples of α particles) like $^{16}O$, $^{24}Mg$, etc. are observed to be overabundant relative to iron at low metallicities, due to their primary production in short-lived massive stars ending as Type II supernovae, whereas some one third of solar iron is made by longer-lived binary stars ending as Type Ia supernovae. Note that there is also likely to be local chemical heterogeneity in the interstellar medium due to the stochastic nature of star formation and mixing processes in the Galaxy (Lugaro et al. 1999; Nittler 2005). Thus, there is certainly scatter around average trends of isotopic ratios with metallicity. We will refer to this process as inhomogeneous GCE.

When a low- or intermediate-mass star ($<8M_\odot$) runs out of H fuel in its core, it expands and cools, becoming a red giant. Shortly after ascending the red giant branch (RGB) on a Hertzsprung-Russell (H-R) diagram, the star undergoes an episode of deep convection, the first dredge-up. The first dredge-up mixes the ashes of partial H burning from deep layers of the star into the envelope, changing its composition. Because H burning by the CNO cycles enriches $^{17}O$, but destroys $^{18}O$, the surface $^{17}O/^{16}O$ ratio increases and the $^{18}O/^{16}O$ ratio decreases, relative to the initial composition of the star (Boothroyd & Sackmann 1999; Boothroyd, Sackmann, & Wasserburg 1994; Dearborn 1992). Detailed calculations indicate that the final $^{17}O/^{16}O$ ratio depends sensitively on the initial mass of the star. For stars of mass 1- $2.5M_\odot$ the predicted $^{17}O/^{16}O$ ratio increases with increasing mass up to a maximum value of ~0.004. The predicted ratio decreases again for more massive stars down to a value of ~0.0014 for stars of ~9 $M_\odot$. The post-first-dredge-up $^{18}O/^{16}O$ ratio, in contrast, does not depend very strongly on the mass of the star, and is predicted to be ~80% of the initial $^{18}O/^{16}O$ ratio. The latter is largely determined by Galactic chemical evolution and thus reflects the initial metallicity. These predictions are in reasonable agreement with O-isotopic ratios measured spectroscopically in RGB stars, though the measurement uncertainties are large (Harris & Lambert 1984b; Harris, Lambert, & Smith 1988).

Some stellar observations indicate that an additional mixing process beyond the first dredge-up occurs in low-mass (< 2.3 $M_\odot$) stars on the red giant branch (Charbonnel 1994; Denissenkov & Weiss 1996; Nollett, Busso, & Wasserburg 2003; Wasserburg et al. 1995a). For example, this process leads to lower $^{12}C/^{13}C$ ratios than predicted by standard first dredge-up models. The process is usually referred to as "extra mixing" or "cool bottom processing (CBP)." Its physical cause is unknown, though it is almost certainly related to multi-dimensional effects (Denissenkov & Tout 2000; Eggleton, Dearborn, & Lattanzio 2006). Cool bottom processing in stars on the red giant branch is not expected to significantly affect the surface O-isotopic ratios except in very low-metallicity stars (Boothroyd & Sackmann 1999; Wasserburg et al. 1995a).

Following core He burning (during which the star lies on the horizontal branch on the H-R diagram), the star expands again and ascends the asymptotic giant branch (AGB). An AGB star consists of an electron-degenerate white dwarf core, surrounded by thin shells burning He and H and a large convective H-rich envelope (Busso, Gallino, & Wasserburg 1999; Herwig 2005). Dust condenses in the strong winds through which the envelope loses mass. In stars more massive than 3-$4M_\odot$, a second dredge-up episode occurring early in the AGB phase can significantly further affect the surface O-isotopic ratios. The He- and H- burning shells of the AGB star are active in an episodic process where the He shell burns intensely for a short time (thermal pulse) driving vigorous convection in the region between the H and He shells. In the He-rich intershell region, neutron capture reactions occur producing large amounts of the so-called *s*-process heavy elements (e.g., Ba), as well as modifying the isotopic composition of lighter elements (e.g., Mg). Following a thermal pulse, the base of the envelope can reach deep into the inter-shell region, mixing the products of nucleosynthesis to the surface (third dredge-up). The third dredge-up largely brings up $^{12}C$ produced by partial He burning and the products of neutron capture nucleosynthesis, especially *s*-process isotopes. AGB stars are believed to be the major source of *s*-process elements in the Universe (Busso et al. 1999). Repeated thermal pulses and third dredge-up episodes gradually increase the C/O ratio and change the isotopic compositions of many elements in the envelope, though the third dredge-up itself has little effect on the surface O-isotopic composition. While the stellar wind is O-rich, oxides and silicates are expected to condense, whereas once the C/O ratio exceeds unity, the dust is dominated by carbonaceous phases like amorphous carbon, graphite and SiC (Lodders & Fegley 1995).

There is significant evidence, both from stellar observations and presolar grains, that cool bottom processing also occurs in low-mass AGB stars (Boothroyd & Sackmann 1999; Nollett et al. 2003; Wasserburg et al. 1995a). For example, the large $^{18}O$ depletions observed in some AGB stars and in Group 2 presolar oxide grains indicate more $^{18}$O destruction in the envelope than can be explained by standard models only involving core and shell H burning followed by first and third dredge-up, respectively. As discussed further below, the O (and $^{26}$Al) data are best explained by cool bottom processing models in which envelope material is slowly cycled through hot regions near the H shell. Similarly, the C isotopic compositions of presolar SiC grains require CBP during the AGB phase of the parent stars (Alexander & Nittler 1999; Zinner et al. 2006).

In intermediate mass (4 – 7 $M_{\odot}$) AGB stars, the base of the convective envelope reaches temperatures high enough for H-burning reactions to occur directly there, a process known as hot bottom burning (HBB, Boothroyd, Sackmann, & Ahern 1993; Boothroyd, Sackmann, & Wasserburg 1995; Cameron & Fowler 1971; Lugaro et al. 2007). Because the envelope is fully convective, it is completely cycled through the hot region so that the products of H-burning nucleosynthesis are enriched at the surface. In terms of O isotopes, HBB rapidly destroys $^{18}$O and, to a lesser degree, $^{16}$O, and produces $^{17}$O. HBB also produces $^{26}$Al by the Mg-Al chain. As discussed further below, the O-isotopic compositions of most, but not all, Group 2 grains are inconsistent with a HBB origin (Boothroyd et al. 1995; Lugaro et al. 2007).

The above discussion outlined the expectations for surface O-isotopic compositions from the various processes occurring in low-mass red giants and AGB stars. We now consider the expected effects of these processes on the surface isotopic compositions of the other elements involved in this study, namely Mg, Al, K and Ca. As discussed earlier, variations in the initial isotopic compositions of these elements in stars are expected due to the chemical evolution of the Galaxy. The first dredge-up does not reach zones that experienced temperatures high enough to allow nuclear reactions involving these elements, so it is only during the AGB phase that modifications due to nucleosynthesis are expected. Here we will consider both general processes and some new specific quantitative predictions for the envelopes of AGB stars. The models we discuss are described in some detail by Gallino et al. (1998), Straniero et al. (2003) and Zinner et al. (2005; 2006; 2007). The calculations use a postprocessing code to calculate the nucleosynthesis occurring in the He and H shells, and to mix shell material into a mass-losing envelope. Stellar parameters (e.g., envelope and dredge-up masses, temperatures, densities, etc.) are based on results from stellar evolution models using the FRANEC code (Straniero et al. 1997). Note that the stellar evolutionary calculations use the solar (and appropriately scaled solar for low metallicity) C and O abundances from Anders & Grevesse (1989), whereas the postprocessing nucleosynthesis code uses the more recent and significantly lower solar abundances of these elements from Allende Prieto, Lambert, & Asplund (2001; 2002). The main effect of using these abundances is that AGB star envelopes become C-rich after fewer thermal pulses than with the old abundances.

In this paper, we will consider detailed predictions for a limited grid of AGB stars, with masses 1.5, 2, 3 and 5$M_{\odot}$ and two metallicities, solar and one-half solar. Neutron-capture reactions in AGB stars are driven by two neutron sources: $^{13}C(\alpha,n)^{16}O$ and $^{22}Ne(\alpha,n)^{25}Mg$; the former is thought to be responsible for the production of most *s*-process heavy elements and is dependent on a free parameter in the models quantifying the amount of $^{13}$C available to produce neutrons ("$^{13}$C pocket"). However, the elements of interest in this paper are primarily affected by the $^{22}$Ne source, thus we only consider models with the standard $^{13}$C pocket that reproduces the solar *s*-process element distribution (Gallino et al. 1998). The initial isotopic compositions of the stellar models with one-half solar metallicity were assumed to scale with metallicity to reflect Galactic chemical evolution. Specifically, the abundances of the minor isotopes (e.g., $^{25}$Mg, $^{42}$Ca, etc.) were assumed to scale with Fe, while the major, "α-isotopes" ($^{24}$Mg, $^{40}$Ca) were assumed to be enhanced at lower metallicity, according to observations of disk stars (Reddy et al. 2003). These assumptions do not precisely match the isotopic ratio-metallicity dependences predicted by full GCE models (e.g., Timmes, Woosley, & Weaver 1995). For example, the assumption that isotopes like $^{25}$Mg, which are produced as secondary isotopes in massive stars, scale uniformly with Fe, a primary product of both massive stars and Type Ia supernovae, is not necessarily well justified. Nonetheless, the delayed input of Fe from Type I supernovae (due to the long evolutionary timescales of the progenitor stars) may well mimic the delayed input of isotopes like $^{25}$Mg (due to the secondary nature of their synthesis). Moreover, our assumed metallicity scaling for isotopic ratios matches the Si isotopic evolution inferred from the compositions of presolar SiC grains quite well (Zinner et al. 2006).

Karakas & Lattanzio (2003) and Zinner et al. (2005) have reviewed the nucleosynthesis of Mg and Al in AGB stars. Briefly, the two heavy Mg isotopes are produced in the He burning shell primarily by $\alpha$ captures on $^{22}$Ne, made from the abundant $^{14}$N produced by CNO-cycle H burning via $^{14}$N($\alpha$,$\gamma$)$^{18}$O and $^{18}$O($\alpha$,$\gamma$)$^{22}$Ne, and to a lesser extent by neutron captures. Thus, the third dredge-up increases the $^{25}$Mg/$^{24}$Mg and $^{26}$Mg/$^{24}$Mg ratios at the stellar surface where grains can form. The magnitude of the enhancement depends strongly on the stellar mass, so that the predicted isotopic shifts are relatively small (<5%) for 1.5$M_\odot$ stars, but can be much larger (>100%) for intermediate-mass AGB stars. Aluminum-26 is produced in the H shell by proton captures on $^{25}$Mg, but strongly depleted in the He intershell by (n,$\alpha$) and (n,p) reactions. It is mixed to the surface by the second (in massive enough stars) and third dredge-ups. Nucleosynthesis models predict surface $^{26}$Al/$^{27}$Al ratios of up to a few times $10^{-3}$ in AGB stars of <3$M_\odot$ (Forestini, Paulus, & Arnould 1991; Karakas & Lattanzio 2003; Mowlavi & Meynet 2000; Zinner et al. 2005). However, cool bottom processing in AGB stars can produce much higher amounts of $^{26}$Al in the envelope, depending on the temperature experienced by the envelope material that is processed (Nollett et al. 2003; Zinner et al. 2005). Similarly, hot bottom burning in intermediate-mass AGB stars is predicted to produce higher $^{26}$Al/$^{27}$Al ratios, especially in stars with metallicity lower than solar (Karakas & Lattanzio 2003; Lugaro et al. 2007). Note that for a solar Mg/Al ratio of 12.6, the $^{25}$Mg depletion that accompanies production of $^{26}$Al has only a very minor effect on the surface Mg isotopic composition.

Potassium isotopes are affected by neutron captures in the He intershell and the third dredge-up is thus expected to modify the envelope K isotopic composition, namely to increase the $^{40}$K/$^{39}$K and $^{41}$K/$^{39}$K ratios. As for Mg, larger effects are expected for more massive AGB stars. Potassium-40 decays to $^{40}$Ar with a half-life of 1.3 Gyr. Unfortunately, no noble-gas isotope data are available for presolar oxide grains. Our AGB models predict an increase of about 10% for $^{41}$K/$^{39}$K from third dredge-up in a 5$M_\odot$ AGB star, with smaller enhancements expected for lower-mass stars. Figure 4 shows that presolar hibonite grains have much larger enhancements of $^{41}$K than this, up to a factor of 140), supporting the interpretation of these anomalies as being due to *in situ* $^{41}$Ca decay. Moreover, our measurement precision is not sufficient to resolve effects at the few percent level, as predicted for K in AGB stars. We thus do not consider K isotopes further.

Calcium, like Mg and K, is also affected by neutron capture in AGB stars. Significant attention has been paid to the production of $^{41}$Ca since this short-lived isotope has been inferred to have been present in the early solar system (Srinivasan et al. 1994; Wasserburg et al. 2006). It is produced by neutron captures on $^{40}$Ca and destroyed by further neutron captures to produce $^{42}$Ca. The $^{41}$Ca/$^{40}$Ca ratio in the envelope depends on the ratio in the He shell, the amount of material dredged-up to the surface, as well as radioactive decay. Because the neutron capture cross-section of $^{41}$Ca is much higher than that of $^{40}$Ca, the $^{41}$Ca/$^{40}$Ca production ratio is given by the inverse ratio of cross-sections: $^{41}$Ca/$^{40}$Ca ~ $\sigma_{40}/\sigma_{41}$ ~$10^{-2}$ (Busso et al. 1999; Wasserburg et al. 1995b). Based on this production ratio, Wasserburg et al. (1995b) estimated $^{41}$Ca/$^{40}$Ca ratios of 0.3 – 1.5 × $10^{-4}$ in the envelopes of low-mass AGB stars. We will consider our more recent calculations in detail in subsequent sections.

The stable Ca isotopes are affected by *s*-process nucleosynthesis to varying degrees according to the relevant neutron capture cross-sections. Figure 5 shows the Ca isotopic pattern (for atomic masses 42, 43 and 44) predicted for a solar-metallicity 2 $M_\odot$ AGB star. Open symbols indicate the final values predicted after the final thermal pulse, the filled symbols indicate the values at the last pulse when the C abundance is still lower than that of O. Modest and similar enhancements of $^{42}$Ca/$^{40}$Ca and $^{43}$Ca/$^{40}$Ca are expected, with somewhat smaller increases of

$^{44}Ca/^{40}Ca$. As for other light elements, increasing isotopic shifts are predicted for the envelopes of AGB stars of increasing mass. For example, the maximum $\delta^{42}Ca/^{40}Ca$ value predicted for a 1.5 $M_{\odot}$, solar metallicity AGB star is 35 ‰, whereas it is 180 ‰ for a 5 $M_{\odot}$ star (see below). Much larger enhancements are predicted for $^{46}Ca$ than for the other stable Ca isotopes (e.g., $\delta^{46}Ca/^{40}Ca$ ~13,000 ‰ for a 5 $M_{\odot}$ star), but as mentioned earlier, this isotope was not measurable in the presolar hibonite grains, nor was $^{48}Ca$.

Titanium isotopes in the envelope are also affected by neutron capture and third dredge-up in AGB stars (Gallino et al. 1994; Lugaro et al. 1999; Zinner et al. 2007). They are affected by both the $^{13}C$ and the $^{22}Ne$ neutron sources, but only small shifts in the surface abundances are predicted for low-mass AGB stars. For example, variations in $^{46}Ti/^{48}Ti$, $^{47}Ti/^{48}Ti$ and $^{49}Ti/^{48}Ti$ ratios measured in presolar SiC grains from AGB stars are believed to reflect mostly the initial compositions of the parent stars (due to GCE) with only a minor component from nucleosynthesis in the parent stars themselves (Alexander & Nittler 1999; Huss & Smith 2007; Zinner et al. 2007). Because $^{50}Ti$ is neutron magic, it has a low neutron capture cross-section and larger nucleosynthetic effects are expected for $^{50}Ti/^{48}Ti$ ratios following the third dredge-up. Thus, more of the variability in observed $^{50}Ti/^{48}Ti$ ratios reflects nucleosynthetic effects, but a strong Galactic component is still present. In any case, most relevant to the discussion here is the prediction that only minor nucleosynthetic effects on the Ti isotopes are expected for low-mass AGB stars, especially O-rich ones that have not dredged-up significant amounts of material from the He shell.

## *4.2. Comparison of Grain Data and Model Predictions*

### *Oxygen Isotopes*

Figures 7 and 8 compare the expectations discussed above for O isotopes in RGB and AGB stars with the compositions of presolar oxide and silicate grains. The first-dredge-up curve on Figure 7 indicates predictions for stars of 1-7 $M_{\odot}$ and initial solar metallicity (Boothroyd & Sackmann 1999). Stars of different metallicity are expected to have different initial O-isotopic ratios (falling on the GCE line) and, following the first dredge-up, their compositions lie on curves parallel to the shown curve. As has been discussed extensively in previous publications (e.g., Nittler et al. 1997), the distribution of Group 1 and 3 grains is in good agreement with both these models and spectroscopic measurements of O isotopes in RGB and AGB stars (Harris et al. 1988; Smith & Lambert 1990). The wide range of $^{17}O/^{16}O$ ratios indicates a range of masses and that of $^{18}O/^{16}O$ ratios indicates a range of initial compositions of the stellar parents. In particular, the Group 3 grains must have formed in low-mass stars with lower than solar $^{17,18}O/^{16}O$ ratios. Because of their longer evolutionary times, parent stars of lower mass are expected to have formed earlier in Galactic history and thus have, on average, lower metallicity than more massive stars, in agreement with the observed distribution of the sub-solar O-isotopic ratios of these grains.

Boothroyd and co-workers (Boothroyd & Sackmann 1999; Boothroyd et al. 1994) calculated the O-isotopic compositions of red giants for a grid of masses and metallicities. Nittler et al. (1997) showed that a simple combination of these dredge-up calculations with stellar lifetimes and a predicted Galactic age-metallicity relationship (Timmes et al. 1995) reproduced very well the average trend of $^{17}O/^{16}O$ and $^{18}O/^{16}O$ ratios of both Group 1 and 3 grains. This result strongly supports the interpretation of the grains as having originated in low-mass RGB or AGB stars with a range of masses and metallicities. The good agreement between the dredge-up models and the grain data allows the initial mass and metallicity of the parent star of a given grain to be inferred by interpolating the O-isotopic predictions of Boothroyd & Sackmann (1999). However, it must be noted that such inferences are potentially subject to large systematic uncertainties.

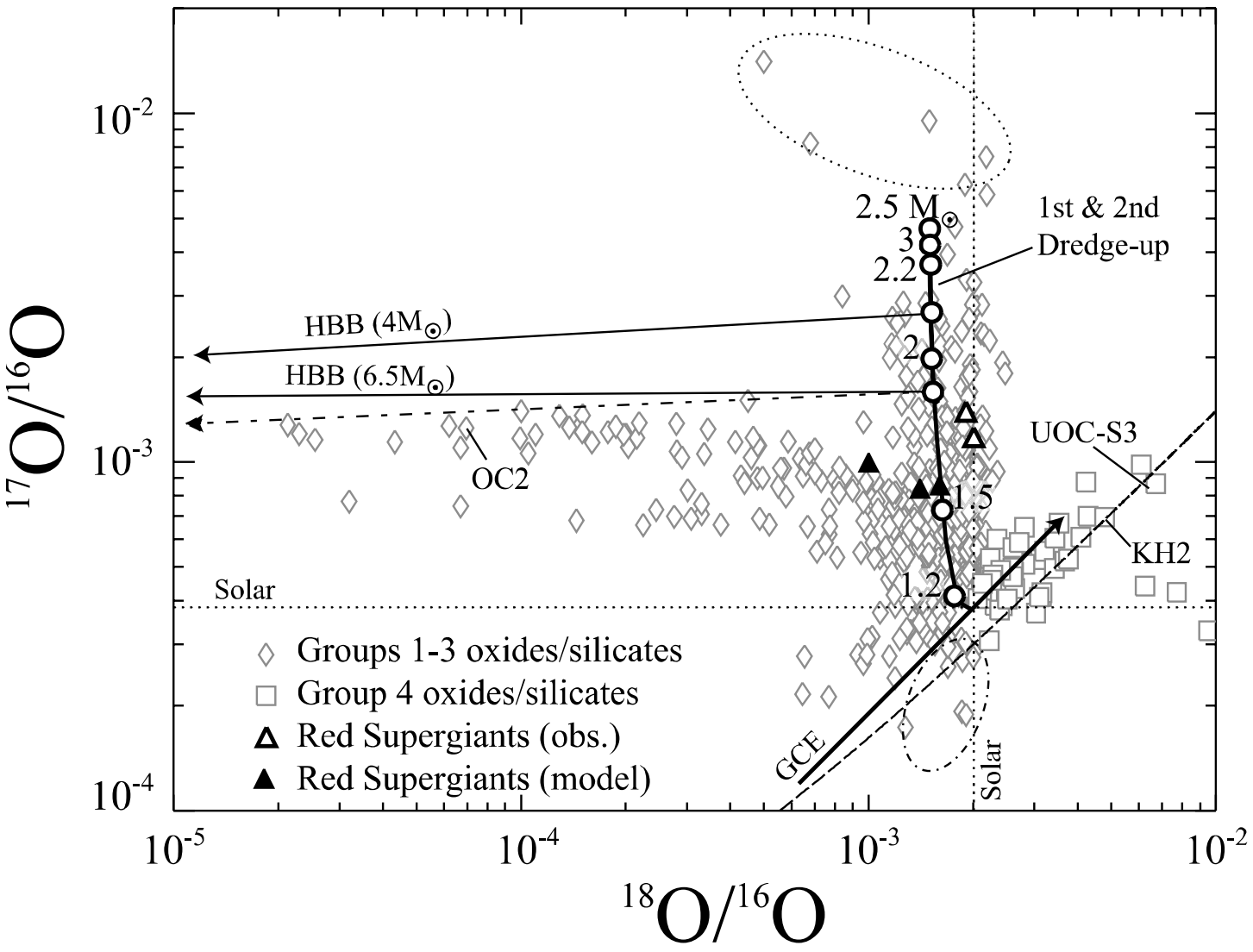

**Figure 7:** O-isotopic ratios of presolar oxide and silicate grains compared to models of the first and second dredge-ups (Boothroyd & Sackmann, 1999) and hot bottom burning (Boothroyd et al. 1995; Lugaro et al. 2007) in low- and intermediate-mass red giants and AGB stars as well as observations (Harris & Lambert 1984a) and predictions (Rauscher et al. 2002) for the surface composition of red supergiants. The solid arrow labeled "GCE" indicates the expected evolution of Galactic O-isotopic ratios with increasing metallicity (Timmes et al. 1995). The dotted ellipse indicates grains with $^{17}O/^{16}O$ ratios higher than predicted for red giants and asymptotic giant branch stars. See caption to Figure 2 for sources of grain data. The long-dashed line connects the $^{18}O$-rich grain KH2 to the average composition of $^{16}O$-rich interior zones of 15 $M_\odot$ supernova (Rauscher et al. 2002). This line passes through many Group 4 grains and Group 3 grains with low $^{17}O/^{16}O$ ratios (dashed-dot ellipse), perhaps reflecting an origin in a single supernova.

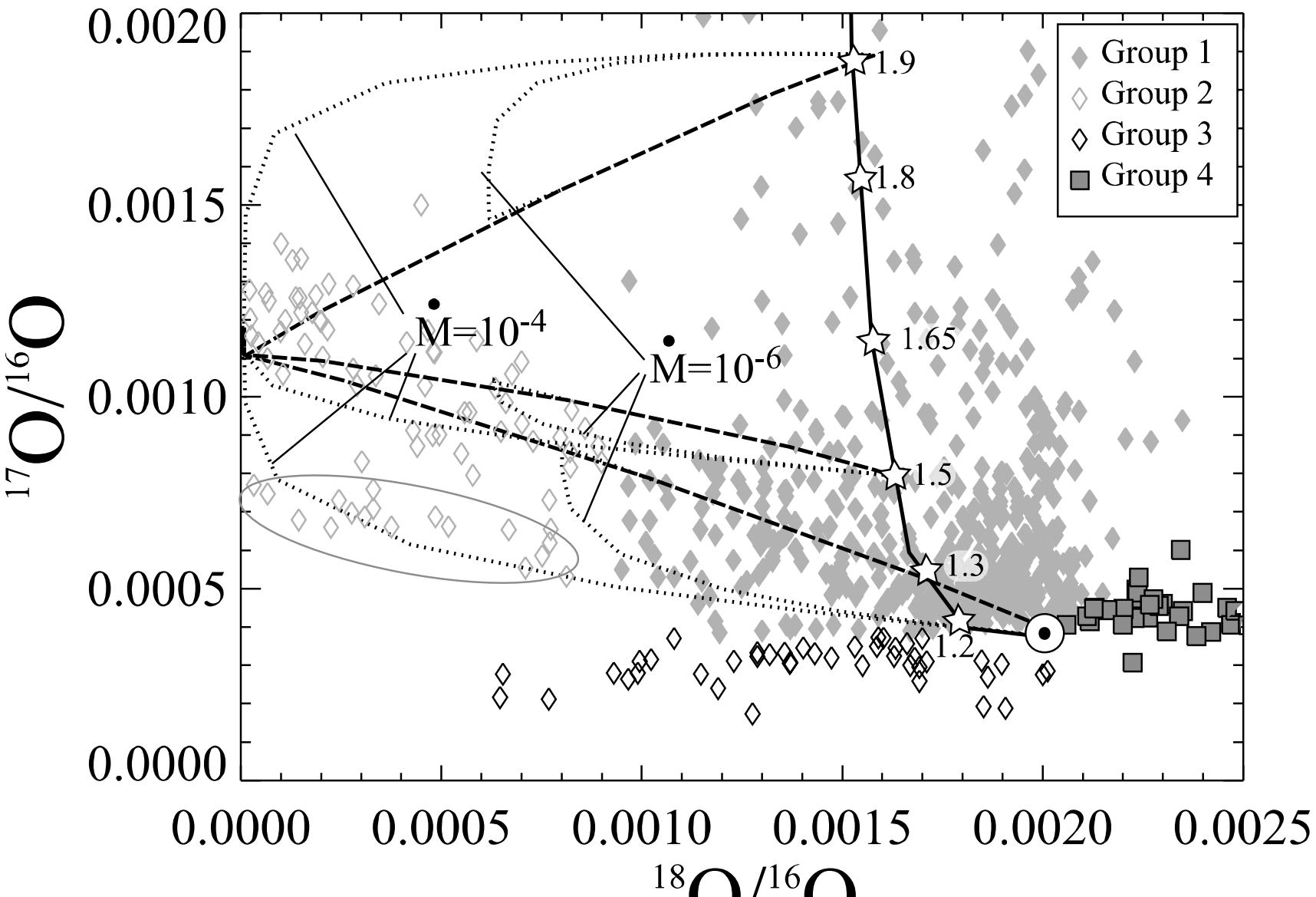

**Figure 8:** O-isotopic ratios of many presolar oxide and silicate grains compared to models of the first dredge-up (open stars, Boothroyd & Sackmann, 1999) and cool bottom processing (CBP, Nollett et al. 2003) in low-mass stars. Extensive CBP at temperatures greater than about 42 × $10^6$ results in compositions lying on mixing lines between the starting composition and a composition with $^{18}O/^{16}O$=0, $^{17}O/^{16}O$=0.0011 (solid dashed lines). Dotted lines indicate compositions for CBP with lower temperatures (T=33–42 × $10^{-6}$ K) and two indicated mass circulation rates. The Group 2 grains and many Group 1 grains are consistent with cool bottom processing in stars of 1.2 - 1.9 $M_\odot$. See caption to Figure 2 for sources of grain data.

Stellar masses are inferred primarily from $^{17}O/^{16}O$ ratios. During first dredge-up in low-mass stars, the envelope dips into a region with steeply increasing $^{17}O/^{16}O$ ratio. The depth reached by the envelope increases with increasing stellar mass. Thus, the post-first-dredge-up envelope $^{17}O/^{16}O$ ratio is a steep function of stellar mass (Boothroyd & Sackmann 1999). Slightly different depths of dredge-up will lead to large differences in the amount of $^{17}O$ brought to the surface and thus the final $^{17}O/^{16}O$ ratio depends on details of the mixing, which will vary according to the details of a given model. Inferred parent star masses for Group 1 and 3 grains found in this study are given in Tables 2 and 3. Despite the potential uncertainties, it should be noted that there is a lower limit on the mass of parent stars of ~1.1 $M_{\odot}$, based on the fact that stars of lower mass would not have evolved to the red giant stage in time to provide grains to the early Solar System 4.6 Gyr ago (e.g., Nittler & Cowsik 1997). Moreover, it seems likely that most of the grains formed in stars of less than ~2.5 $M_{\odot}$. The dredge-up models predict a fairly narrow range of $^{17}O/^{16}O$ ratios for stars between 2.2 and 3 $M_{\odot}$ (Fig. 7). Thus, if the grains came from a continuous distribution of masses up to 3-4 $M_{\odot}$ (above which hot bottom burning would likely drastically change their compositions; see below) one would expect an accumulation of grains with $^{17}O/^{16}O$ ~4–5 × $10^{-3}$, which is not observed.

Metallicities are inferred primarily from $^{18}O/^{16}O$ ratios, which reflect mostly the initial stellar compositions, determined by Galactic chemical evolution processes. Unfortunately, the precise relationship of O-isotopic ratios with metallicity in the Galaxy is not well known (Prantzos, Aubert, & Audouze 1996). Boothroyd & Sackmann (1999) assumed that the initial O-isotopic ratios vary linearly with metallicity, so that a star of half solar metallicity starts with $^{18}O/^{16}O$ and $^{17}O/^{16}O$ ratios that are half of the solar values. Thus, metallicities inferred by interpolating the Boothroyd & Sackmann (1999) models depend on this specific assumption of the isotope-metallicity relationship. Should the true dependence of $^{18}O/^{16}O$ on metallicity be different, inferred metallicity values would be different as well. For this reason, we report in Tables 2 and 3 and discuss further below, the inferred initial $^{18}O/^{16}O$ ratios of the parent stars, normalized to solar, rather than metallicity values. This does not, however, change the general correlation of $^{18}O/^{16}O$ and metallicity, and the inferred initial $^{18}O/^{16}O$ ratios for different grains can still be used to investigate GCE effects. An important additional caveat, discussed further below, is that cool bottom processing could have affected to some extent the $^{18}O/^{16}O$ ratios of some Group 1 and 3 grains and the inferred initial ratios are thus lower limits.

In contrast to the Group 1 and 3 grains, it has long been recognized that the Group 2 grains have $^{18}O/^{16}O$ ratios that are too low to be plausibly explained by first dredge-up in red giant stars with metallicities in the range expected for disk stars in the solar neighborhood 4.6 Gyr ago. Hot bottom burning (HBB) in intermediate mass AGB stars was initially suggested as an explanation for these grains (Nittler et al. 1994). However, detailed calculations of HBB (Boothroyd et al. 1995) indicated that it could not be the explanation for most Group 2 grains, both because HBB so efficiently destroys $^{18}O$ that the intermediate $^{18}O/^{16}O$ ratios ($10^{-4}$ – $10^{-3}$) observed in many grains cannot be explained, and because it predicts $^{17}O/^{16}O$ ratios higher than observed. The solid HBB curves on Figure 7 indicate predictions for HBB in AGB stars of 4 and 6.5 $M_{\odot}$ (Boothroyd et al. 1995; Lugaro et al. 2007); these models rapidly reach $^{18}O/^{16}O$ ratios of $10^{-6}$ – $10^{-7}$. Recently, Zinner et al. (2005) and Lugaro et al. (2007) argued for HBB as an explanation for the unusual O and Mg isotopic composition of the presolar spinel grain OC2. Lugaro et al. (2007) showed that the $^{17}O/^{16}O$ ratio and Mg isotopic composition of OC2 could be explained by HBB in an intermediate mass AGB star if certain reaction rates involving the O isotopes are adjusted within their experimental uncertainties (long-short-dashed curve on Figure 7). HBB predicts lower $^{18}O/^{16}O$ ratios than that observed in OC2, but surface contamination on sample mounts makes it likely that the measured $^{18}O/^{16}O$ ratios for the most $^{18}O$-depleted Group 2 grains ($^{18}O/^{16}O$ <$10^{-4}$) are in fact upper limits. Even with the revised reaction rates that permit a solution for OC2's composition, HBB cannot explain most of the other extreme Group 2 grains, since these have $^{17}O/^{16}O$ ratios even lower than that of OC2.

The intermediate $^{18}O/^{16}O$ ratios of many Group 2 presolar oxide and silicate grains are better explained by a slow mixing of material in the parent stars' envelopes through regions hot enough to partially process it by H-burning nuclear reactions. As discussed in the previous section, this "cool bottom processing" is also required to explain the low observed $^{12}C/^{13}C$ ratios in low-mass red giants. Wasserburg et al. (1995a) first showed that CBP can plausibly explain the O-isotopic compositions and relatively high inferred $^{26}Al/^{27}Al$ ratios of the Group 2 grains. The CBP must have occurred while the parent stars were on the AGB, because temperatures during the RGB phase are not high enough to significantly affect the O isotopes (Wasserburg et al. 1995a). More recently, Nollett et al. (2003) performed a detailed study of CBP in AGB stars and compared the results to the isotopic compositions of presolar oxide and SiC grains. Their model has two main parameters: the rate at which material gets circulated through the hot radiative region below the stellar envelope and the maximum temperature experienced by the circulating material. They found that the final $^{18}O/^{16}O$ ratio depends mostly on the mass circulation rate, while the $^{26}Al/^{27}Al$ ratio depends on the maximum temperature.

Figure 8 shows CBP calculations of Nollett et al. (2003, from their Fig. 7), overlain on the O-isotopic data for most of the presolar oxide and silicate grains. The thick-dashed lines indicate two-component mixing lines between three assumed starting compositions and the equilibrium composition reached if CBP continues for a long time at temperatures of $\sim 10^{7.7}$ ($50 \times 10^6$) K ($^{17}O/^{16}O$ =0.0011, $^{18}O/^{16}O$ =0). This equilibrium composition represents the region near the H-burning shell through which the envelope material is cycled during CBP. The assumed starting compositions correspond to solar and the post first dredge-up composition of 1.5 $M_\odot$ and 1.9 $M_\odot$ stars. Extensive CBP with temperature log T>7.62 ($42 \times 10^6$) K of material with a given starting composition will result in final compositions along such mixing lines. The final position along the line depends on the mass circulation rate. The dotted curves indicate incomplete processing with log T = 7.53 to 7.62 (T = 34 – 42 × $10^6$ K) and two different mass circulation rates. This plot makes it clear that the Group 2 grains' compositions can be explained by CBP if one assumes a range of parent star masses, mass circulation rates and degrees of processing. The majority of Group 2 grains has O-isotopic compositions reasonably well-reproduced by the calculated CBP mixing lines corresponding to solar initial composition and a 1.5 $M_\odot$ dredge-up composition. Interestingly, a fraction of grains seem to define a lower, separate trend (grey ellipse), but their compositions are also explained by the partial-processing initially solar-composition CBP model. Note that it is highly unlikely that the parent stars of any of the presolar grains had solar O-isotopic ratios at the start of the AGB, because the first dredge-up should modify the composition of stars with even the minimum mass, 1.1 $M_\odot$, allowed by the age of the Galaxy. However, it is clear from Figure 8 that the solar-initial-composition CBP model of Nollett et al. (2003) intersects the solar-metallicity first dredge-up curve (solid line with stars) for stars of mass 1.3 $M_\odot$ (mixing line) and 1.2 $M_\odot$ (incomplete processing curves). Thus, comparison of these model curves with the grain data suggests that the Group 2 grains formed in AGB stars with masses in the approximate range from 1.2 to 1.8 $M_\odot$, with many coming from stars with masses of less than ~1.6 $M_\odot$, and mass circulation rates higher than $10^{-6}$ solar masses per year. Moreover, and perhaps more importantly, Figure 8 shows that the $^{18}O/^{16}O$ compositions of many grains classified as belonging to Group 1 may also have been affected by CBP. For example, the parent stars of all of the grains currently classified as Group 1 with $^{18}O/^{16}O$ < 0.0015 could have begun the AGB with compositions along the indicated solar-metallicity first dredge-up curve and had some $^{18}O$ subsequently destroyed by CBP. This obviously affects the use of $^{18}O/^{16}O$ ratios in the grains to infer initial metallicities of the parent stars and indicates that initial $^{18}O/^{16}O$ ratios (and hence metallicities) inferred by interpolating first dredge-up models should be considered as lower limits. We note that Nittler (2005) already argued, based on comparisons with presolar SiC data, that some of the spread in $^{18}O/^{16}O$ ratios in Group 1 grains must reflect CBP.

A possible test of whether CBP has affected the composition of a given grain is whether or not it contains extinct $^{26}$Al. About one third of Group 1 grains analyzed for Mg isotopes show no evidence for having $^{26}$Al, indicating that they formed either in RGB or early AGB stars that had not yet experienced significant third dredge-up episodes. Because CBP that affects O isotopes is assumed to occur along with thermal pulses and third dredge-up in AGB stars (Nollett et al. 2003), the parent stars of these $^{26}$Al-free grains most likely did not experience CBP. We note that such grains span the full range of $^{18}$O/$^{16}$O ratios observed in Group 1 grains, and thus some of the lower $^{18}$O/$^{16}$O ratios in presolar grains are likely due to metallicity effects and not CBP. Additional isotopic data for other elements might in principle help better distinguish whether a given grain originated in a low-metallicity star or one that experienced CBP, but as discussed further below, the case is not that clear for the elements analyzed here.

We have assumed thus far that all of the Group 1, 2 and 3 grains originated in low- and intermediate mass stars. However, massive stars (e.g., >10M$_\odot$) in the red supergiant phase are observed and predicted to have $^{17}$O enrichments and $^{18}$O depletions, similar to those seen in Group 1 oxide grains. Moreover, a recent estimate of dust production in the Galaxy suggests that as much as 2/3 of the O-rich interstellar dust is originally produced by red supergiants (Kemper, Vriend, & Tielens 2004), though this is much higher than other estimates (e.g., Tielens, Waters, & Bernatowicz 2005). Shown on Figure 7 are the observed O-isotopic ratios of two red supergiants (open triangles, Harris & Lambert 1984a) and those predicted for the envelopes of massive stars of mass 15–25 M$_\odot$ (filled triangles, Rauscher et al. 2002). Red supergiants are also expected to have $^{26}$Al/$^{27}$Al ratios in the range of $10^{-4}$ to $10^{-2}$. About 15% of the presolar oxides have similar O-isotopic compositions and inferred $^{26}$Al/$^{27}$Al ratios and thus plausibly could have formed in red supergiants. Note also that Wolf-Rayet winds from very massive (>40M$_\odot$) stars during the WN phase cannot be ruled out as sources of some of the presolar oxide grains, especially the more extremely $^{18}$O-depleted Group 2 grains (Arnould, Meynet, & Paulus 1997). However, there is very little evidence for dust formation in WN winds (van der Hucht, Williams, & Morris 2001; Barniske et al. 2006). The above considerations confirm that presolar oxide grains are dominantly from stars of relatively low mass with only a fraction plausibly having originated in supergiants or Wolf-Rayet stars.

Indicated by a dotted ellipse on Figure 7 are a few Group 1 presolar oxide and silicate grains with relatively high $^{17}$O/$^{16}$O ratios, ranging from ~0.006 to ~0.014, and $^{18}$O/$^{16}$O ratios ranging from about 25% to 1 times the solar ratio of 0.002. These $^{17}$O/$^{16}$O ratios are higher than the maximum values predicted for first dredge-up in stars of near solar metallicity (Boothroyd & Sackmann 1999). Stars of much lower metallicity (e.g., $Z$<0.001 or 20 times lower than solar) are predicted to reach $^{17}$O/$^{16}$O as high as those observed in these $^{17}$O-rich grains, but these stars are expected to have much lower $^{18}$O/$^{16}$O ratios than are observed in the grains. Moreover, such low metallicity would be very unusual for stars in the disk of the Milky Way near the time that the Sun formed. The approximately solar $^{18}$O/$^{16}$O ratios of some of these grains suggests an origin in stars of somewhat higher than solar metallicity ($Z$~1.25 solar), if they formed in RGB and/or AGB stars. Boothroyd et al. (1994) calculated first dredge-up compositions for stars with metallicity of 1.25 times solar and did not find appreciably higher $^{17}$O/$^{16}$O ratios than was predicted for solar metallicity ones, but the detailed calculations of Boothroyd & Sackmann (1999) did not include super-solar metallicities. Additional calculations of first dredge-up in high-metallicity stars are desirable.

It is possible, of course, that these grains did not form in RGB or AGB stars. For example, Nittler & Hoppe (2005) recently proposed that the most $^{17}$O-rich grain (T54, Nittler et al. 1997) might have formed in a nova explosion, since its composition matches that predicted by some recent nova models (José et al. 2004). Such a source is less likely for the grains with $^{17}$O/$^{16}$O <0.01, however, but binary systems (believed to produce novae) might play a role in explaining these grains. For example, mass-loss to a binary companion by a main sequence star might significantly diminish its envelope so that by the time of the first dredge-up, the $^{17}$O mixed to the surface is less diluted, resulting in a higher $^{17}$O/$^{16}$O ratio. Alternatively, mass transfer from an evolved stellar companion (e.g.,an AGB star or ejecta from a classical nova) could conceivably enrich a main-sequence star's envelope with significant amounts of $^{17}O$. In fact, Marks, Sarna, & Prialnik (1997) modeled this process for close-binary systems undergoing nova outbursts and predicted very high surface $^{17}O/^{16}O$ ratios (>0.01 in some cases) for secondary stars following re-accretion of nova ejecta. Clearly additional modeling to fully explore these scenarios is desirable.

### *$^{26}Al$ and $^{41}Ca$*

Figure 9 compares the inferred $^{26}Al/^{27}Al$ and $^{41}Ca/^{40}Ca$ ratios and $\delta^{25}Mg/^{24}Mg$ values for the presolar hibonite grains with predictions for AGB star nucleosynthesis. Each plotted symbol for an AGB model indicates the predicted envelope composition after a thermal pulse. Smaller symbols indicate pulses for which the resulting composition has O>C, larger symbols indicate C-rich compositions. The left-hand panels indicate models for which the $^{41}Ca$ was not allowed to decay during the star's evolution through the AGB phase, the right hand panels include $^{41}Ca$ decay according to terrestrial lifetime. Since $^{41}Ca$ decays by electron capture, its lifetime is likely to be greatly increased while it resides in hot and dense regions of the convective envelope where it becomes fully ionized. Thus, these cases represent limiting ones for the maximum and minimum $^{41}Ca/^{40}Ca$ ratios expected after a given pulse in a given AGB model. Although the two extinct isotopes are made in different places in our AGB models (the $^{26}Al$ by proton capture in the H-burning shell, the $^{41}Ca$ by neutron capture in the He-rich intershell) in these models they are both mixed to the surface by third dredge-up episodes, so that their surface abundances are predicted to be correlated with each other. Note also that while the predicted $^{26}Al/^{27}Al$ ratio is mostly independent of metallicity, $^{41}Ca/^{40}Ca$ ratios are typically somewhat higher for stars of lower metallicity than for solar-metallicity ones.

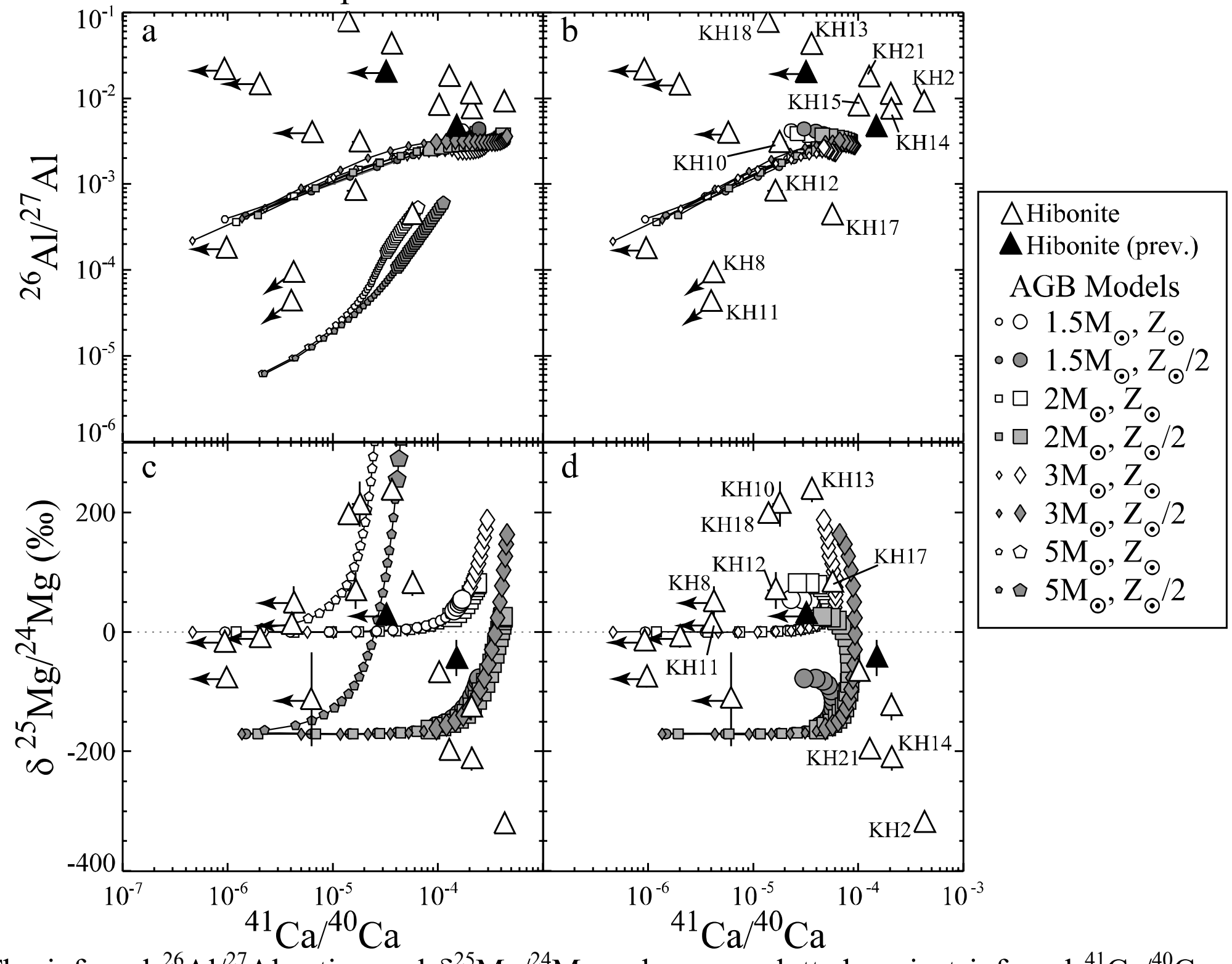

**Figure 9:** The inferred $^{26}Al/^{27}Al$ ratios and $\delta^{25}Mg/^{24}Mg$ values are plotted against inferred $^{41}Ca/^{40}Ca$ ratios for presolar hibonite grains and envelope models of AGB stars. Arrows indicate upper limits. In panels (a) and (c), radioactive decay of $^{41}Ca$ is suppressed in the AGB envelope, whereas $^{41}Ca$ decays with its terrestrial lifetime in panels (b) and (d). Small model symbols indicate compositions while the stellar envelope is O–rich (O>C), when oxides are most likely to form; larger model symbols indicate C-rich conditions. Previous hibonite data are from Choi et al. (1999).

The AGB calculations shown in Figure 9 predict $^{41}Ca/^{40}Ca$ ratios in the range of $10^{-6}$ to $10^{-4}$ for low-mass O-rich AGB stars, comparable with the previous estimate of 0.3 – 1.5 × $10^{-4}$ by Wasserburg et al. (1995b). If $^{41}Ca$ decay in the envelope is suppressed, the envelope ratios can reach ~4×$10^{-4}$ after many thermal pulses, long after the envelope has become C-rich. If $^{41}Ca$ is allowed to decay according to its terrestrial lifetime, predicted $^{41}Ca/^{40}Ca$ ratios do not exceed ~$10^{-4}$, even after many thermal pulses. We will discuss the composition of the $^{18}O$-rich grain KH2, which has the highest inferred $^{41}Ca/^{40}Ca$ ratio, further below in the context of a supernova origin. The other hibonite grains have inferred $^{41}Ca/^{40}Ca$ ratios (including upper limits for grains without $^{41}K$ excesses) in the predicted range, though the five grains with $^{41}Ca/^{40}Ca$ >$10^{-4}$ seem to require that $^{41}Ca$ decay is suppressed in the AGB envelopes. These results indicate that the nucleosynthesis of $^{41}Ca$ in AGB stars is reasonably well understood. The grains with upper limits on $^{41}Ca/^{40}Ca$ ratios must have formed prior to or very early in the AGB phase, otherwise they would have higher ratios.

In contrast to the good agreement for $^{41}Ca/^{40}Ca$ ratios, most of the measured hibonite grains have $^{26}Al/^{27}Al$ ratios that are higher than the maximum value of ~3 × $10^{-3}$ predicted for standard models of shell H-burning and dredge-up in AGB stars. This is also true of the $Al_2O_3$ grains we have analyzed here (Fig. 3 and Table 2). Some of the discrepancy might be attributed to uncertainties in reaction rates. Notably, the very large experimental uncertainty in the $^{26}Al(p,\gamma)^{27}Si$ rate allows an uncertainty of more than two orders of magnitude in computed $^{26}Al/^{27}Al$ ratios for AGB envelopes (Izzard et al. 2007). However, most of this uncertainty is in the downward direction, that is, it is not clear that the reaction rate uncertainties would allow for more than a factor of a few increase in the predicted $^{26}Al/^{27}Al$ ratios, which is still insufficient to explain $^{26}Al/^{27}Al$ ratios of > 0.01 measured in some of the grains. More likely, as has been previously discussed by several authors, the higher $^{26}Al/^{27}Al$ ratios of many presolar oxide grains are the result of efficient cool bottom processing in the parent stars (Nittler 1997; Nollett et al. 2003; Zinner et al. 2005). The parameterized model of Nollett et al. (2003) indicates that production of envelope $^{26}Al/^{27}Al$ ratios as high as those observed can be explained as long as the envelope material is circulated to regions with T>~50 × $10^6$ K, though again these results depend on the uncertain reaction rates.

The $^{25}Mg/^{24}Mg$ ratios of the presolar grains will be discussed in more detail in the following section. However, the bottom panels of Figure 9 reveal an interesting trend. The hibonite grains with $^{41}Ca/^{40}Ca$ ratios higher than $10^{-4}$ all have $^{25}Mg$ depletions, ranging from 5 to 25%, relative to solar. Again, the highly $^{25}Mg$ depleted grain KH2 most likely formed in a supernova and will be discussed separately. The five other grains with the highest $^{41}Ca/^{40}Ca$ ratios also have high $^{26}Al/^{27}Al$ ratios requiring CBP. At first glance, this suggests that the $^{25}Mg$ depletions simply reflect destruction of $^{25}Mg$ during CBP production of $^{26}Al$. However, this would require extremely low Mg/Al ratios (≈0.1 compared to the solar ratio of 12.6) for which there would be no obvious explanation. Moreover, $^{25}Mg$ synthesized by α and neutron capture in the He intershell is also brought to the surface along with $^{41}Ca$ during the third dredge-up (Zinner et al. 2005), effectively canceling the destruction by proton capture during CBP. More likely, the $^{25}Mg$ depletions indicate that these grains formed in stars with lower than solar metallicity and their relatively higher $^{41}Ca/^{40}Ca$ ratios are consistent with this as well. The low-metallicity AGB models shown on Figure 9 were assumed to start with Mg isotopic compositions that are depleted in $^{25}Mg$ by ~200‰ relative to solar. Figure 9c shows that the Mg isotopes and $^{41}Ca/^{40}Ca$ ratios of the $^{41}Ca$-rich grains can be reasonably well reproduced by the low-mass AGB models, provided that $^{41}Ca$ decay is suppressed in the AGB envelope. Note that according to the models, the AGB envelope should be C-rich by the time these grain compositions are reached, in contrast to the requirement that O>C for oxides like hibonite to condense (Lodders & Fegley 1995). CBP most likely provides the solution to this discrepancy as well, since CBP can destroy the $^{12}C$ brought up by the third dredge-up following thermal pulses, keeping the envelope O-rich (Nollett et al. 2003). Whether this explanation can be made to work in quantitative detail, however, remains to be seen. For example, the CBP model of Nollett et al. (2003) suggests that a mass circulation rate larger than ~$6\times10^{-7}$ $M_{\odot}$/yr is needed to ensure that C/O remains lower than unity. At these rates, their model predicts $^{18}O/^{16}O$ ratios lower than ~0.001, but of the $^{25}Mg$-depleted, $^{41}Ca$-rich hibonite grains, only KH15 has an $^{18}O/^{16}O$ ratio this low.

In contrast to the grains just discussed, hibonite grains with $^{41}Ca/^{40}Ca$ ratios in the range of $10^{-5}$ to $10^{-4}$ all have $^{25}Mg$ enrichments, relative to solar. Several of these grains lie along the 5$M_{\odot}$ AGB model predictions in Figure 9c. However, it is unlikely that these grains formed in such massive AGB stars since their $^{17}O/^{16}O$ ratios are significantly lower than those predicted for the envelopes of such stars after first dredge-up. The $\delta^{25}Mg/^{24}Mg$ value and inferred $^{41}Ca/^{40}Ca$ ratio of grain KH13 can almost be reached by the 3$M_{\odot}$ AGB models in Figure 9d, where $^{41}Ca$ is allowed to decay in the envelope. Of course, this composition is only reached after many thermal pulses when the envelope is predicted to have a high C/O ratio, and therefore hibonite is not expected to condense. However, as discussed above, the high $^{26}Al/^{27}Al$ ratio of grain KH13 points to CBP and this process could also maintain a low C/O ratio. In any case, if this is indeed the origin of grain KH13 (though its O-isotopic composition argues for a parent star of lower mass), this indicates a range of AGB envelope conditions affecting the ionization state of $^{41}Ca$ and hence its lifetime against electron capture. The $^{25}Mg$-rich compositions of grains KH10 and KH18 are harder to explain. As discussed by Zinner et al. (2005) and in the next section, it is difficult to disentangle the effects of nucleosynthesis and GCE in interpreting $^{25}Mg/^{24}Mg$ ratios in presolar oxide grains. High-metallicity stars are predicted to have excesses in $^{25}Mg$ due to GCE (Fenner et al. 2003) and we will argue below for an origin in a high-metallicity star for grain KH10. However, an explanation for the $^{25}Mg$ excess and corresponding relatively low $^{41}Ca/^{40}Ca$ ratio of grain KH18 is lacking.

## *Mg and Stable Ca Isotopes: Nucleosynthesis*

The Mg and stable Ca isotopic compositions at the surfaces of AGB stars are determined both by Galactic chemical evolution, which sets the initial composition of a star, and by nucleosynthesis in the He intershell followed by third dredge-up episodes. Zinner et al. (2005) extensively discussed the synthesis of Mg isotopes in AGB stars in the context of presolar spinel data. In terms of $^{25}Mg$, the AGB models predict only modest increases in $^{25}Mg/^{24}Mg$ in low-mass O-rich AGB stars. For example, the third dredge-up in stars of ≤3$M_{\odot}$ and solar metallicity is expected to increase the surface $^{25}Mg/^{24}Mg$ ratio by a maximum value of ~50‰ while the star is O-rich. Larger excesses are expected during later thermal pulses when C>O, up to 125‰ for a 2$M_{\odot}$ star and up to 180‰ for a 3 $M_{\odot}$ star of solar metallicity (Figs. 9c,d). For a given mass, somewhat larger $^{25}Mg$ enhancements are predicted for stars of lower metallicities, but these are also assumed to form with lower initial $^{25}Mg/^{24}Mg$ ratios. Much larger enhancements are expected for intermediate-mass stars, but these are unlikely sources of most grains discussed here, since hot bottom burning in such stars will rapidly destroy all of the $^{18}O$ (Lugaro et al. 2007).

The range of $^{25}Mg/^{24}Mg$ ratios observed in most of the presolar grains reported here is generally compatible with expectations for low-mass AGB stars, if we assume a relatively limited range of initial compositions. However, $^{25}Mg$ excesses larger than ~150 ‰ observed in a few grains are harder to reconcile with the AGB models. We consider four grains in particular (Fig. 10): 3 hibonite grains (KH13, KH18 and UOC-H2) and one spinel grain (UOC-S2). These show $\delta^{25}Mg/^{24}Mg$ values from ≈200–300 ‰, similar Group 2 O isotope signatures, with $^{18}O/^{16}O = 5–7 \times 10^{-4}$ and $^{17}O/^{16}O \approx 1 \times 10^{-3}$, and very high $^{26}Al/^{27}Al$ ratios of 0.04–0.08. This O composition lies very close to that expected for cool bottom processing in a 1.5$M_{\odot}$ AGB star with a mass circulation rate of $10^{-6}$ $M_{\odot}yr^{-1}$ (Fig. 8 and Nollett et al. 2003) and the high $^{26}Al$ points to extensive cool bottom processing as well. The $^{25}Mg$ excesses in these grains are much larger than can be explained by the nucleosynthesis models for such low-mass AGB stars, however and would require stars with masses of at least 3$M_{\odot}$ (Fig. 9). Lugaro et al. (2007) suggested a modification of the rates for the reactions $^{16}O(p,\gamma)^{17}F$ and $^{17}O(p,\alpha)^{14}N$, leading to lower $^{17}O$ abundances during hot bottom burning in AGB stars, in order to explain the composition of presolar spinel grain OC2. If the true values for these rates are indeed close to what they suggested, this most likely also results in a reduction in the predicted $^{17}O/^{16}O$ ratios after first dredge-up, perhaps by as much as a factor of two (M. Lugaro, pers. comm.). However, 3-4$M_{\odot}$ stars are predicted to have $^{17}O/^{16}O$ ratios following the first dredge-up more than 3 times as high as the value observed for these unusual grains (Boothroyd & Sackmann 1999). Thus, it is highly unlikely that these grains formed in AGB stars as massive as 3 $M_{\odot}$, and we must seek an alternate explanation for their $^{25}Mg$ excesses.

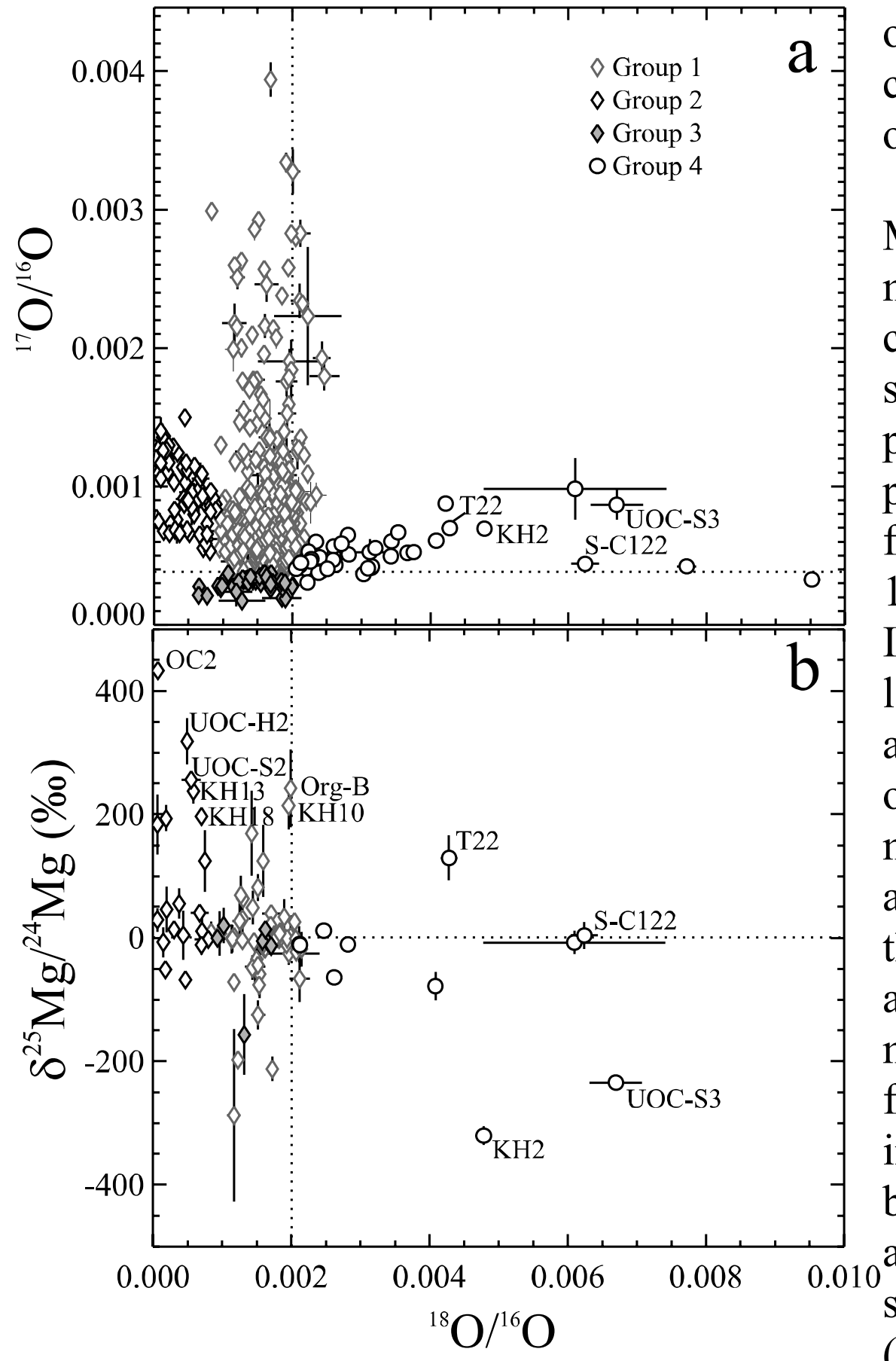

**Figure 10:** a) O-isotopic ratios of presolar oxides and silicates, plotted on a linear scale. b) $\delta^{25}Mg/^{24}Mg$ values plotted against $^{18}O/^{16}O$ ratios for presolar oxide and silicate grains. For clarity, only grains with $\delta^{25}Mg/^{24}Mg$ errors smaller than 50‰ or more than 2σ away from solar are plotted. Most $^{18}O$-rich (Group 4) grains have solar or sub-solar $^{25}Mg/^{24}Mg$ ratios. Dotted lines indicate solar ratios. See caption to Figure 2 for sources of grain data.

One obvious possibility is that the third dredge-up enriches the surface of low-mass AGB stars with more $^{25}Mg$ than is predicted by the models. However, the existence of many other presolar grains with O-isotopic compositions indicating an origin in low-mass AGB stars without large $^{25}Mg$ excesses suggests that the explanation lies elsewhere. Another possibility is that the grains formed in high-metallicity stars for which GCE has led to high initial $^{25}Mg/^{24}Mg$ ratios. In this case, the initial $^{18}O/^{16}O$ ratios of the parent stars would have been decreased by CBP. However, as discussed in the next section, we would also expect initial excesses in $^{42}Ca$ and $^{43}Ca$ in high-metallicity stars and these are not observed in these grains (Table 4). We thus consider it unlikely that the high $^{25}Mg/^{24}Mg$ ratios observed in these grains are due to GCE.

An alternative explanation is that the unusual Mg isotopic compositions of these grains reflect mass transfer from intermediate-mass AGB companions in binary star systems. Examples of such systems are barium stars, CH stars and Tc-poor S stars, all of which have chemical peculiarities explained by transfer of material from an evolved binary companion (Busso et al. 1992; Busso et al. 1995; Jorissen et al. 1998). Intermediate mass AGB stars, especially those of low metallicity, produce large amounts of $^{25}Mg$ and $^{26}Mg$ (Karakas & Lattanzio 2003). Transfer of envelope material from such stars to lower-mass companions could thus explain high $^{25}Mg$ and $^{26}Mg$ abundances in low-mass stars prior to their reaching the AGB phase. Although Busso et al. (1995) argued on the basis of Rb/Sr ratios that most barium stars are contaminated by material from low-mass AGB stars, transfer from intermediate mass AGB stars has been suggested by Yong et al. (2003) to explain high $^{25}Mg/^{24}Mg$ and $^{26}Mg/^{24}Mg$ ratios observed in several cool stars, including BD +5°3640, a low-metallicity ([Fe/H]= −1.34) CH star with *s*-process enrichments due to mass transfer from a companion. To test this suggestion, we consider yields predicted for a 5$M_{\odot}$, $Z$=0.4 $Z_{\odot}$ AGB star (Karakas et al. 2006; Lugaro et al. 2007). We assume that the AGB star transferred material to a low-mass binary companion with initially the same composition and compute the compositions of mixtures of the initial composition with varying amounts of material ejected during the AGB phase. We find that a mixture containing 22% by mass of the AGB ejecta and 78% of the initial composition has a $^{25}Mg/^{24}Mg$ ratio 30% higher than solar, as needed to explain the grains. With this fraction of AGB material, the $^{26}Mg/^{24}Mg$ and $^{17}O/^{16}O$ ratios are also enhanced, by 70% and 45%, respectively. The $^{17}O/^{16}O$ and $^{26}Mg/^{24}Mg$ ratios measured in the $^{25}Mg$-rich grains are much more enriched than this, so these ratios do not argue against the mass-transfer hypothesis. Note also that the AGB ejecta would diminish the $^{18}O/^{16}O$ ratio of the mixture by ~20%. We would not be able to distinguish this effect from those of Galactic chemical evolution or cool bottom processing. Using yields from a different AGB star model with the same mass and metallicity, but computed with a different set of reaction rates and a different assumed initial elemental and isotopic composition (Izzard et al. 2007), we find a similar result, only in this case just 10% of the AGB material is required to be transferred to explain a 30% $^{25}Mg$ enrichment. These simple calculations indicate that the proposed binary star scenario to explain the $^{25}Mg$-rich grains is plausible, but clearly much more work needs to be done to investigate this thoroughly. Note that AGB stars of lower mass and/or higher metallicity will not work as well, because such stars produce less $^{25}Mg$. Conversely, more massive AGB stars would need to contribute a smaller amount of material as these have larger enrichments of the neutron-rich Mg isotopes.

Figure 11 shows 3-isotope plots for the stable Ca isotopes measured in presolar hibonite grains, compared with the AGB star predictions. The predictions for $\delta^{43}Ca/^{40}Ca$ values as a function of $\delta^{42}Ca/^{40}Ca$ values are similar for AGB stars of mass 1.5-3 $M_{\odot}$, so for clarity, we only indicate the ranges of the predictions in panel a, distinguishing models with O>C from those with C>O. The 5 $M_{\odot}$ AGB stars (not shown in panel a) are predicted to reach more extreme compositions with a slightly steeper slope on this plot, but as discussed earlier are highly unlikely sources for most of the grains. Long-dashed grey arrows indicate the Galactic chemical evolution trends calculated by Timmes et al. (1995), normalized to reproduce solar isotopic ratios at solar metallicity. Both the AGB and the GCE models predict that the $\delta^{42}Ca/^{40}Ca$ and $\delta^{43}Ca/^{40}Ca$ values should be correlated with each other along a slope of ~1. The hibonite grains roughly follow this trend, but errors for many are relatively large. The fact that the total spread in the grain data is larger than is expected for nucleosynthesis in a single O-rich AGB star suggests that much of the range indeed reflects variations in initial compositions due to GCE. This is analogous to the case for Si and Ti isotopes in presolar SiC grains (Alexander & Nittler 1999; Zinner et al. 2007) and for $^{18}O/^{16}O$ ratios in Group 1 and 3 oxide and silicate grains (see above). Grain KH10 has a highly unusual Ca composition compared with the other grains, with very large excesses of both $^{42}Ca$ and $^{43}Ca$. Its O-isotopic composition indicates an origin in a star with higher than solar metallicity (1.2 $Z_{\odot}$ based on the $^{18}O/^{16}O$-metallicity relationship assumed by Boothroyd & Sackmann, 1999) and its large $^{25}Mg$ excess is consistent with this as well. Also, the $^{26}Al/^{27}Al$ and $^{41}Ca/^{40}Ca$ ratios inferred for this grain are close to the predictions for low-mass AGB stars without any cool bottom processing (Fig. 9). All of these data taken together suggest that the excesses of $^{42}Ca$ and $^{43}Ca$ observed in KH10 reflect the initial composition of its parent star and are higher than solar due to GCE. If so, this grain's composition suggests that the true GCE $\delta^{43}Ca/^{40}Ca$ versus $\delta^{42}Ca/^{40}Ca$ trend is slightly flatter than predicted by Timmes et al. (1995), or that the trend flattens out for higher than solar metallicity.

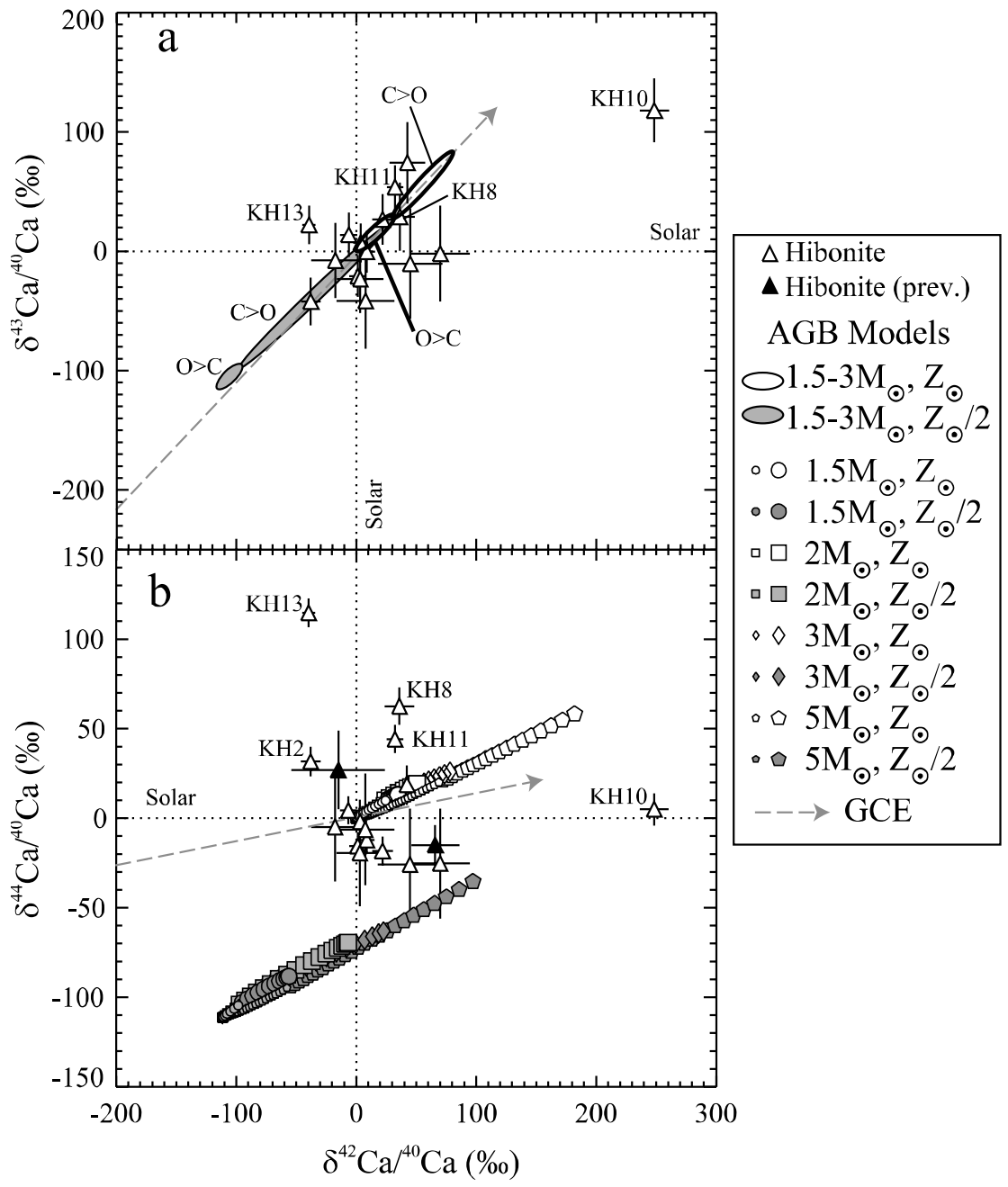

**Figure 11:** Calcium isotopic ratios of presolar hibonite grains are compared with nucleosynthetic predictions for AGB stars. a) $\delta^{43}Ca/^{40}Ca$ values plotted versus $\delta^{42}Ca/^{40}Ca$ values, b) $\delta^{44}Ca/^{40}Ca$ values plotted versus $\delta^{42}Ca/^{40}Ca$ values. Normalizing ratios are given in Table 4 and dotted lines indicate solar ratios. On panel (a), ranges of predictions for 1.5 – 3 $M_{\odot}$ AGB stars are shown. Results for individual thermal pulses are shown in panel (b). Grey dashed arrows indicate predicted Galactic evolution of interstellar Ca isotopic ratios (Timmes et al. 1995). Small model symbols indicate compositions while the stellar envelope is O–rich (O>C), when oxides are most likely to form; larger model symbols indicate C-rich conditions. Previous hibonite data are from Choi et al. (1999).

The case of $^{44}Ca/^{40}Ca$ ratios is more complicated. As shown in Figure 11b, AGB nucleosynthesis is predicted to produce smaller $^{44}Ca$ excesses than those of $^{42}Ca$ or $^{43}Ca$, resulting in trends with slopes of about 1/3 on a $\delta^{44}Ca/^{40}Ca$ versus $\delta^{42}Ca/^{40}Ca$ plot. The GCE model of Timmes et al. (1995) predicts an even shallower slope of ~0.12. Note that this is very different from the GCE trend we assumed for determining the initial composition of the half-solar metallicity AGB models. This discrepancy arises because we assumed that $^{44}Ca$ scales with Fe rather than with the α-elements like $^{28}Si$ or $^{40}Ca$. However, $^{44}Ca$ is made in supernovae as radioactive $^{44}Ti$ during an α-rich freeze-out from nuclear statistical equilibrium. It thus behaves somewhat as an α-element and the Galactic $^{44}Ca/^{40}Ca$ should be relatively constant with metallicity. In contrast with the case for $^{43}Ca$, for the most part the $^{44}Ca$ vs $^{42}Ca$ grain data do not lie close to the predicted trends for GCE or AGB nucleosynthesis. For example, several of the hibonite grains show moderate depletions of $^{44}Ca$, relative to solar, and lie below the Timmes et al. (1995) GCE trend. AGB dredge-up shifts compositions upward along a steeper slope than this GCE line. Therefore, if the latter correctly describes the evolution of Ca isotopes in the Galaxy, this makes it impossible to explain the $^{44}Ca$-depleted grains as arising from GCE plus third dredge-up in AGB stars. This suggests a steeper GCE trend, perhaps closer to the one we assumed. However, the observation of ~solar $^{44}Ca/^{40}Ca$ in grain KH10, for which other isotopic ratios indicate an origin in a high metallicity star, supports a relatively flat $^{44}Ca/^{40}Ca$ GCE trend. Alternatively, we will argue in the next section that relatively high $\delta^{42}Ca/^{40}Ca$ and $\delta^{43}Ca/^{40}Ca$ values observed in many of the grains suggests that the Sun is slightly depleted in these isotopes, relative to typical stars of its metallicity. If this is the case, but the Sun is not also depleted in $^{44}Ca$, the GCE trend would be parallel to the plotted line on Figure 11b, but displaced to lower $\delta^{44}Ca/^{40}Ca$ values, and could thus pass through many of the hibonite data points. Explaining the hibonite grains with $^{44}Ca$ excesses is not any easier. For example, although grains KH8 and KH11 plot roughly on extensions of our assumed $^{43}Ca$-$^{42}Ca$ and $^{44}Ca$-$^{42}Ca$ GCE trends, their O compositions do not suggest an origin in higher than solar metallicity stars. The highly $^{44}Ca$-enriched grain KH13 lies far from expectations for GCE and AGB nucleosynthesis on Figure 11b. We consider the most likely explanation for the scatter in $\delta^{44}Ca/^{40}Ca$ values for the hibonite grains to be locally inhomogeneous GCE, as spelled out further below.

## *Mg and Stable Ca Isotopes: Galactic Chemical Evolution*

As discussed earlier, the $^{18}O/^{16}O$ ratios of presolar grains belonging to Groups 1 and 3 are believed to largely reflect GCE effects, and the preceding discussions argue that some of the variation in Mg and Ca isotopic compositions seen in presolar hibonite grains reflects GCE as well. To better deconvolve the effects of GCE from those of nuclear processes in the grains' parent stars, and to investigate how homogeneously isotopic ratios evolve in the Galaxy, in Figure 12, we show Mg and Ca isotopic ratios of Group 1 and 3 presolar oxide grains, plotted against the solar-normalized inferred initial $^{18}O/^{16}O$ ratios. The initial $^{18}O/^{16}O$ ratio of the parent star of a given grain was determined by projecting the grain's O-isotopic composition back to the GCE line of the O isotopes along a trajectory backtracking the evolution of the O isotopes in the star's envelope during the first dredge-up. The trajectory was obtained by interpolating the models of Boothroyd & Sackmann (1999). Since the O-isotopic ratios are assumed to evolve linearly with the metallicity, the initial $^{18}O/^{16}O$ ratio of a grain gives the metallicity of its parent star. Recall that some of these initial $^{18}O/^{16}O$ ratios are probably lower limits, due to the possibility that cool bottom processing has depleted their parent stars' $^{18}O$, relative to their original first dredge-up values. Also shown on Figure 12 are the predicted GCE trends of Timmes et al. (1995, dashed grey curves) and Fenner et al. (2003, dashed-dot grey curve), though for these the plotted value along the x-axis is solar-normalized metallicity, not $^{18}O/^{16}O$ ratio. Except as discussed below, the GCE models are also normalized to pass through the solar composition at solar metallicity. The main difference between the Timmes et al. and Fenner et al. calculations is that the latter includes the contribution of AGB stars to the Galactic Mg isotopic evolution.

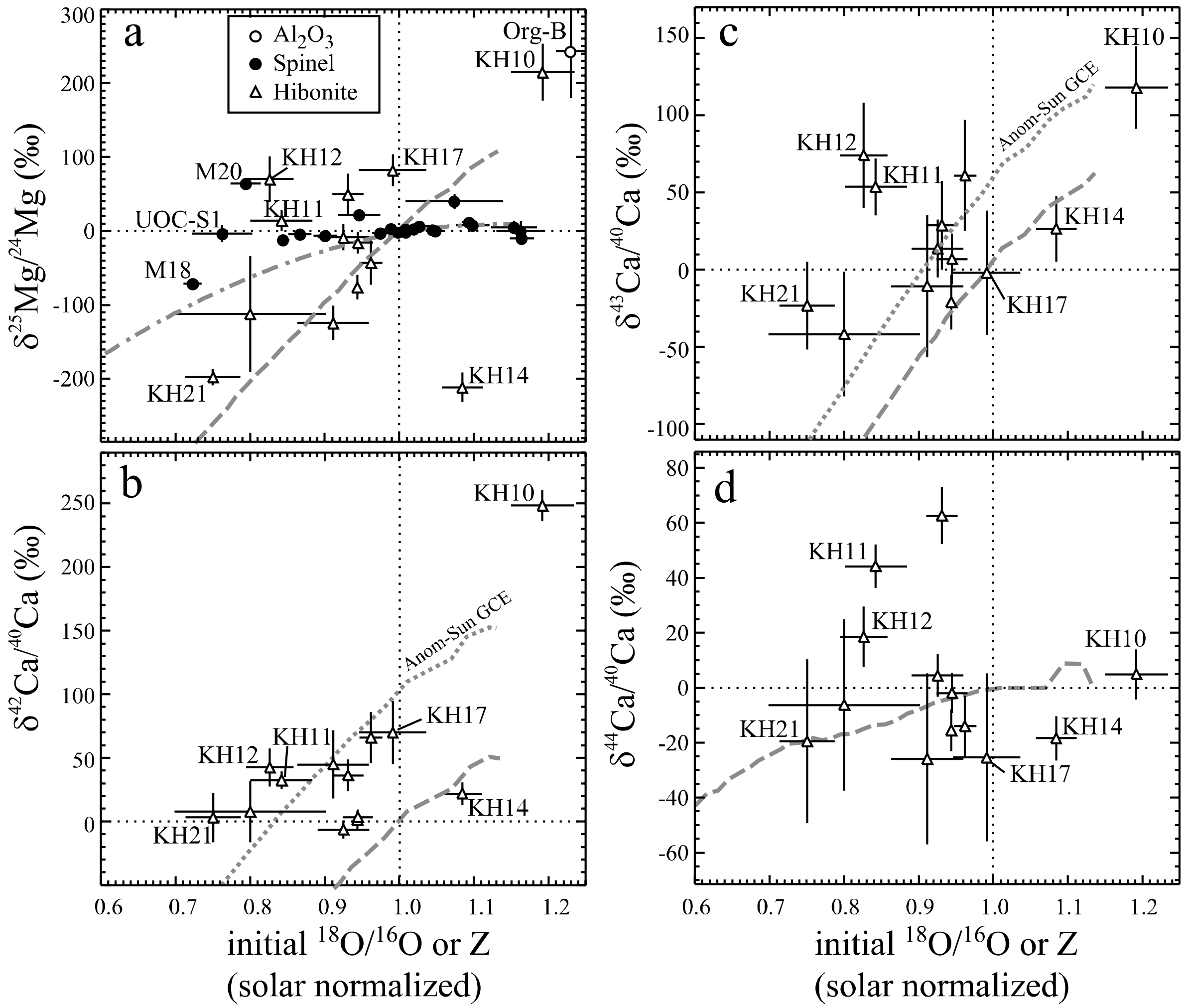

**Figure 12:** Mg and Ca isotopic ratios of Group 1 and 3 presolar oxides plotted against inferred initial $^{18}O/^{16}O$ ratios, normalized to solar. The $^{18}O/^{16}O$ ratios are determined by subtracting a first dredge-up component (Boothroyd & Sackmann, 1999) from measured O-isotopic compositions. Dotted lines indicate solar isotopic ratios. Grey curves indicate predictions for the Galactic chemical evolution of isotopic ratios as a function of metallicity, normalized to solar. a) Spinel data are taken from Zinner et al (2005) and references therein and the present work. Long-dashed curve is GCE model of Timmes et al. (1995); dot-dashed curve is that of Fenner et al. (2003), which includes the contribution of AGB stars as well as supernovae. (b-d) Long-dashed curves are GCE predictions of Timmes et al. (1995); dotted curves are GCE curves shifted on the assumption that the Sun has slightly anomalous $^{42}Ca/^{40}Ca$ and $^{43}Ca/^{40}Ca$ ratios for its metallicity (see text). Data are from this work, Huss et al. (1995), Choi et al. (1999) and Zinner et al. (2005).

Figure 12a shows $\delta^{25}Mg/^{24}Mg$ data for the Group 1 and 3 presolar spinel and hibonite grains reported here and by Zinner et al. (2005) and Choi et al. (1999), as well as the $^{25}Mg$-rich $Al_2O_3$ grain Orgueil-B (Huss, Fahey, & Wasserburg 1995). If the Mg isotopic compositions of the grains' parent stars were solely determined by homogeneous GCE and not subsequently modified by nucleosynthesis in the parent stars, one would expect the data to lie along a monotonic curve on this plot. This is clearly not the case. A majority of the spinel grains have $^{25}Mg/^{24}Mg$ ratios close to the solar value, independent of their inferred initial $^{18}O/^{16}O$ ratios and hence the relative metallicities of their parent stars. This observation extends to a new spinel grain, UOC-S1, which has a terrestrial $^{25}Mg/^{24}Mg$ ratio, but O isotopes suggesting an origin in a star with ~80% of solar metallicity. Spinel grain M18 lies near the GCE prediction of Fenner et al (2003). This observation and the fact that most of the spinel grains with inferred $^{18}O/^{16}O$ ratios higher than solar are not highly enriched in $^{25}Mg$ led Zinner et al. (2005) to argue that AGB stars are indeed an important contributor to the Galactic inventory of Mg isotopes. Also, the $^{25}Mg$ enrichment observed in grain M20 was argued not to reflect dredge-up in the parent star, but rather to reflect an unusual initial composition. The observation of several low-metallicity stars with high $^{25}Mg/^{24}Mg$ ratios (Yong et al. 2003) supports the possibility that the parent stars of some presolar grains were enriched in $^{25}Mg$, relative to the mean Galactic chemical evolution, perhaps through the binary mass transfer process discussed earlier.

Addition of the hibonite data to Figure 12a complicates the story. In contrast to the spinel grains, most of the Group 1 and 3 hibonite grains have non-solar $^{25}Mg/^{24}Mg$ ratios and several of the grains lie close to the Timmes et al. (1995) GCE prediction. Moreover, grain KH10 has, within errors, the same composition as $Al_2O_3$ grain Orgueil-B, consistent with an extension of the Timmes et al. (1995) GCE model to higher metallicity. Since dredge-up in parent stars and/or mass transfer from AGB companion stars will only lead to higher $^{25}Mg/^{24}Mg$ ratios, the hibonite data suggest that the GCE model not including AGB contributions (e.g., Timmes et al. 1995) is a better description of Mg isotope evolution in the Galaxy. This conclusion is also supported by recent observations of Mg isotopes in halo stars (Meléndez & Cohen 2007). However, even if this is the case, the wide range of $^{25}Mg$ values uncorrelated with inferred initial $^{18}O/^{16}O$ ratios clearly requires additional processes to be at work. Population synthesis models that track the isotopic compositions of stars (especially O and Mg) and take into account the possibility of mass transfer in binary systems could be very helpful to making further progress in understanding Mg isotopes both in presolar grains and stars.

One hibonite grain, KH14, falls far from the other data on Figure 12a, showing a large $^{25}Mg$ depletion despite having an O-isotopic composition that suggests an origin in a relatively high-metallicity star. The $^{42}Ca$ and $^{43}Ca$ enrichments observed in this grain (Figs. 12b,c) are also consistent with a high-metallicity origin, so the large $^{25}Mg$ depletion is difficult to understand. Since there is no obvious way to significantly decrease $^{25}Mg/^{24}Mg$ ratios at the surfaces of low-mass AGB stars, it most likely reflects an anomalous starting composition of the parent star, perhaps due to local heterogeneities in the interstellar medium arising from incomplete mixing of supernova ejecta (e.g., Lugaro et al. 1999; Nittler 2005). In any case, the unusual composition of this grain is a further indication that we are far from a satisfactory understanding of $^{25}Mg/^{24}Mg$ ratios in presolar grains.

An additional important puzzle posed by Figure 12a is why so many of the spinel grains have terrestrial $^{25}Mg/^{24}Mg$ ratios, in contrast to the hibonite grains. One possibility is that the grains have undergone isotopic exchange either in space or on the meteorite parent bodies. Presolar spinel is highly abundant in CM2 chondrites that have seen essentially no thermal metamorphism (Zinner et al. 2003). In ordinary chondrites, the abundance of presolar spinel, at least for micron-sized grains, appears to drop precipitously with increasing metamorphism. For example, spinel makes up a significant fraction of presolar oxides found in our OC and UOC samples (Table 1), which include several Type 3.0 meteorites, but only a tiny fraction of presolar oxides in Tieschitz (3.8) and Krymka (3.1). This observation indicates that even a modest degree of thermal metamorphism completely erases the O-isotopic signatures of presolar spinel grains so that they can no longer be recognized as presolar. Experiments indicate that Mg volume diffusion is very slow in spinel (Sheng, Wasserburg, & Hutcheon 1992). However, Greenwood et al. (2000) found spinel in a metamorphosed calcium-aluminum-rich inclusion in the CK chondrite Karoonda that clearly had exchanged Fe and Mg but retained its original $^{16}$O-rich composition. This suggests that a very small amount of heating might allow for sufficient Mg isotope exchange to equilibrate the $^{25}$Mg/$^{24}$Mg ratio of a sub-micron spinel grain without erasing the larger anomalous signatures of its O isotopes. The high abundance ratio of presolar hibonite to presolar spinel in Krymka also suggests that hibonite is more resistant to thermal alteration and hence Mg isotopic exchange than is spinel. A difficulty with this scenario is that a significant number of presolar spinel grains with normal $^{25}$Mg/$^{24}$Mg ratios were found in Murray, which is not thought to have seen temperatures higher than 20 to 50°C. This suggests that if the spinel grains' $^{25}$Mg/$^{24}$Mg ratios have been modified by isotope exchange, it happened in space prior to accretion into parent bodies, either in the interstellar medium or in the solar nebula. This is supported by observations of calcium-aluminum-rich inclusions in meteorites that show Mg isotopic fractionation in their interiors that becomes more normal toward the surface, indicating equilibration with a gaseous reservoir (Caillet Komorowski et al. 2007; Fahey et al. 1987; Simon et al. 2005).

In contrast to the $\delta^{25}$Mg/$^{24}$Mg values of presolar oxides, the $\delta^{42}$Ca/$^{40}$Ca and $\delta^{43}$Ca/$^{40}$Ca values of the hibonite grains are better correlated with inferred initial $^{18}$O/$^{16}$O ratios (Figs. 12b,c), indicating a significant influence of GCE on the grains' Ca isotopic compositions. In both plots the grain correlations are roughly parallel to, but displaced from, the solar-normalized model GCE curves. In particular, all of the grains with inferred initial $^{18}$O/$^{16}$O ratios lower than solar have $\delta^{42}$Ca/$^{40}$Ca values close to or higher than solar, rather than lower than solar as would be expected. The same is true of many of the $\delta^{43}$Ca/$^{40}$Ca values. This is puzzling, especially since many of these grains have $^{25}$Mg depletions as expected for low-metallicity stars. One possible explanation for this could be that the Sun is somewhat depleted in $^{42}$Ca and $^{43}$Ca relative to the average stars of solar metallicity, just as we suggested above that the solar $^{44}$Ca/$^{40}$Ca ratio is anomalously high. This assumption would result in the GCE curves in Figure 12b,c being shifted upwards, achieving a better match with the data. Example curves shifted by 100 ‰ in $\delta^{42}$Ca/$^{40}$Ca and 50 ‰ in $\delta^{43}$Ca/$^{40}$Ca are labeled "Anom-Sun GCE" on Figure 12b,c. A plausible explanation for this anomaly in solar composition could be local chemical heterogeneities in the interstellar medium due to inhomogeneous GCE. The scatter in the data around the GCE curves also points to inhomogeneous GCE. We note that an analogous suggestion has been made for the Si isotopic composition of the Sun, based on Si and Ti isotopic data in presolar SiC grains (Alexander & Nittler 1999; Clayton & Timmes 1997).

We showed earlier that the $\delta^{44}$Ca/$^{40}$Ca values for the presolar hibonite grains are not easily understood in the context of the GCE and dredge-up models we have been considering (Fig. 11b). This is shown by Figure 12d as well. The $\delta^{44}$Ca/$^{40}$Ca values are essentially uncorrelated with the initial $^{18}$O/$^{16}$O ratios for the grains. Several of the grains, especially the most $^{18}$O-rich (KH10) and $^{18}$O-poor (KH21) are consistent with the relatively flat Timmes et al. (1995) GCE trend, but there are grains with large excursions from this trend, to both positive and negative $\delta^{44}$Ca/$^{40}$Ca values. To better understand these data, let us consider the nucleosynthetic sources of the Ca isotopes in the Galaxy. The isotopes $^{40}$Ca, $^{42}$Ca and $^{43}$Ca are all made primarily by both hydrostatic and explosive oxygen-burning (and silicon-burning for $^{40}$Ca) in Type II supernovae. In contrast, $^{44}$Ca is produced in Type II supernovae mainly as radioactive $^{44}$Ti during the $\alpha$-rich freeze-out from high-temperature burning near Nuclear Statistical Equilibrium (e.g., The et al. 2006; Woosley, Arnett, & Clayton 1973). However, a rare class of Type Ia supernovae, those occurring on sub-Chandrasekhar mass white dwarfs, are also predicted to produce very large amounts of $^{44}$Ti (Woosley & Weaver 1994). The et al. (2006) have discussed in detail γ-ray observations of $^{44}$Ti in the Galaxy and concluded that supernovae that produce the most $^{44}$Ti (and hence $^{44}$Ca) might be atypical and rare. If this is indeed the case, then one would expect much more variability in $^{44}$Ca/$^{40}$Ca ratios in the Galaxy, compared to the other stable Ca isotope ratios. The scatter in $^{44}$Ca/$^{40}$Ca ratios observed in presolar hibonite grains, larger than expected for third dredge-up in the parent stars and largely uncorrelated with inferred initial metallicity, certainly seems to point in this direction.

We have argued that locally inhomogeneous GCE might play a role in establishing the isotopic compositions of the grains, both in introducing scatter in some isotopic ratios about expected mean GCE trends, and in possibly leading to an unusual composition for the Sun itself. Timmes & Clayton (1996) first suggested that inhomogeneous GCE might explain the distribution of Si isotopes in mainstream presolar SiC grains from AGB stars. Lugaro et al. (1999) followed up on this suggestion with a Monte Carlo model of inhomogeneous mixing of supernova ejecta into localized regions of the interstellar medium (ISM). These authors showed that such a model could easily explain the range and correlation of Si isotopic ratios in mainstream SiC grains. However, Nittler (2005) extended this model to Ti and O isotopes and showed that the high degree of correlation between Si and Ti isotopic ratios in the SiC grains argues against the bulk of the isotopic variation being caused by inhomogeneous GCE. The Si-Ti isotope correlation indicates that diverse supernova ejecta are well mixed in the ISM such that residual chemical heterogeneity is of order a few %. In addition, with this high degree of mixing, the expected range of $^{18}$O/$^{16}$O ratios due to inhomogeneous GCE is much smaller than that observed in the grains in Figure 12, so the grain data are almost certainly dominated by homogeneous GCE.

A detailed extension of this Monte Carlo model to Mg and Ca isotopes is beyond the scope of this paper, but we can make some general comments. First, the Mg and Si isotope yields from Type II supernovae are roughly similar to each other (Woosley & Weaver 1995). Thus, inhomogeneous mixing of supernova ejecta into the ISM should lead to a similar range of $^{25}$Mg/$^{24}$Mg ratios as for $^{29}$Si/$^{28}$Si ratios, which is limited by the Si-Ti correlation in SiC to be ~5% (Nittler 2005). This is much smaller than the scatter in $^{25}$Mg/$^{24}$Mg ratios observed in the presolar spinel and hibonite grains (Fig. 12a). In principle, massive AGB stars could variably enrich portions of the interstellar medium with $^{25}$Mg leading to a larger scatter in $^{25}$Mg/$^{24}$Mg ratios than predicted by the Monte Carlo model (that only includes supernovae). However, the relatively small amount of material ejected by a given AGB star means that many stars would have to contribute to a region of the ISM in order to have an appreciable effect on the isotopic composition, and statistical fluctuations would be very small. We thus think that binary mass transfer is a more likely explanation for the observed range in $^{25}$Mg/$^{24}$Mg ratios, but clearly additional modeling is needed. Second, the Type II supernova yields of $^{42}$Ca and $^{43}$Ca are highly correlated with each other (they are after all made by the same process), but uncorrelated with that of $^{44}$Ca (Woosley & Weaver 1995). Thus, inhomogeneous mixing of supernova ejecta would be expected to lead to more scatter on a $\delta^{44}$Ca/$^{40}$Ca vs $\delta^{42}$Ca/$^{40}$Ca plot, compared to a $\delta^{43}$Ca/$^{40}$Ca vs $\delta^{42}$Ca/$^{40}$Ca plot, as is observed (Fig. 11). Third, although the scatter in $^{44}$Ca/$^{40}$Ca ratios about the expected GCE trend is significant, it is not grossly larger than that of the other Ca isotopic ratios. This does not seem to support the idea that $^{44}$Ti (and its decay product $^{44}$Ca) is dominantly produced by rare sub-Chandrasekhar mass Type Ia supernova events happening very infrequently but ejecting up to 100 times more $^{44}$Ti than a typical Type II supernova. If the latter was the case, one might expect larger variations in $^{44}$Ca/$^{40}$Ca ratios than observed, as argued with regards to the presolar SiC grains by The et al. (2006), though detailed modeling is needed to thoroughly address this issue.

## 5. Group 4 ($^{18}$O-rich) grains

Some 10% of presolar oxide and silicate grains have $^{18}$O enrichments and are hence assigned to Group 4 (Nittler et al. 1997). Most Group 4 grains with $^{18}$O/$^{16}$O>0.003 (1.5 ×solar) define a linear trend with the magnitude of $^{17}$O enrichment (relative to solar) being approximately equal to the $^{18}O$ enrichment (i.e., ~slope 1 line on a $\delta^{18}O/^{16}O$ vs $\delta^{17}O/^{16}O$ plot). Four $^{18}O$-rich grains differ strikingly from this trend. One $Al_2O_3$ grain and two silicate grains have very large $^{18}O$ enrichments ($^{18}O/^{16}O$=0.006 to 0.0095), but close to solar $^{17}O/^{16}O$ ratios (Bland et al. 2007; Choi et al. 1998; Mostefaoui & Hoppe 2004). An additional olivine grain, found in an interplanetary dust particle (Messenger, Keller, & Lauretta 2005), has an extreme $^{18}O$ enrichment and $^{17}O$ depletion ($^{18}O/^{16}O$=0.027, $^{17}O/^{16}O=1.1\times 10^{-4}$, Fig. 2). Note that a few grains with small $^{18}O$ enrichments, but much larger $^{17}O$ enrichments, have been classified as Group 1 grains (e.g., Fig. 7), underscoring the difficulty in classifying some grains at the borders of group definitions.

The origin or origins of the Group 4 grains has been enigmatic. Nittler et al. (1997) discussed two possible sources. First these authors proposed that $^{18}O$ produced by the $^{14}N+\alpha$ reaction during He burning could be mixed to the surface during very early thermal pulses and third dredge-up events in AGB stars. Because later thermal pulses would destroy the $^{18}O$, this scenario requires dredge-up to occur during the earliest pulses to produce any $^{18}O$ enrichment in the envelope. In fact, subsequent calculations by Boothroyd & Sackmann (1999) did predict some $^{18}O$ enrichment during the second dredge-up in ~7$M_{\odot}$ stars of low metallicity. However, to our knowledge, no models have ever predicted $^{18}O$ enrichments in low-mass AGB stars due to dredge-up of partially He-burnt material. Moreover, AGB models do not predict third dredge-up to occur during the earliest thermal pulses, as required to produce large surface enhancements of $^{18}O$ (e.g., Stancliffe & Jeffery 2007). Also, the strong correlation between $^{17}O/^{16}O$ and $^{18}O/^{16}O$ observed in Group 4 grains would not be expected in this scenario. For these reasons, we thus consider early dredge-up in AGB stars to be an unlikely explanation for the Group 4 grains. Second, the high $^{18}O/^{16}O$ ratios could be due to GCE effects and indicate that the grains formed in high metallicity red giants and/or AGB stars. The high $^{25}Mg/^{24}Mg$ and $^{26}Mg/^{24}Mg$ ratios measured in one Group 4 $Al_2O_3$ grain (T22 with $\delta^{25}$Mg= 130±36 ‰, $\delta^{26}$Mg= 236±38 ‰, Nittler et al. 1997) was taken as supporting evidence for this suggestion. We also consider this scenario highly unlikely, however, both because grains from such stars would be expected to have significantly higher $^{17}O/^{16}O$ ratios than observed, due to the first dredge-up, and because very high metallicities (e.g., > $2Z_{\odot}$) would be required, higher than would be common at the time of Solar System formation.

With the discovery of the unusual $Al_2O_3$ grain S-C122, Choi et al. (1998) raised the possibility of a Type II supernova origin for $^{18}O$-rich grains. Just prior to exploding as supernovae, massive stars can be viewed as being comprised of concentric zones of material that have different chemical compositions due to their different nuclear burning histories (Meyer, Weaver, & Woosley 1995). Mixing of these zones during the expansion of the supernova ejecta is expected on theoretical grounds due to hydrodynamic instabilities (e.g., Kifonidis et al. 2003) and observed in supernova remnants (Hughes et al. 2000). Moreover, the isotopic compositions of presolar grains of SiC, graphite and $Si_3N_4$ from supernovae indicate extensive and heterogeneous mixing of supernova zones (Hoppe et al. 2000; Nittler et al. 1995; Travaglio et al. 1999; Yoshida & Hashimoto 2004; Yoshida, Umeda, & Nomoto 2005). Choi et al. (1998) showed that the isotopic composition of grain S-C122 could be explained by a mixture of a small amount of material from the He-burning zone (rich in $^{18}O$) with material from the H-rich envelope of 15 $M_{\odot}$ supernova. The $^{18}O$-rich and $^{17}O$-poor grain reported by Messenger et al. (2005) also most likely formed in a supernova. Although these grains differ in $^{17}O/^{16}O$ ratios from the bulk of Group 4 grains, these results suggest that other Group 4 grains could have originated in supernovae as well. Isotopic data for two Group 4 grains of the present study, hibonite KH2 and spinel UOC-S3, strongly support this conclusion.

Hibonite grain KH2 plots towards the $^{17}O$-rich and $^{18}O$-rich end of the main Group 4 O isotope trend (Fig. 2). It has the highest inferred $^{41}Ca/^{40}Ca$ ratio of any of the measured grains, $4 \times 10^{-4}$, a relatively high inferred initial $^{26}Al/^{27}Al$ ratio of ~0.01, a 30% depletion in $^{25}$Mg, minor depletions in $^{42}$Ca and $^{43}$Ca and a small excess of $^{44}$Ca. Spinel UOC-S3 also lies at the isotopically heavy end of the Group 4 trend and like KH2 shows a strong (~20%) depletion of $^{25}$Mg and $^{26}$Mg excess (~30%). If we assume that the $^{26}$Mg excess in this grain is due to extinct $^{26}$Al, the inferred initial $^{26}$Al/$^{27}$Al ratio is $\sim 1.5 \times 10^{-2}$. The large $^{25}$Mg depletions observed in these two grains argue strongly against a high-metallicity AGB star origin for these grains, since excesses in $^{25}$Mg would have been expected in that case. For example, the high $^{18}$O/$^{16}$O ratios of these two grains, if caused by GCE, would indicate parent metallicities of ~2-4 times $Z_{\odot}$, implying in turn extremely high $^{25}$Mg/$^{24}$Mg ratios. In fact, solar or sub-solar $^{25}$Mg/$^{24}$Mg ratios seems to be a common characteristic of Group 4 grains (Fig. 10b). Of ten $^{18}$O-rich grains analyzed for $^{25}$Mg/$^{24}$Mg ratios, only one ($Al_2O_3$ T22, Nittler et al. 1997) has a $^{25}$Mg excess and 4 have resolvable ($>2\sigma$) $^{25}$Mg depletions.

To investigate whether supernovae can account for the isotopic characteristics of these grains, we follow Meyer et al. (1995) and divide supernova ejecta into zones labeled by their most abundant elements (e.g., the O/C zone consists mostly of oxygen and carbon). We will consider a specific supernova model below, but for now we will begin with some general characteristics common to most or all models of Type II supernova nucleosynthesis. In terms of O isotopes and moving inwards, the H-rich envelope and He/N zone have $^{17}$O enrichments and $^{18}$O depletions due to H burning, the He/C zone shows low $^{17}$O/$^{16}$O and high $^{18}$O/$^{16}$O ratios from partial He burning, and the O/C, O/Ne and O/Si zones have very high $^{16}$O abundances from complete He burning. The inner Si/S and Ni zones have little O, but could contribute other elements to mixtures. The highest $^{26}$Al/$^{27}$Al ratios are found in the He/N zone, but there is $^{26}$Al throughout the ejecta including the envelope. Calcium-41 is made by He-burning and O-burning and is present in the He/C and O-rich zones. The C/O zone has very high $^{25,26}$Mg/$^{24}$Mg and $^{42,43,44}$Ca/$^{40}$Ca ratios from He burning, but these ratios are low in the inner zones. Based on these characteristics, one might qualitatively explain the Group 4 oxide data by requiring mixtures of the H envelope, He/N zone and He/C zone (to get high $^{17}$O/$^{16}$O, $^{18}$O/$^{16}$O and $^{26}$Al/$^{27}$Al ratios) with inner zone material (to get $^{25}$Mg, $^{42}$Ca and $^{43}$Ca depletions and high $^{41}$Ca/$^{40}$Ca ratios). Let us now examine this more quantitatively.

We consider the 15$M_{\odot}$ supernova model of Rauscher et al. (2002), specifically the model 's15a28c' available from the World Wide Web site www.nucleosynthesis.org. The model provides the abundances of several hundred isotopes throughout the ejecta (following passage of the supernova shock wave, but prior to any significant mixing that might occur). We integrated the abundances for all calculated isotopes over each of the zones (e.g., He/C) outlined above and then mixed these in varying amounts to attempt to reproduce grain compositions. The mass boundaries for the zones we defined are given in Table 5. Aluminum-26 (and $^{44}$Ti for grain KH2) was assumed to condense into the grains according to the measured Al/Mg (and Ti/Ca) ratios. This mixing exercise is essentially the same procedure used by many others in investigating presolar supernova grains (Choi et al. 1998; Hoppe et al. 2000; Travaglio et al. 1999; Yoshida & Hashimoto 2004). We did not attempt to mathematically find a best-fit match to the composition of any grain, for example by minimizing $\chi^2$, because significant uncertainties in supernova nucleosynthesis calculations preclude ascribing too much reality to the specific quantitative results. Our goal is primarily to show that the compositions of the grains can be best explained by a supernova origin, not to link a specific supernova model to a specific grain.

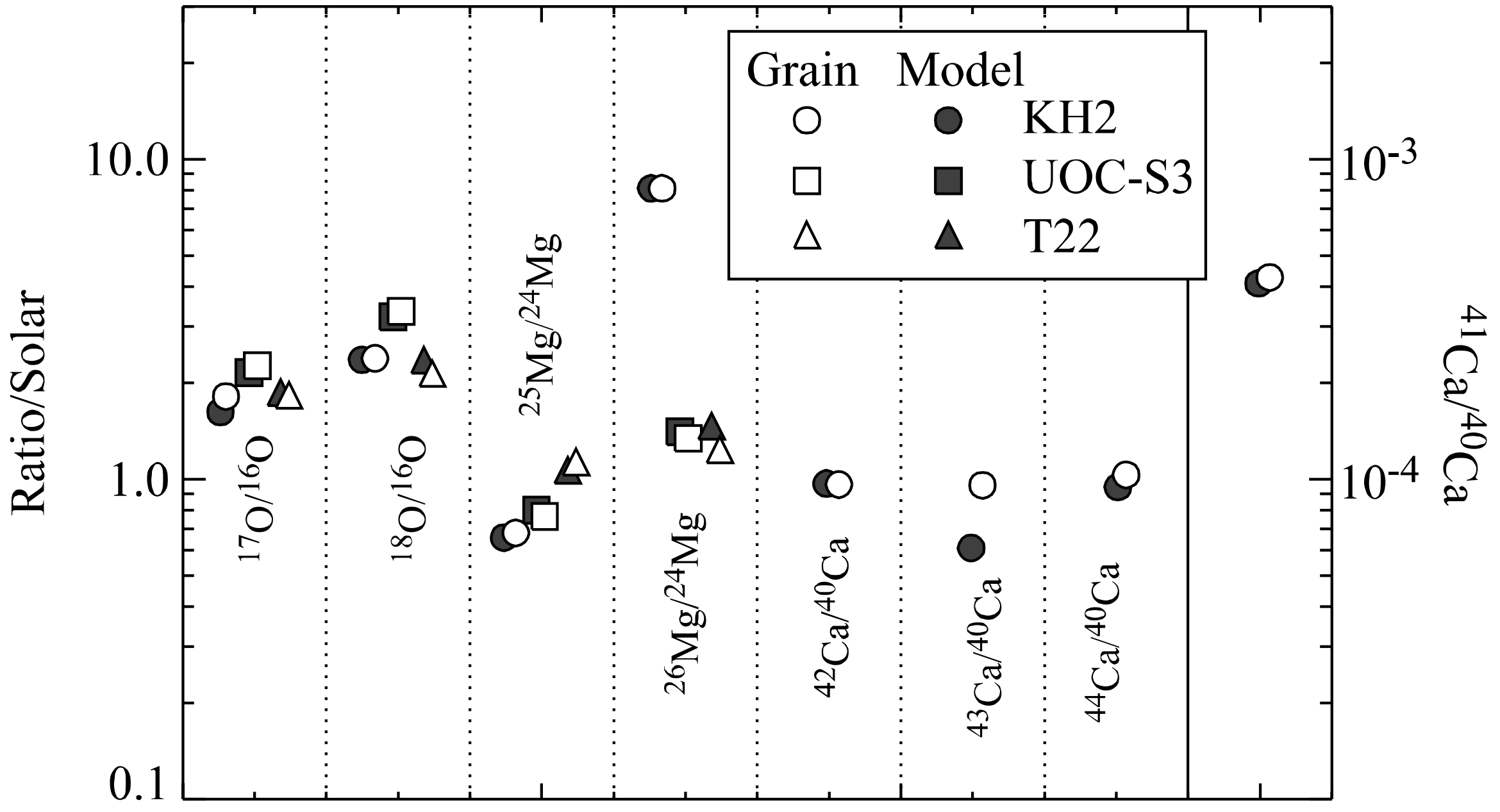

**Figure 13:** Comparison of isotopic ratios measured in three Group 4 oxide grains with those predicted by mixing of different zones of a 15 $M_\odot$ supernova. The absolute $^{41}Ca/^{40}Ca$ ratio is shown; all other ratios are normalized to solar values. With the exception of $^{43}Ca/^{40}Ca$ in Grain KH2, all ratios are reproduced within 10% by the mixing models, lending strong support to the idea that the Group 4 grains originated in supernovae. The $^{26}Mg/^{24}Mg$ and $^{44}Ca/^{40}Ca$ ratios include the effects of $^{26}Al$ and $^{44}Ti$ incorporation into the grains. Mixing fractions from the various zones are given in Table 5. Grain T22 was reported by Nittler et al. (1997).

Grain KH2 has the largest number of measured isotopic compositions and hence provides the tightest constraints on the supernova zones that can contribute to its composition. We found an excellent fit to the composition of this grain could be obtained by mixing the zones in the proportions given in Table 5. This mixture is dominated by material from the envelope (93%) and He/N zone (5%) and results in a C/O ratio of 0.26. As shown in Figure 13, this mixture reproduces the measured isotopic composition of this grain within 5-10%, with the exception of the $^{43}Ca/^{40}Ca$ ratio. Note that only 1% of material from the He/C zone is sufficient to explain the $^{18}O$ enrichment of this grain. Material from the O/Ne and O/C zones is excluded because inclusion of these zones leads to $^{25}Mg/^{24}Mg$ ratios that are much too high. Also, this mixture predicts a $^{26}Mg/^{24}Mg$ ratio of 0.1231 (~88% of solar), but reproduces the grain's higher $^{26}Mg/^{24}Mg$ ratio by including Al with $^{26}Al/^{27}Al=1.1 \times 10^{-2}$. Similarly, the grain's $^{44}Ca/^{40}Ca$ ratio is matched by including significant $^{44}Ti$ from the Ni zone. The inferred initial $^{44}Ti/^{48}Ti$ ratio is 0.13, comparable to that inferred for several presolar SiC and graphite grains believed to have formed from supernovae (Besmehn & Hoppe 2003; Hoppe et al. 2000; Nittler et al. 1996). Without this $^{44}Ti$, the predicted $^{44}Ca/^{40}Ca$ ratio is 0.016, or about 60% of solar, similar to the predicted $^{43}Ca/^{40}Ca$ ratio. We note that supernova models underproduce $^{43}Ca$, relative to its solar abundance (Timmes et al. 1995), and the discrepancy between the mixing model and grain KH2 for $^{43}Ca/^{40}Ca$ might well reflect this problem. The very good agreement between the mixing model and the grain composition for 7 measured isotopic ratios spanning three orders of magnitude in absolute value argues strongly that this grain formed in a Type II supernova that experienced extensive and heterogeneous mixing of various zones in the ejecta.

Table 5 and Figure 13 also give results for two other grains, UOC-S3 and T22 (Nittler et al. 1997), though these have fewer isotopic ratios to match than KH2. The O and Mg isotopic compositions of UOC-S3 are reproduced within 5% by a zone mixture quite similar to the one that matches KH2, except that a smaller portion of the mix is from the inner O-rich zones (to explain the smaller $^{25}$Mg depletion) and no material from the Ni zone is included. However, material could be included from the Ni zone without significantly affecting the O and Mg isotopes, so we cannot exclude the possibility that material from this zone was mixed into the region from which this grain condensed. We also attempted to match grain T22 because its $^{25}$Mg/$^{24}$Mg ratio is so different from those of the other Group 4 grains (Fig. 10b). We found that the composition of this grain can also be matched by an appropriate mixture of zones (Table 5). In contrast to the mixtures described above, this mixture does not include any material from zones interior to the O/C zone as these would lead to $^{25}$Mg depletions. The small amount of material from the O/C zone is sufficient to reproduce the $^{25}$Mg/$^{24}$Mg and $^{26}$Mg/$^{24}$Mg ratios and material from the He/N zone must be excluded to avoid high $^{26}$Al abundances that would raise the $^{26}$Mg/$^{24}$Mg ratio too much.

We consider two additional Group 4 grains. As discussed earlier, Choi et al. (1998) argued that mixing of the H envelope with a small amount of material from the He/C zone of a $15M_\odot$ supernova can quantitatively explain the composition of the $Al_2O_3$ grain S-C122. However, these authors did not consider $^{17}$O in their mixing calculation, because reliable $^{17}$O yields from supernovae were not available at the time. Using the Rauscher et al. (2002) model, which includes updated reaction rates involving $^{17}$O, we find that no mixture of He/C zone and envelope material can simultaneously explain the $^{17}$O/$^{16}$O and $^{18}$O/$^{16}$O ratios observed in this grain. A small amount of material from the inner $^{16}$O-rich zones is required to match the grain's O-isotopic composition. We have not found a specific mixture of zones that simultaneously reproduces this grain's O-isotopic composition and its mostly solar isotopic composition for Ca and Ti. However, since the errors on the Ca and Ti isotopic ratios are large, it is likely that a supernova mixing model that satisfactorily matches the data could be found with additional effort. We thus concur with Choi et al. (1998) that a supernova is the most likely source for this grain.

The pyroxene grain 6495A (Mostefaoui & Hoppe 2004) has an O–isotopic composition reminiscent of that of S-C122 ($^{18}$O/$^{16}$O= 3.8 × solar, normal $^{17}$O/$^{16}$O) as well as small enrichments in $^{29}$Si and $^{54}$Fe (7% and 12%, respectively). Mostefaoui & Hoppe (2004) argued that a high-metallicity red giant or AGB star was most likely the source of this grain. However, it is difficult to explain the grain's unusual O-isotopic composition with such an origin. If the very high $^{18}$O/$^{16}$O ratio is due to GCE, this would imply a parent star with metallicity several times higher than solar. Aside from the fact that such stars are exceedingly rare and probably were even more so at the time of Solar System formation, such a star is expected to have a much higher a $^{17}$O/$^{16}$O ratio than is observed, both because of GCE and the first dredge-up. Moreover, a much larger enrichment in $^{29}$Si would also be expected. While it is true that very large variations in Si and Fe isotopic abundances are expected in different layers of a Type II supernova, these variations are mainly found in the inner zones. As discussed for the other Group 4 grains above, the O isotopes of the Group 4 grains point to supernova mixtures dominated by material from the envelope, He/N and He/C zones, with only a minor component from the inner $^{16}$O-rich zones. Because these zones are so much dominated by $^{16}$O, a small admixture from them is enough to match grain 6495A without significantly affecting the Si or Fe isotopic composition. For example, we find that a mixture of the $15M_\odot$ zones (O/C):(He/C):(H) of about 1:3:96% can match this grain's O and Si isotopic composition within 13%. With these mixing parameters, the $^{54}$Fe/$^{56}$Fe ratio is close to that of the envelope, which is assumed to be solar in the supernova calculation. The fact that this ratio is slightly higher than solar in the grain would be easily explained if the parent massive star had an initial metallicity slightly higher than solar, as the $^{54}$Fe/$^{56}$Fe ratio is expected to increase with metallicity (Timmes et al. 1995). We thus think that the most likely origin of grain 6495A, as well as most or all of the other $^{18}$O-rich grains, was in the expanding ejecta of a Type II supernova.

Note that the GCE line shown on Figure 7 is based on the assumption that the $^{17}$O/$^{18}$O ratio (the ratio of two secondary isotopes) is constant and equal to the solar ratio of 0.19 throughout Galactic history. The majority of Group 3 ($^{17,18}$O-depleted) presolar grains plot above this GCE line, but a fraction plots below the line (dash-dot ellipse on Fig. 7). Because the first dredge-up increases $^{17}$O and decreases $^{18}$O, these grains' compositions cannot be explained by dredge-up in red giants or AGB stars unless those stars formed with $^{17}$O/$^{18}$O ratios much lower than solar and hence below the assumed GCE line. However, the solar $^{17}$O/$^{18}$O ratio is already lower by a factor of ~1.5 than that of typical molecular clouds in the Milky Way (Wilson & Rood 1994) and one would not expect even lower ratios for the initial compositions of parent stars of presolar grains. Interestingly, several of these grains appear to be co-linear on an O 3-isotope plot with many of the $^{17}$O- and $^{18}$O-rich Group 4 grains. For example, the long-dashed line on Figure 7 connects hibonite grain KH2 with the average composition of the $^{16}$O-enriched inner supernova zones of the 15M$_{\odot}$ model of Rauscher et al. (2002); this line also passes through or near several other Group 4 and low $^{17}$O/$^{18}$O Group 3 grains. Thus the O-isotopic compositions of these grains are consistent with mixing of $^{16}$O-rich material from a supernova interior with an end-member mixture of the He/C, He/N and H envelope. This points to supernovae as the sources of the unusual Group 3 grains with $^{17}$O/$^{18}$O ratios below the GCE line. Additional isotopic data for grains like these are highly desirable to test this suggestion.

We have shown above that the Group 4 grains (and perhaps some Group 3 grains) can clearly be explained as originating from supernovae. However, this identification raises some interesting puzzles. First it is somewhat surprising that the known supernova oxides (and silicates) seem to be dominated by $^{17}$O- and $^{18}$O-rich compositions. Most of the O ejected by supernovae is $^{16}$O synthesized in the inner zones, and one might expect a priori that supernova oxides should be mainly highly $^{16}$O-rich. However only one such grain has been found, T84 (Nittler et al. 1998), compared with several tens of Group 4 grains. Second, mixing of the different supernova zones in variable proportions would be expected to lead to highly variable O-isotopic compositions, yet most of the Group 4 grains appear to be consistent with a rather narrow range of mixing parameters. Mixing in supernova ejecta is observed to be highly heterogeneous and variable from remnant to remnant (Hwang et al. 2004; Park et al. 2004). This is consistent with the wide range of isotopic compositions observed in single presolar SiC and graphitic grains from supernovae (Hoppe et al. 2000; Travaglio et al. 1999). Given the extreme diversity in supernova compositions as a function of mass and heterogeneous mixing in ejecta, it is difficult to envision a realistic scenario in which different supernovae ended up producing grains primarily along a single mixing line as observed in the O-rich grains. Thus, if the Group 4 grains indeed have a supernova origin, the data point to a single source for the majority of them. For example, if the main Group 4 trend is indeed a mixing line, this might be explained by a single jet of inner zone material passing through and mixing to varying degrees with a pre-existing mixture of the outer layers (the $^{17,18}$O-rich end-member). What process might be responsible for this unique mixture of outer zone material is not known. In any case, this result is in contrast to the supernova-derived SiC and graphite grains, for which no evidence of a single, dominant source has been identified. Grains coming from a unique and rather arbitrary mixture of zones of a single supernova could be evidence for seeding of the early Solar System by a nearby explosion (e.g., Ouellette, Desch, & Hester 2007). This is clearly speculative and it remains to be seen whether such a picture could simultaneously explain the oxide grain data and the data for short-lived radionuclides in the early Solar System.

## 6. Presolar Titanium Oxide

Of 105 presolar oxide grains identified in this study, four turned out to be titanium oxide (Table 1). This identification is based solely on the presence of only Ti and O x-ray peaks in SEM-EDS spectra. Unfortunately, we do not have any crystallographic data for these grains that would allow us to determine their specific mineralogy (rutile, anatase, …) or even precise enough Ti/O ratios to determine whether the grains are stoichiometric $TiO_2$. The identified presolar Ti oxide grains differ from the other presolar oxides in the same residues in that they are on average smaller (≤0.8 μm compared to ~1 – 1.5μm) and in that they show a remarkably limited range of isotopic compositions compared to the other grains (see Figs. 2 and 6). Three of the grains belong to Group 1, with $^{17}O/^{16}O$ between 1.19 and 1.64 times solar, $^{18}O/^{16}O$ between 0.8 and 1 times solar and Ti isotopic compositions close to solar. The other grain, KT1, is a moderate Group 4 grain with $^{18}O/^{16}O$=1.2 times solar and normal $^{17}O/^{16}O$. Like the other Group 1 presolar grains, the $^{17}O$-rich Ti oxide grains most likely condensed in low-mass red giants or AGB stars. Their O-isotopic compositions indicate that their parent stars were of low mass (1.1-1.4 $M_{\odot}$) and had metallicities close to solar or slightly higher (Table 3). The origin of KT1 is less clear. Its $^{18}O$ enrichment puts it into the Group 4 class, but it is isotopically much less extreme than the Group 4 grains for which we argued a supernova origin in the previous section. It could plausibly have originated in a low-mass star with higher than solar $^{18}O/^{16}O$ ratio and somewhat lower than solar $^{17}O/^{16}O$ ratio, perhaps due to local heterogeneities in the interstellar medium. In any case, without additional isotopic data beyond O, it is difficult to say any more about this grain and we hope to find additional ones like it in the future.

Based on equilibrium thermodynamics, Ti oxides are expected to condense from a solar-composition gas at temperatures in the 1300-1400 K range, a few hundred degrees below the condensation temperatures of Al2O3 and hibonite (e.g., Ebel & Grossman 2000). On the basis of classical nucleation theory, Gail & Sedlmayr (1998) and Jeong et al. (1999) argued that $TiO_2$ is the first stable condensate in O-rich AGB stars, serving as a seed for the heterogeneous nucleation of $Al_2O_3$. However, no evidence for $TiO_2$-rich cores was found in a study of the microstructures of two presolar $Al_2O_3$ grains (Stroud et al. 2004). Moreover, Posch et al. (1999) showed that composite $TiO_2/Al_2O_3$ grains do not provide a good match to the infrared spectra of dusty O-rich AGB stars. Also, a recent experimental study by Demyk et al. (2004) indicates that Al oxide clusters are stable and can serve as seeds for $Al_2O_3$ growth, without the need for $TiO_2$ clusters. Thus, there is no compelling evidence that Ti oxides condense prior to the more refractory phases $Al_2O_3$ and hibonite.

The abundance ratio of Ti oxides to Al-rich grains identified in this study is about 0.04 (Table 1), comparable to the solar Ti/Al ratio of 0.029 (Lodders 2003). Thus, the low relative abundance of Ti oxide grains is easily explained on elemental abundance grounds alone. The low Ti/Al ratio would also lead to the expectation that Ti oxide condensates should be smaller on average than Al oxides, as observed, since there are fewer Ti atoms to add to growing grains (e.g., Bernatowicz et al. 2005). Finally, although the statistics are severely limited and we are biased in this study towards grains larger than ≈500 nm, the apparent bias towards Ti oxide grains from low-mass stars of relatively high metallicity can also be understood on the basis of kinetics arguments. For example, although concerned with grain growth in C stars, the calculations of Bernatowicz et al. (2005) suggest that larger grains form in lower-mass stars. Stars with low metallicity have fewer Ti atoms available and there might be a threshold metallicity below which there is insufficient Ti to grow micron-sized grains. If this is indeed the case, NanoSIMS studies of smaller Ti oxide grains in our residues might identify presolar grains with a wider range of O-isotopic compositions, indicating a wider range of masses and metallicities of parent stars.

## 7. Summary and Conclusions

We have presented isotopic data for some 100 new presolar oxide grains that we identified in acid-resistant residues of several primitive ordinary chondrite meteorites. In addition to the well-known $Al_2O_3$ and spinel ($MgAl_2O_4$) phases, we have greatly increased the database on presolar hibonite ($CaAl_{12}O_{19}$) and have reported the first discovery of Ti oxide as a presolar grain phase. We have also made a detailed comparison of the isotopic data for the presolar grains with theoretical models of stellar evolution and nucleosynthesis in both AGB stars and supernovae as well as Galactic chemical evolution. These discussions allow us to come to the following conclusions:

1) Presolar hibonite grains span a comparable range of O-isotopic compositions as presolar $Al_2O_3$ and spinel grains studied previously, and include members of the four grain Groups previously defined by Nittler et al. (1994; 1997). The O-isotopic ratios of Groups 1, 2 and 3 grains are most consistent with an origin in low-mass (<2.5 $M_{\odot}$) red giants and AGB stars, though red supergiants and Wolf-Rayet stars cannot be ruled out as the sources of a fraction of Group 1 and Group 2 grains, respectively.

2) Cool bottom processing, a mixing process modeled as circulation of envelope material near the H-burning shell of AGB stars, is required to explain strong $^{18}$O depletions in Group 2 grains and high inferred $^{26}$Al/$^{27}$Al ratios in many Group 1 and 2 grains. A range of CBP temperatures and mass circulation rates are required to reproduce the isotopic data. Comparison of the data with predictions of the CBP model of Nollett et al. (2003) indicates that the Group 2 grains formed in stars of mass 1.2 – 1.8$M_{\odot}$. Moreover, many grains classified as belonging to Group 1 likely experienced CBP as well.

3) A few grains have $^{17}$O/$^{16}$O ratios higher than can be explained by models of dredge-up in red giant stars. These grains might have originated in binary star systems where the parent stars' surfaces were polluted by $^{17}$O-rich material from a companion intermediate-mass AGB star or nova outburst.

4) Half of the hibonite grains analyzed for Ca-K isotopic compositions have $^{41}$K excesses attributable to *in situ* decay of now-extinct $^{41}$Ca. The inferred $^{41}$Ca/$^{40}$Ca ratios are in good agreement with those predicted for the envelopes of low-mass AGB stars. However, the highest observed ratios (>$10^{-4}$) require that the radioactive decay of $^{41}$Ca is suppressed in the envelopes of AGB stars, indicating that Ca is fully ionized in the envelope, increasing the lifetime against electron capture.

5) The presolar hibonite grains show a wide range of $^{25}$Mg/$^{24}$Mg ratios, as previously observed in presolar spinel grains (Zinner et al. 2005). However, in comparison to presolar spinel, fewer hibonite grains have solar $^{25}$Mg/$^{24}$Mg ratios. This result suggests that many presolar spinel grains have undergone Mg isotopic exchange, most likely in space, with material of solar composition. The range of $^{25}$Mg/$^{24}$Mg ratios in presolar hibonite is larger than can be explained by dredge-up of He-shell material in low-mass O-rich AGB stars and thus points to a large range of initial compositions of the parent stars. The hibonite data are more consistent with Galactic chemical evolution models that do not include a significant contribution of AGB stars to the Galactic Mg isotope inventory (Timmes et al. 1995) than models that do (Fenner et al. 2003). However, several grains whose O-isotopic ratios indicate low-metallicity parent stars have large excesses of $^{25}$Mg, not expected from GCE models. These data are plausibly explained by mass-transfer from an intermediate-mass AGB star companion in binary star systems. This result indicates the need for more consideration of the role of binary systems in studies of nucleosynthesis and GCE.

6) Presolar hibonite grains show a wide range of patterns of stable Ca-isotopic ($^{42,43,44}$Ca/$^{40}$Ca) ratios. Comparison with models indicates that these ratios largely reflect the initial compositions of parent stars, with only a minor contribution from nucleosynthesis in the parent stars themselves. The inferred initial Ca isotopic compositions are generally consistent with expectations for GCE with some scatter from local heterogeneities in the interstellar medium. However, the data suggest that the Sun has slightly (5-10%) anomalous $^{42}$Ca/$^{40}$Ca and $^{43}$Ca/$^{40}$Ca ratios for its metallicity. Also, the scatter in $^{44}$Ca/$^{40}$Ca is not significantly larger than that for the other Ca isotopic ratios as has been predicted based on models of $^{44}$Ca nucleosynthesis in the Galaxy (The et al. 2006).

7) Isotopic data for two new grains, one hibonite (KH2) and one spinel (UOC-S3), are strong evidence for a supernova origin of the enigmatic Group 4 ($^{18}$O-enriched) grains. Both grains show strong $^{25}$Mg depletions, relative to solar, inconsistent with a high-metallicity source for Group 4 grains, and high inferred $^{26}$Al/$^{27}$Al and $^{41}$Ca/$^{40}$Ca ratios point to a supernova source. The isotopic compositions of these, and other Group 4 grains, can be very well reproduced by mixing of different layers of a 15$M_{\odot}$ supernova model (Rauscher et al. 2002). Moreover, the linear trend of O-isotopic ratios observed for most Group 4 and several Group 3 grains suggests that these grains might have mainly formed in a single supernova.

8) Three of the identified Ti oxide grains have O- and Ti-isotopic compositions indicating that they formed in low-mass AGB stars of close-to-solar metallicity. The fourth grain is slightly $^{18}O$-rich and its origin is unknown. The rarity and smaller size of presolar Ti oxide, compared to presolar $Al_2O_3$, are explained by the low solar Ti/Al ratio and are consistent with expectations for condensation in AGB outflows.

***Acknowledgements***
We thank Amanda Karakas, Alexander Heger and Andy Davis for providing stellar yield data. We thank Don Clayton, Amanda Karakas, John Lattanzio, Maria Lugaro, Ken Nollett and Rhonda Stroud for helpful discussions, Glenn MacPherson and Rhonda Stroud for use of scanning electron microscopes early in this project, and Lara Stroud for providing the Amazonite standard. We are grateful to Sasha Krot for a constructive and positive review. This work has been supported by several NASA grants to C.A., L.N. and E. Z. R.G. acknowledges support by the Italian MIUR-FIRB 2006 Project "Final Phases of Stellar Evolution, Nucleosynthesis in Supernovae, AGB stars, Planetary Nebulae" and A.N. acknowledges support of the Carnegie Institution of Washington.

Table 1: Analyzed sample mounts.

| Mount | Meteorite | #grains analyzed (# good[a]) | Presolar Grains | | | | | |
|---|---|---|---|---|---|---|---|---|
| | | | $Al_2O_3$ | Spinel | Ti Oxide | Hibonite | Unknown | Total |
| OC1 | OCs[b] | 2170 (666) | 4 | 6 | 1 | | | 11 |
| OC2 | OCs[b] | 834 (430) | 2 | 1 | | | 1 | 4 |
| TC301 | Tieschitz | 2287(1259) | 20 | | | | | 20 |
| KR1 | Krymka | 1769(1717) | 6 | | | 6 | | 12 |
| KR3 | Krymka | 7073(5492) | 27 | 2 | 2 | 15 | 3 | 49 |
| UOC1 | UOCs[c] | 4116 (1914) | 1 | 3 | 1 | 2 | | 7 |

[a]Number of grains remaining after removal of grains that had completely sputtered away or had enormous error bars.
[b]Mixture of primitive ordinary chondrites: Semarkona, Tieschitz, Bishunpur, and Krymka.
[c]Mixture of primitive Antarctic ordinary chondrites: QUE 97008, WSG 95300 and MET 00452.

Table 2: Oxygen and Mg-Al isotopic data for presolar $Al_2O_3$, spinel and hibonite grains (1σ errors).

| Grain | Phase | Group | $^{17}O/^{16}O$ ± 1σ | $^{18}O/^{16}O$ ± 1σ | $\delta^{25}Mg/^{24}Mg^a$ ± 1σ | $\delta^{26}Mg/^{24}Mg^a$ ± 1σ | $Al/^{24}Mg$ | $^{26}Al/^{27}Al$ ± 1σ | Initial $^{18}O/^{16}O^b$ (rel. Solar) | Mass$^b$ ($M_\odot$) |
|---|---|---|---|---|---|---|---|---|---|---|
| | | | | | Solar | | | | | |
| | | | 0.0003829 | 0.0020052 | ≡0 | ≡0 | | | | |
| | | | | | Mixed OCs | | | | | |
| OC4 | $Al_2O_3$ | 1 | $9.46 \pm 1.47 \times 10^{-4}$ | $1.51 \pm 0.22 \times 10^{-3}$ | | | | | 0.94 | 1.6 |
| OC5 | $Al_2O_3$ | 1 | $5.29 \pm 0.23 \times 10^{-4}$ | $1.94 \pm 0.12 \times 10^{-3}$ | | | | | 1.12 | 1.3 |
| OC6 | $Al_2O_3$ | 1 | $4.68 \pm 0.10 \times 10^{-4}$ | $1.69 \pm 0.06 \times 10^{-3}$ | | | | | 0.98 | 1.2 |
| OC8 | Unknown | 2 | $9.10 \pm 1.12 \times 10^{-4}$ | $6.49 \pm 3.29 \times 10^{-4}$ | | | | | | |
| OC9 | $Al_2O_3$ | 1 | $5.40 \pm 0.12 \times 10^{-4}$ | $1.46 \pm 0.05 \times 10^{-3}$ | -6 ± 13 | 9770 ± 140 | 157 | $8.6 \pm 0.1 \times 10^{-3}$ | 0.87 | 1.4 |
| OC12 | $Al_2O_3$ | 1 | $5.23 \pm 0.22 \times 10^{-4}$ | $1.82 \pm 0.06 \times 10^{-3}$ | | | | | 1.07 | 1.3 |
| OC14 | $Al_2O_3$ | 2 | $5.32 \pm 0.19 \times 10^{-4}$ | $8.12 \pm 0.41 \times 10^{-4}$ | -80 ± 60 | 440 ± 180 | 96 | $7.2 \pm 1.6 \times 10^{-4}$ | | |
| UOC-S1 | Spinel | 1 | $5.35 \pm 0.36 \times 10^{-4}$ | $1.29 \pm 0.08 \times 10^{-3}$ | -4 ± 11 | 241 ± 14 | 2.8 | $1.18 \pm 0.07 \times 10^{-2}$ | 0.76 | 1.4 |
| UOC-S2 | Spinel | 2 | $8.52 \pm 1.33 \times 10^{-4}$ | $5.49 \pm 1.40 \times 10^{-3}$ | 256 ± 10 | 1020 ± 17 | 2.5 | $4.8 \pm 0.3 \times 10^{-2}$ | | |
| UOC-S3 | Spinel | 4 | $8.67 \pm 1.07 \times 10^{-4}$ | $6.70 \pm 0.38 \times 10^{-3}$ | -235 ± 11 | 342 ± 17 | 2.7 | $1.8 \pm 0.1 \times 10^{-2}$ | | |
| UOC-C1 | $Al_2O_3$ | 1 | $1.11 \pm 0.05 \times 10^{-3}$ | $1.86 \pm 0.09 \times 10^{-3}$ | | | | | 1.16 | 1.7 |
| UOC-H1 | Hibonite | 4 | $4.48 \pm 0.13 \times 10^{-4}$ | $2.13 \pm 0.06 \times 10^{-3}$ | -12 ± 14 | 14900 ± 150 | 160 | $1.3 \pm 0.8 \times 10^{-3}$ | | |
| UOC-H2 | Hibonite | 2 | $9.01 \pm 0.63 \times 10^{-4}$ | $4.95 \pm 0.66 \times 10^{-4}$ | 318 ± 37 | 20660 ± 360 | 49 | $5.9 \pm 0.3 \times 10^{-2}$ | | |
| | | | | | Tieschitz | | | | | |
| T95 | $Al_2O_3$ | 1 | $3.94 \pm 0.12 \times 10^{-3}$ | $1.69 \pm 0.05 \times 10^{-3}$ | 23 ± 83 | 36 ± 80 | 1000 | $< 2.8 \times 10^{-5}$ | 1.11 | 2.5 |
| T96 | $Al_2O_3$ | 1 | $1.96 \pm 0.04 \times 10^{-3}$ | $1.59 \pm 0.05 \times 10^{-3}$ | 125 ± 59 | 619 ± 67 | 1360 | $6.3 \pm 0.9 \times 10^{-5}$ | 1.04 | 1.9 |
| T97 | $Al_2O_3$ | 1 | $1.54 \pm 0.02 \times 10^{-3}$ | $1.53 \pm 0.03 \times 10^{-3}$ | 37 ± 60 | $4.1 \pm 0.3 \times 10^{4}$ | 1700 | $3.3 \pm 0.5 \times 10^{-3}$ | 0.99 | 1.8 |
| T98 | $Al_2O_3$ | 1 | $1.37 \pm 0.08 \times 10^{-3}$ | $1.68 \pm 0.12 \times 10^{-3}$ | | | | | 1.07 | 1.7 |
| T99 | $Al_2O_3$ | 1 | $5.31 \pm 0.30 \times 10^{-4}$ | $1.08 \pm 0.06 \times 10^{-3}$ | 102 ± 70 | 380 ± 100 | 260 | $2.0 \pm 0.6 \times 10^{-4}$ | 0.65 | 1.4 |
| T100 | $Al_2O_3$ | 3 | $3.33 \pm 0.18 \times 10^{-4}$ | $1.29 \pm 0.05 \times 10^{-3}$ | -18 ± 41 | -4 ± 40 | 500 | $< 2.2 \times 10^{-5}$ | 0.73 | 1.2 |
| T101 | $Al_2O_3$ | 4 | $4.15 \pm 0.09 \times 10^{-4}$ | $2.12 \pm 0.05 \times 10^{-3}$ | -10 ± 29 | 98 ± 34 | 1160 | $1.2 \pm 0.4 \times 10^{-5}$ | | |
| T102 | $Al_2O_3$ | 1 | $1.90 \pm 0.09 \times 10^{-3}$ | $1.96 \pm 0.45 \times 10^{-3}$ | -26 ± 19 | -24 ± 18 | 81 | $< 2.2 \times 10^{-5}$ | 1.28 | 2.0 |
| T103 | $Al_2O_3$ | 1 | $1.55 \pm 0.07 \times 10^{-3}$ | $1.30 \pm 0.11 \times 10^{-3}$ | 31 ± 22 | 640 ± 110 | 318 | $2.8 \pm 0.8 \times 10^{-4}$ | 0.84 | 1.8 |
| T104 | $Al_2O_3$ | 1 | $1.59 \pm 0.04 \times 10^{-3}$ | $1.96 \pm 0.06 \times 10^{-3}$ | -11 ± 14 | 2360 ± 150 | 600 | $5.4 \pm 0.9 \times 10^{-4}$ | 1.26 | 1.9 |
| T105 | $Al_2O_3$ | 1 | $1.30 \pm 0.04 \times 10^{-3}$ | $9.69 \pm 0.54 \times 10^{-4}$ | 7 ± 36 | 11770 ± 440 | 440 | $3.7 \pm 0.6 \times 10^{-3}$ | 0.62 | 1.7 |
| T106 | $Al_2O_3$ | 2 | $1.10 \pm 0.06 \times 10^{-3}$ | $2.03 \pm 0.49 \times 10^{-4}$ | -3 ± 42 | 20890 ± 980 | 291 | $1.00 \pm 0.16 \times 10^{-2}$ | | |
| T107 | $Al_2O_3$ | 2 | $9.59 \pm 0.51 \times 10^{-4}$ | $5.58 \pm 1.92 \times 10^{-4}$ | 32 ± 42 | 44 ± 42 | 84 | $< 2.1 \times 10^{-4}$ | | |
| T108 | $Al_2O_3$ | 1 | $7.49 \pm 0.87 \times 10^{-4}$ | $1.26 \pm 0.11 \times 10^{-3}$ | 27 ± 31 | 24 ± 30 | 141 | $< 8.2 \times 10^{-5}$ | 0.77 | 1.5 |
| T109 | $Al_2O_3$ | 3 | $2.64 \pm 0.20 \times 10^{-4}$ | $9.67 \pm 0.74 \times 10^{-4}$ | 8 ± 42 | 940 ± 130 | 194 | $6.8 \pm 1.2 \times 10^{-4}$ | 0.55 | 1.1 |
| T110 | $Al_2O_3$ | 1 | $4.05 \pm 0.26 \times 10^{-4}$ | $1.38 \pm 0.11 \times 10^{-3}$ | | | | | 0.80 | 1.2 |
| T111 | $Al_2O_3$ | 1 | $5.10 \pm 0.25 \times 10^{-4}$ | $1.65 \pm 0.10 \times 10^{-3}$ | -49 ± 47 | 16810 ± 540 | 1300 | $1.8 \pm 0.3 \times 10^{-3}$ | 0.97 | 1.3 |
| T112 | $Al_2O_3$ | 1 | $9.55 \pm 0.44 \times 10^{-4}$ | $1.71 \pm 0.17 \times 10^{-3}$ | 35 ± 42 | 46 ± 41 | 280 | $< 6.4 \times 10^{-5}$ | 1.06 | 1.6 |
| T113 | $Al_2O_3$ | 4 | $4.25 \pm 0.15 \times 10^{-4}$ | $2.26 \pm 0.06 \times 10^{-3}$ | 43 ± 44 | 15 ± 43 | 555 | $< 2.5 \times 10^{-5}$ | | |
| T114[c] | $Al_2O_3$ | 1 | $6.41 \pm 0.34 \times 10^{-4}$ | $1.47 \pm 0.05 \times 10^{-3}$ | | | | | 1.75 × 10-3 | 1.4 |
| | | | | | Krymka | | | | | |
| KC1[c] | $Al_2O_3$ | 2 | $7.30 \pm 0.19 \times 10^{-4}$ | $2.45 \pm 0.22 \times 10^{-4}$ | | | | | | |
| KC2[c] | $Al_2O_3$ | 1 | $1.66 \pm 0.03 \times 10^{-3}$ | $1.55 \pm 0.02 \times 10^{-3}$ | | | | | 1.00 | 1.8 |
| KC3[c] | $Al_2O_3$ | 1 | $5.11 \pm 0.05 \times 10^{-4}$ | $1.51 \pm 0.01 \times 10^{-3}$ | | | | | 0.89 | 1.3 |
| KC4[c] | $Al_2O_3$ | 1 | $3.34 \pm 0.07 \times 10^{-3}$ | $1.91 \pm 0.05 \times 10^{-3}$ | | | | | 1.25 | 2.2 |
| KC5[c] | $Al_2O_3$ | 1 | $2.58 \pm 0.03 \times 10^{-3}$ | $1.95 \pm 0.02 \times 10^{-3}$ | | | | | 1.27 | 2.1 |
| KC6 | $Al_2O_3$ | 1 | $8.82 \pm 0.40 \times 10^{-4}$ | $1.38 \pm 0.08 \times 10^{-3}$ | | | | | 0.86 | 1.5 |
| KC7 | $Al_2O_3$ | 2 | $1.17 \pm 0.06 \times 10^{-3}$ | $2.16 \pm 0.46 \times 10^{-4}$ | | | | | | |
| KC8 | $Al_2O_3$ | 1 | $1.35 \pm 0.08 \times 10^{-3}$ | $1.63 \pm 0.12 \times 10^{-3}$ | | | | | 1.04 | 1.7 |
| KC9 | $Al_2O_3$ | 1 | $2.38 \pm 0.04 \times 10^{-3}$ | $1.85 \pm 0.04 \times 10^{-3}$ | | | | | 1.21 | 2.1 |
| KC10 | $Al_2O_3$ | 2 | $1.20 \pm 0.03 \times 10^{-3}$ | $2.30 \pm 0.94 \times 10^{-5}$ | | | | | | |
| KC11 | $Al_2O_3$ | 2 | $6.62 \pm 0.54 \times 10^{-4}$ | $5.18 \pm 0.69 \times 10^{-4}$ | | | | | | |
| KC12 | $Al_2O_3$ | 1 | $1.76 \pm 0.11 \times 10^{-3}$ | $1.92 \pm 0.16 \times 10^{-3}$ | | | | | 1.25 | 1.9 |
| KC13 | $Al_2O_3$ | 2 | $1.30 \pm 0.07 \times 10^{-3}$ | $2.19 \pm 0.59 \times 10^{-4}$ | | | | | | |
| KC14 | $Al_2O_3$ | 1 | $5.71 \pm 0.50 \times 10^{-4}$ | $1.12 \pm 0.07 \times 10^{-3}$ | | | | | 0.68 | 1.4 |
| KC15 | $Al_2O_3$ | 1 | $2.18 \pm 0.14 \times 10^{-3}$ | $1.17 \pm 0.17 \times 10^{-3}$ | | | | | 0.77 | 1.9 |
| KC16 | $Al_2O_3$ | 1 | $5.04 \pm 0.59 \times 10^{-4}$ | $1.25 \pm 0.13 \times 10^{-3}$ | | | | | 0.75 | 1.3 |
| KC17 | $Al_2O_3$ | 1 | $1.63 \pm 0.04 \times 10^{-3}$ | $1.58 \pm 0.07 \times 10^{-3}$ | | | | | 1.02 | 1.8 |
| KC18 | $Al_2O_3$ | 1 | $5.71 \pm 0.45 \times 10^{-4}$ | $1.40 \pm 0.10 \times 10^{-3}$ | | | | | 0.84 | 1.4 |
| KC19 | $Al_2O_3$ | 2 | $1.26 \pm 0.10 \times 10^{-3}$ | $1.87 \pm 0.32 \times 10^{-4}$ | | | | | | |
| KC20 | $Al_2O_3$ | 1 | $5.88 \pm 0.39 \times 10^{-4}$ | $1.66 \pm 0.13 \times 10^{-3}$ | | | | | 0.99 | 1.4 |
| KC21 | $Al_2O_3$ | 1 | $3.28 \pm 0.16 \times 10^{-3}$ | $2.01 \pm 0.11 \times 10^{-3}$ | | | | | 1.31 | 2.2 |
| KC22 | $Al_2O_3$ | 2 | $8.99 \pm 0.57 \times 10^{-4}$ | $4.78 \pm 0.27 \times 10^{-4}$ | | | | | | |
| KC23 | $Al_2O_3$ | 1 | $5.85 \pm 0.18 \times 10^{-3}$ | $2.19 \pm 0.06 \times 10^{-3}$ | 45 ± 35 | 6 ± 35 | 769 | $< 1.1 \times 10^{-5}$ | | |
| KC24 | $Al_2O_3$ | 2 | $6.54 \pm 0.54 \times 10^{-4}$ | $6.67 \pm 0.48 \times 10^{-4}$ | | | | | | |
| KC25 | $Al_2O_3$ | 2 | $1.06 \pm 0.03 \times 10^{-3}$ | $1.05 \pm 0.11 \times 10^{-4}$ | 86 ± 159 | $1.20 \pm 0.06 \times 10^{6}$ | 16500 | $1.01 \pm 0.05 \times 10^{-2}$ | | |
| KC26 | $Al_2O_3$ | 1 | $6.54 \pm 0.29 \times 10^{-4}$ | $1.15 \pm 0.05 \times 10^{-3}$ | 243 ± 221 | 150 ± 210 | 16100 | $<4.4 \times 10^{-6}$ | 0.70 | 1.4 |
| KC27[c] | $Al_2O_3$ | 1 | $1.70 \pm 0.02 \times 10^{-3}$ | $1.38 \pm 0.08 \times 10^{-3}$ | 0 ± 53 | 24090 ± 580 | 997 | $3.37 \pm 0.08 \times 10^{-3}$ | 0.90 | 1.8 |
| KC28 | $Al_2O_3$ | 2 | $5.85 \pm 0.44 \times 10^{-4}$ | $7.51 \pm 0.47 \times 10^{-4}$ | 125 ± 50 | $6.45 \pm 0.10 \times 10^{5}$ | 1425 | $6.3 \pm 0.1 \times 10^{-2}$ | | |
| KC29 | $Al_2O_3$ | 1 | $4.80 \pm 0.46 \times 10^{-4}$ | $1.17 \pm 0.09 \times 10^{-3}$ | -288 ± 140 | $5.09 \pm 0.26 \times 10^{5}$ | 5130 | $1.38 \pm 0.07 \times 10^{-2}$ | 0.70 | 1.3 |
| KC30 | $Al_2O_3$ | 2 | $6.61 \pm 0.49 \times 10^{-4}$ | $3.76 \pm 0.78 \times 10^{-4}$ | 55 ± 24 | 8470 ± 130 | 190 | $6.2 \pm 0.1 \times 10^{-3}$ | | |
| KC31 | $Al_2O_3$ | 2 | $1.26 \pm 0.06 \times 10^{-3}$ | $1.37 \pm 0.26 \times 10^{-4}$ | -76 ± 93 | $3.92 \pm 0.12 \times 10^{5}$ | 1580 | $3.4 \pm 0.1 \times 10^{-2}$ | | |
| KC32 | $Al_2O_3$ | 2 | $1.17 \pm 0.04 \times 10^{-3}$ | $9.92 \pm 1.48 \times 10^{-5}$ | -41 ± 245 | $1.76 \pm 0.13 \times 10^{6}$ | 22700 | $1.08 \pm 0.08 \times 10^{-2}$ | | |
| KC33[c] | $Al_2O_3$ | 1 | $8.22 \pm 0.06 \times 10^{-3}$ | $6.80 \pm 0.80 \times 10^{-4}$ | | | | | | |
| KU1 | Unknown | 1 | $4.66 \pm 0.18 \times 10^{-4}$ | $1.91 \pm 0.05 \times 10^{-3}$ | | | | | 1.08 | 1.2 |
| KU2 | Unknown | 1 | $1.80 \pm 0.10 \times 10^{-3}$ | $2.47 \pm 0.22 \times 10^{-3}$ | | | | | 1.61 | 2.0 |
| KU3 | Unknown | 1 | $1.25 \pm 0.13 \times 10^{-3}$ | $2.08 \pm 0.12 \times 10^{-3}$ | | | | | 1.31 | 1.7 |
| KH1[c] | Hibonite | 2 | $6.59 \pm 0.11 \times 10^{-4}$ | $2.25 \pm 0.38 \times 10^{-4}$ | | | | | | |
| KH2[c] | Hibonite | 4 | $6.95 \pm 0.09 \times 10^{-4}$ | $4.78 \pm 0.04 \times 10^{-3}$ | -320 ± 15 | 7090 ± 120 | 109 | $9.1 \pm 0.2 \times 10^{-3}$ | | |
| KH3[c] | Hibonite | 1 | $7.47 \pm 0.11 \times 10^{-4}$ | $1.18 \pm 0.02 \times 10^{-3}$ | | | | | 0.73 | 1.5 |
| KH4[c] | Hibonite | 1 | $8.70 \pm 0.07 \times 10^{-4}$ | $1.53 \pm 0.01 \times 10^{-3}$ | -77 ± 17 | 166 ± 22 | 130 | $1.8 \pm 0.2 \times 10^{-4}$ | 0.95 | 1.5 |
| KH5[c] | Hibonite | 1 | $5.03 \pm 0.08 \times 10^{-4}$ | $1.49 \pm 0.01 \times 10^{-3}$ | | | | | 0.88 | 1.3 |
| KH6[c] | Hibonite | 1 | $5.77 \pm 0.08 \times 10^{-4}$ | $1.58 \pm 0.02 \times 10^{-3}$ | | | | | 0.94 | 1.4 |
| KH7 | Hibonite | 1 | $4.64 \pm 0.08 \times 10^{-4}$ | $1.63 \pm 0.05 \times 10^{-3}$ | -17 ± 14 | 13120 ± 190 | 85 | $2.16 \pm 0.03 \times 10^{-2}$ | 0.95 | 1.2 |
| KH8 | Hibonite | 1 | $1.75 \pm 0.03 \times 10^{-3}$ | $1.44 \pm 0.04 \times 10^{-3}$ | 49 ± 28 | 31 ± 28 | 105 | $< 9.3 \times 10^{-5}$ | 0.93 | 1.8 |
| KH9 | Hibonite | 1 | $4.51 \pm 0.14 \times 10^{-4}$ | $1.60 \pm 0.08 \times 10^{-3}$ | -9 ± 18 | 12900 ± 200 | 130 | $1.43 \pm 0.02 \times 10^{-2}$ | 0.93 | 1.2 |
| KH10 | Hibonite | 1 | $8.39 \pm 0.39 \times 10^{-4}$ | $1.96 \pm 0.09 \times 10^{-3}$ | 214 ± 38 | 4040 ± 110 | 180 | $3.09 \pm 0.09 \times 10^{-3}$ | 1.19 | 1.5 |
| KH11 | Hibonite | 3 | $3.24 \pm 0.30 \times 10^{-4}$ | $1.63 \pm 0.10 \times 10^{-3}$ | 13 ± 14 | -28 ± 16 | 81 | $< 4.3 \times 10^{-5}$ | 0.84 | 1.0 |
| KH12 | Hibonite | 1 | $2.00 \pm 0.07 \times 10^{-3}$ | $1.27 \pm 0.06 \times 10^{-3}$ | 69 ± 30 | 880 ± 47 | 150 | $8.4 \pm 0.4 \times 10^{-4}$ | 0.83 | 1.9 |
| KH13 | Hibonite | 2 | $1.15 \pm 0.04 \times 10^{-3}$ | $5.87 \pm 0.20 \times 10^{-4}$ | 237 ± 20 | 41340 ± 570 | 135 | $4.27 \pm 0.06 \times 10^{-2}$ | | |
| KH14 | Hibonite | 1 | $1.21 \pm 0.04 \times 10^{-3}$ | $1.72 \pm 0.05 \times 10^{-3}$ | -212 ± 20 | 10910 ± 190 | 200 | $7.5 \pm 0.1 \times 10^{-3}$ | 1.09 | 1.7 |
| KH15 | Hibonite | 2 | $1.17 \pm 0.04 \times 10^{-3}$ | $4.66 \pm 0.17 \times 10^{-4}$ | -68 ± 14 | 2882 ± 55 | 49 | $8.2 \pm 0.2 \times 10^{-3}$ | 0.00 | |
| KH16 | Hibonite | 1 | $6.20 \pm 0.67 \times 10^{-4}$ | $1.52 \pm 0.10 \times 10^{-3}$ | -125 ± 24 | 18960 ± 260 | 235 | $1.12 \pm 0.02 \times 10^{-2}$ | 0.91 | 1.4 |
| KH17 | Hibonite | 1 | $2.92 \pm 0.06 \times 10^{-3}$ | $1.52 \pm 0.09 \times 10^{-3}$ | 82 ± 22 | 234 ± 25 | 89 | $4.4 \pm 0.5 \times 10^{-4}$ | 0.99 | 2.1 |
| KH18 | Hibonite | 2 | $9.31 \pm 0.19 \times 10^{-4}$ | $7.02 \pm 0.22 \times 10^{-4}$ | 197 ± 15 | 30490 ± 290 | 66 | $7.76 \pm 0.07 \times 10^{-2}$ | | |
| KH19 | Hibonite | 1 | $7.61 \pm 0.77 \times 10^{-4}$ | $1.30 \pm 0.21 \times 10^{-3}$ | -113 ± 78 | 5600 ± 310 | 238 | $4.0 \pm 0.2 \times 10^{-3}$ | 0.80 | 1.5 |
| KH20 | Hibonite | 1 | $7.15 \pm 0.22 \times 10^{-4}$ | $1.96 \pm 0.08 \times 10^{-3}$ | | | | | 1.17 | 1.4 |
| KH21[c] | Hibonite | 1 | $6.84 \pm 0.16 \times 10^{-4}$ | $1.23 \pm 0.08 \times 10^{-3}$ | -198 ± 11 | 13900 ± 140 | 132 | $1.78 \pm 0.02 \times 10^{-2}$ | 0.75 | 1.5 |
| KS1 | Spinel | 3 | $3.31 \pm 0.82 \times 10^{-4}$ | $1.43 \pm 0.10 \times 10^{-3}$ | | | | | 0.80 | 1.1 |

| | | | | | | | | | |
|---|---|---|---|---|---|---|---|---|---|
| KS2 | Spinel | 3 | $3.05 \pm 0.40 \times 10^{-4}$ | $1.37 \pm 0.14 \times 10^{-3}$ | | | | 0.75 | 1.1 |
| Murray | | | | | | | | | |
| M16[d] | Spinel | 2 | $1.10 \pm 0.49 \times 10^{-3}$ | $1.93 \pm 0.40 \times 10^{-4}$ | 192 ± 21 | 719 ± 25 | 3.0 | $2.5 \pm 0.3 \times 10^{-2}$ | |

[a] See Figure 3 for normalizing ratios.
[b] Inferred by interpolation of dredge-up models of Boothroyd and Sackmann (1999).
[c] O-isotopic ratios based on NanoSIMS re-measurement of grains identified by ims-6f.
[d] Re-measurement of grain reported by Zinner et al. (2005).

Table 3: Oxygen and Titanium isotopic compositions of presolar titanium oxide grains. Normalization ratios for δ-values are given in parenthesis.

| GRAIN | Group | $^{17}O/^{16}O$ ± 1σ | $^{18}O/^{16}O$ ± 1σ | $\delta^{46}Ti/^{48}Ti$ (0.108548) ± 1σ | $\delta^{47}Ti/^{48}Ti$ (0.0993150) ± 1σ | $\delta^{49}Ti/^{48}Ti$ (0.0744630) ± 1σ | $\delta^{50}Ti/^{48}Ti$ (0.0724180) ± 1σ | Initial $^{18}O/^{16}O$[a] (rel. Solar) | Mass[a] ($M_\odot$) |
|---|---|---|---|---|---|---|---|---|---|
| OC13 | 1 | $4.54 \pm 0.09 \times 10^{-4}$ | $1.97 \pm 0.07 \times 10^{-3}$ | -14 ± 5 | -9 ± 3 | 5 ± 3 | 11 ± 5 | 1.10 | 1.1 |
| KT1 | 4 | $4.05 \pm 0.51 \times 10^{-4}$ | $2.47 \pm 0.11 \times 10^{-3}$ | | | | | | |
| KT2 | 1 | $4.84 \pm 0.28 \times 10^{-4}$ | $1.90 \pm 0.12 \times 10^{-3}$ | -7 ± 12 | -9 ± 10 | 13 ± 10 | 8 ± 13 | 1.09 | 1.2 |
| UOC-T1 | 1 | $6.29 \pm 0.74 \times 10^{-4}$ | $1.67 \pm 0.13 \times 10^{-3}$ | 4 ± 16 | 12 ± 15 | -30 ± 14 | 10 ± 14 | 1.00 | 1.4 |

[a] Inferred by interpolation of dredge-up models of Boothroyd and Sackmann (1999)

Table 4: Calcium isotopic compositions of presolar hibonite grains. Normalization ratios for δ-values are given in parenthesis.

| Grain | $\delta^{42}Ca/^{40}Ca \pm 1\sigma$ (0.00662121) | $\delta^{43}Ca/^{40}Ca \pm 1\sigma$ (0.00137552) | $\delta^{44}Ca/^{40}Ca \pm 1\sigma$ (0.021208) | $^{41}Ca/^{40}Ca \pm 1\sigma$ |
|---|---|---|---|---|
| KH2 | -38 ± 8 | -42 ± 20 | 58 ± 8 | $4.27 \pm 0.07 \times 10^{-4}$ |
| KH4 | 1 ± 7 | -21 ± 18 | 11 ± 8 | $< 9.7 \times 10^{-7}$ |
| KH7 | 3 ± 6 | 7 ± 16 | 25 ± 7 | $< 9.3 \times 10^{-7}$ |
| KH8 | 36 ± 13 | 29 ± 29 | 89 ± 10 | $< 4.2 \times 10^{-6}$ |
| KH9 | -7 ± 7 | 14 ± 19 | 31 ± 8 | $< 2.0 \times 10^{-6}$ |
| KH10 | 249 ± 12 | 118 ± 27 | 31 ± 9 | $1.8 \pm 0.3 \times 10^{-5}$ |
| KH11 | 32 ± 7 | 54 ± 18 | 70 ± 8 | $< 4.0 \times 10^{-6}$ |
| KH12 | 42 ± 15 | 74 ± 34 | 45 ± 11 | $1.6 \pm 0.3 \times 10^{-5}$ |
| KH13 | -40 ± 6 | 22 ± 16 | 141 ± 8 | $3.6 \pm 0.1 \times 10^{-5}$ |
| KH14 | 22 ± 9 | 27 ± 21 | 8 ± 8 | $2.09 \pm 0.05 \times 10^{-4}$ |
| KH15 | 8.7 ± 7 | -1 ± 18 | 15 ± 8 | $1.02 \pm 0.02 \times 10^{-4}$ |
| KH16 | 45 ± 27 | -11 ± 46 | -26 ± 31 | $2.07 \pm 0.12 \times 10^{-4}$ |
| KH17 | 70 ± 25 | -2 ± 40 | -25 ± 31 | $5.7 \pm 0.5 \times 10^{-5}$ |
| KH18 | -18 ± 20 | -8 ± 31 | -5 ± 30 | $1.40 \pm 0.14 \times 10^{-5}$ |
| KH19 | 8 ± 24 | -42 ± 40 | -6 ± 31 | $<6.0 \times 10^{-6}$ |
| KH21 | 3 ± 20 | -23 ± 28 | -20 ± 30 | $1.28 \pm 0.03 \times 10^{-4}$ |

Table 5: Details of supernova mixing models. Based on $15M_\odot$ supernova nucleosynthesis model of Rauscher et al. (2002). The bottom row indicates C/O ratio of the three mixtures.

| Zone | Mass Range[a] ($M_\odot$) | Mix Fraction[b] | | |
|---|---|---|---|---|
| | | KH2 | UOC-S3 | T22 |
| Ni | 1.683 - 1.785 | 0.49 | 0.00 | 0.00 |
| Si/S | 1.785 - 1.920 | 0.09 | 0.01 | 0.00 |
| O/Si | 1.920 - 2.200 | 0.38 | 0.04 | 0.00 |
| O/Ne | 2.200 - 2.630 | 0.00 | 0.00 | 0.19 |
| O/C | 2.630 - 3.050 | 0.00 | 0.00 | 0.06 |
| He/C | 3.050 - 3.800 | 1.07 | 1.09 | 0.93 |
| He/N | 3.800 - 4.233 | 4.99 | 5.82 | 0.00 |
| H envelope | 4.233 - 12.612[c] | 92.98 | 93.03 | 98.82 |
| C/O ratio | | 0.26 | 0.35 | 0.31 |

[a]Range of radial mass in supernova over which zone is defined.
[b]Mixing mass fraction (in %) of each zone that reproduces isotopic composition of given presolar grain.
[c]About 2.4 $M_\odot$ was lost by stellar winds prior to the supernova explosion.